\documentclass[11pt,a4paper]{article}
\usepackage[T1]{fontenc}
\usepackage[utf8]{inputenc}
\usepackage{tikz}
\usepackage{color}
\usepackage{amsmath,empheq}

\usetikzlibrary{arrows.meta,calc,positioning}

\makeatletter

\def\cN{{\cal N}}
\def\cO{{\cal O}}

\def\cC{{\cal C}}

\newcommand{\be}{\begin{equation}}
\newcommand{\ee}{\end{equation}}
\newcommand{\ba}{\begin{aligned}}
\newcommand{\ea}{\end{aligned}}
\newcommand{\bea}{\begin{eqnarray}}
\newcommand{\eea}{\end{eqnarray}}

\newcommand{\mathd}{\mathrm{d}}
\newcommand{\mathe}{\mathrm{e}}
\newcommand{\mathi}{\mathrm{i}}

\newcommand{\e}{\epsilon}
\newcommand{\lam}{\lambda}
\newcommand{\BV}{\mathbb{V}}

\newcommand{\BC}{\mathbb{C}}

\newcommand{\dvol}{\delta\mathrm{Vol}}

\newcommand{\ccz}{\widetilde{L^{020}}}
\newcommand{\cczp}{\widetilde{L^{020}}'}
\newcommand{\fccz}{\mathrm{f}\widetilde{L^{020}}}
\newcommand{\fcczp}{\mathrm{f}\widetilde{L^{020}}'}

\usepackage{comment}

\pdfoutput=1

\usepackage{jheppub}

\usepackage{braketmod}

\global\long\def\bbra#1{\Bbra{#1}}%
\global\long\def\ket#1{\Ket{#1}}%
\global\long\def\kket#1{\Kket{#1}}%
\global\long\def\braket#1{\Braket{#1}}%
\global\long\def\bbraket#1{\Bbraket{#1}}%
\global\long\def\brakket#1{\Brakket{#1}}%
\global\long\def\bbrakket#1{\Bbrakket{#1}}%

\makeatletter
\gdef\@fpheader{}
\makeatother

\begin{document}
\title{$S^3$ partition functions and Equivariant CY$_4$/CY$_3$ correspondence from Quantum curves}

\author[a,b]{Kiril Hristov,}
\author[c]{Naotaka Kubo}
\author[d,e]{and Yi Pang}

\affiliation[a]{Faculty of Physics, Sofia University, J. Bourchier Blvd. 5, 1164 Sofia, Bulgaria}
\affiliation[b]{INRNE, Bulgarian Academy of Sciences, Tsarigradsko Chaussee 72, 1784 Sofia, Bulgaria}
\affiliation[c]{RIKEN Center for Interdisciplinary Theoretical and Mathematical Sciences (iTHEMS), RIKEN, Wako 351-0198, Japan}
\affiliation[d]{Center for Joint Quantum Studies and Department of Physics, School of Science,\\ Tianjin University, Tianjin 300350, China}
\affiliation[e]{Peng Huanwu Center for Fundamental Theory, Hefei, Anhui 230026, China}

\emailAdd{khristov@phys.uni-sofia.bg}
\emailAdd{naotaka.kubo@riken.jp}
\emailAdd{pangyi1@tju.edu.cn}

\preprint{KUNS-3094, USTC-ICTS/PCFT-26-17 }

\note{For N.K.: Work performed at the Department of Physics, Kyoto University.}

\abstract{
We study the perturbative large-$N$ expansion of the round three-sphere partition function in a class of M2-brane theories, including flavored SYM and ABJM theories as well as more general 3d theories admitting dual $(p,q)$ 5-brane web descriptions. 
Using the Fermi gas formalism and quantum curve techniques, we derive the Airy-function representation of the partition function and find exact agreement with predictions based on equivariant constant maps in topological string theory proposed in \cite{Cassia:2025jkr}. 
In particular, we provide affirmative tests of this proposal for the toric geometries $\mathbb{C} \times \cC$ (the conifold), the cone over the Sasakian space $Q^{1,1,1}$, and $\mathbb{C} \times \mathrm{SPP}$ (the suspended pinch point). 
Motivated by a recent conjecture in \cite{Kubo:2025jxi}, we further propose a novel equivariant correspondence between distinct toric Calabi–Yau manifolds of the form $\mathrm{CY}_4 \leftrightarrow \mathbb{C} \times \mathrm{CY}_3$, arising from relations between the corresponding quantum curves under specific constraints. 
This correspondence suggests an equivariant extension and points toward a geometric origin of the topological string/spectral theory (TS/ST) correspondence, while offering new insight into the structure of the holographic duality.}

\setcounter{tocdepth}{2}
\maketitle \flushbottom

\section{Introduction}\label{sec:Introduction}

One of the most profound insights of string theory is the AdS/CFT correspondence \cite{Maldacena:1997re}, which proposes a duality between gravitational theories in anti-de Sitter (AdS) space and conformal field theories (CFTs) defined on their boundaries. A particularly rich setting for precision tests of this duality is the AdS$_4$/CFT$_3$ correspondence, realized in Chern–Simons–matter theories such as ABJM \cite{Aharony:2008ug} and their M-theory duals. The ability to perform exact computations in these models arises from the method of supersymmetric localization \cite{Nekrasov:2002qd,Pestun:2007rz,Kapustin:2009kz,Pestun:2016zxk}, which employs tools from equivariant cohomology \cite{Duistermaat:1982vw,Berline1982classes,Atiyah:1984px} to reduce infinite-dimensional path integrals to finite-dimensional matrix integrals.

This formalism has enabled the exact computation of a variety of supersymmetric observables on the CFT$_3$ side. Landmark results include the round and squashed three-sphere partition functions, Wilson loop expectation values, and the superconformal and topologically twisted indices \cite{Kapustin:2009kz, Kim:2009wb, Drukker:2010nc, Jafferis:2010un, Hama:2010av, Imamura:2011su, Hama:2011ea, Benini:2015noa, Benini:2015eyy}. These observables serve as benchmarks for holography, allowing highly non-trivial tests of the dual gravitational description.

In particular, the large-$N$ limit of the sphere partition function reproduces the celebrated $N^{3/2}$ scaling of the holographic free energy on AdS$_4$ \cite{Drukker:2010nc, Herzog:2010hf, Marino:2011eh, Santamaria:2010dm, Jafferis:2011zi,Gulotta:2011si, Amariti:2012tj}, and the twisted index precisely captures the entropy of AdS$_4$ black holes \cite{Benini:2015eyy,Benini:2016rke}. Beyond leading order, the correspondence extends to perturbative $1/N$ corrections, \cite{Bobev:2020egg,Bobev:2022jte,Hristov:2022lcw}, and non-perturbative effects \cite{Gautason:2023igo,Beccaria:2023ujc,Gautason:2025per,Gautason:2025plx,vanMuiden:2026nsp}, where worldsheet and membrane instantons in M-theory match exactly the structure of the resummed CFT partition function \cite{Drukker:2010nc, Hatsuda:2012dt, Marino:2011eh, Calvo:2012du, Honda:2014ica, Grassi:2014cla}. These results stand among the most precise and quantitative confirmations of holography to date.

While the aforementioned results already constitute remarkable tests of AdS/CFT and shed light on the structure of quantum gravity, it is desirable to place AdS$_4$/CFT$_3$ on an even firmer footing regarding supersymmetric observables. Ideally, one would like to develop a comprehensive framework encompassing all supersymmetric instances of holographic duality, with exact matches of BPS observables across the correspondence. Such a framework would provide unprecedented control over quantum gravity in sectors protected by supersymmetry.

We believe this ambitious goal is within reach, thanks to several converging lines of progress that extend the scope of holography through alternative dual descriptions. Central to the present work are three key ingredients: 
\begin{itemize}
\item The $(p,q)$ 5-brane web constructions \cite{Aharony:1997ju, Aharony:1997bh, Leung:1997tw} offer a geometric engineering framework for supersymmetric gauge theories.
Furthermore, inserting probe D3-branes into the $(p,q)$ 5-brane web induces 3d $\mathcal{N}=2$ Chern–Simons–matter quiver theories \cite{Hanany:1996ie, Aharony:1997ju, Bergman:1999na, Jensen:2009xh, Cremonesi:2010ae}. 
A complementary realization arises in M-theory via compactification on Calabi–Yau fourfold cones over Sasaki–Einstein seven-manifolds (SE$_7$), which also engineer similar quiver gauge theories \cite{Martelli:2008rt, Jafferis:2008qz, Hanany:2008cd,Ueda:2008hx,Hanany:2008fj,Franco:2008um,Benini:2009qs, Jafferis:2009th, Amariti:2012tj}. Both descriptions rely on the development of toric geometry and toric diagrams, as we explain in due course.

\item The quantum curve perspective on the three-sphere partition function was initially realized via the Fermi-gas formalism 
\cite{Marino:2011eh} 
and further developed through its deep connection to the topological string/spectral theory (TS/ST) correspondence \cite{Grassi:2014cla, Grassi:2014zfa,Marino:2015ixa,Kashaev:2015wia,Codesido:2015dia}, in which the partition function is interpreted as a spectral determinant of a quantum operator encoding the mirror curve of a toric Calabi–Yau (CY) threefold. In this framework, the quantization of mirror curves provides a bridge between exact results in supersymmetric gauge theories and topological string theory. Recent advances, such as 
\cite{Honda:2014npa,Drukker:2015awa,Kashaev:2015wia,Hatsuda:2015lpa,Nosaka:2015iiw,Kubo:2018cqw,Kubo:2019ejc,Kubo:2020qed,Kubo:2025jxi,Kubo:2021enh}, 
have further extended the role of quantum curves to broader classes of 3d gauge theories with intricate quiver structures.

\item In an a priori unrelated development, the equivariant extension of constant maps in topological string theory on toric Calabi–Yau fourfolds was proposed as a geometric dual to M2-brane partition functions \cite{Cassia:2025aus,Cassia:2025jkr}, building on earlier supergravity-based extremization principles \cite{Martelli:2005tp,Butti:2005vn,Butti:2005ps,Martelli:2006yb,Amariti:2011uw,Couzens:2018wnk,Gauntlett:2018dpc,Hosseini:2019use,Hosseini:2019ddy,Gauntlett:2019roi,Kim:2019umc,Martelli:2023oqk,Colombo:2023fhu}. Via a change of ensemble with respect to the redundant K\"ahler parameters in the geometric description, the topological string partition function was then proposed to agree with the dual (in general squashed) three-sphere partition function of the dual field theory.
\end{itemize}

Together, these developments point toward a unifying picture where supersymmetric observables in AdS$_4$/CFT$_3$ can be systematically understood and exactly computed.~\footnote{From this point onward, we adopt the notion of holography implicitly used in \cite{Martelli:2005tp,Butti:2005vn,Butti:2005ps,Martelli:2006yb,Amariti:2011uw,Couzens:2018wnk,Gauntlett:2018dpc,Hosseini:2019use,Hosseini:2019ddy,Gauntlett:2019roi,Kim:2019umc,Martelli:2023oqk,Colombo:2023fhu,Cassia:2025aus,Cassia:2025jkr}. In particular, we focus on the topology of the transverse space in the brane construction, rather than on the AdS factor itself. In practice, this means that our bulk analysis is carried out at the level of geometry, rather than full (super)gravity dynamics. We return to this point in the Discussion section.} We devote the next section to a more comprehensive review of these three main developments, which constitute our main calculational tools in this work.

In this paper we focus on the exact perturbative expansion in the gauge group rank $N$ of the round three-sphere partition function in the above setting, utilizing all these tools to test the geometric result of \cite{Cassia:2025aus,Cassia:2025jkr} on the field theory side for a number of new examples, as well as to propose a new equivariant CY$_4$/CY$_3$ correspondence. Below, we present a qualitative overview of these results before turning to the main body of the paper.

\subsection{Summary of results and propositions}

The results in the present work can be roughly divided into the following two (closely interrelated) topics, for which we briefly summarize the novel contributions.\footnote{
In this footnote, we briefly clarify the status of the statements made under these two topics. On the field-theory side we obtain the relevant quantum curves from the localized matrix models, following the construction of Ref.~\cite{Kubo:2025jxi} with the generalizations described in section \ref{sec:Review} and appendix \ref{sec:BCtoQC}, and use them to compute the Airy coefficients $C$ and $B$. On the geometric side, we evaluate the corresponding equivariant-volume and Chern-number data, while their relation to M2-brane partition functions is the holographic prescription proposed in \cite{Cassia:2025aus,Cassia:2025jkr} and is not derived here. Section \ref{sec:Examples} tests this prescription through exact algebraic agreement of $C$ and $B$ in the examples considered, under the stated parameter constraints, whereas section \ref{sec:correspondence} formulates a conjectural CY$_4$/CY$_3$ relation and tests it in four examples. Throughout this paper, ``exact'' refers to exact expressions and identities for $C$ and $B$ within the perturbative Airy sector, not to the full finite-$N$ partition function including the $N$-independent prefactor and exponentially suppressed corrections.}

\subsection*{$S^3$ partition function}

For three-dimensional SCFTs that admit a Fermi gas description \cite{Marino:2011eh}, it is known that, up to an overall constant prefactor of order $\cO(N^0)$ and exponentially suppressed non-perturbative corrections, the $S^3$ partition function assumes a universal Airy-function form,
\be
\label{eq:S3pf}
Z_{S^3}^\text{pert} (\Delta, N) \simeq {\rm Ai} \Big[ \left( C(\Delta) \right)^{-1/3} \left(N - B(\Delta) \right) \Big]\ ,
\ee
where $N$ denotes the rank of the gauge group and $\Delta$ collectively represents supersymmetric deformation parameters, such as complexified mass parameters or trial R-charge assignments. The functions $C(\Delta)$ and $B(\Delta)$ encode detailed information about the underlying SCFT and fully determine the perturbative $1/N$ expansion in the large-$N$ limit.

In this work, we extend the quantum curve formalism to a new class of three-dimensional $\cN=2$ quiver gauge theories, which can be viewed as flavored generalizations of ABJM and SYM theories. Within this framework, we explicitly derive the Airy-function structure \eqref{eq:S3pf} and compute the associated functions $C(\Delta)$ and $B(\Delta)$ from first principles, building on recent progress in \cite{Kubo:2025jxi}.

Our results provide a non-trivial test of the AdS$_4$/CFT$_3$ correspondence for this class of theories. On the gravity side, we reproduce the same Airy-function form of the partition function, including the exact coefficients $C$ and $B$, by computing the equivariant volume of the Calabi–Yau cone over the corresponding Sasaki–Einstein seven-manifold ($SE_7$). This bulk computation follows the equivariant extension of topological string constant map contributions developed in \cite{Cassia:2025aus,Cassia:2025jkr}. From the geometric perspective, the deformation parameters $\Delta_i$ are naturally identified with equivariant parameters $\e_i$ associated with the toric action on the Calabi–Yau manifold,
\be
	\Delta_i = \e_i\ , \qquad \sum_i \Delta_i = \sum_i \e_i = 2\ ,
\label{eq:eqParaCond}
\ee
with the latter condition enforced by supersymmetry.

For the theories considered here, the applicability of the Fermi gas approximation and the quantum curve construction requires additional constraints on the parameters $\Delta_i$. While these restrictions depend on the specific field theory and will be discussed in detail later (see section \ref{subsec:BC}), they admit a simple geometric interpretation in terms of the structure of the underlying CY$_4$ toric diagram, see the left panel of figure \ref{fig:CY43-ABJM}. In all examples studied in this work, the toric diagram consists of two layers, which we may place at $z=0$ and $z=1$ without loss of generality. The corresponding constraint on the equivariant parameters, and hence on the R-charges of the dual field theory, takes the form
\begin{equation}
\sum_{i\in P_{z=0}}\epsilon^{(z=0)}_{i}=\sum_{j \in P_{z=1}}\epsilon^{(z=1)}_{j}=1\ ,
\label{eq:CY43-Rcond}
\end{equation}
where $P_{z=0}$ and $P_{z=1}$ denote the two distinct $(x,y)$ planes at fixed $z$, and $\e^{(z)}$ denote the associated equivariant parameters. While these constraints clearly imply \eqref{eq:eqParaCond}, they are more restrictive. Therefore the quantum curve construction, which is central in this work, corresponds to a one-parameter restriction with respect to the most general set of R-charges of the 3d SCFT and, in turn, equivariant parameters on the toric geometry side.

Under these conditions, we find precise agreement between the quantum curve computation of the $S^3$ partition function and the geometric predictions of \cite{Cassia:2025aus,Cassia:2025jkr} for several non-trivial backgrounds corresponding to flavored SYM and ABJM theories. These include $\mathbb{C} \times \cC$, with $\cC$ the conifold; the cone over $Q^{1,1,1}$; and $\mathbb{C} \times \mathrm{SPP}$, the suspended pinch point. Compared to the simpler case of $\mathbb{C}^4$ and its orbifolds, these geometries exhibit qualitatively new features, which we analyze in detail in the remainder of the paper.

\subsection*{Equivariant CY$_4$ / CY$_3$ correspondence}

\begin{figure}
\begin{centering}
\includegraphics[scale=0.5]{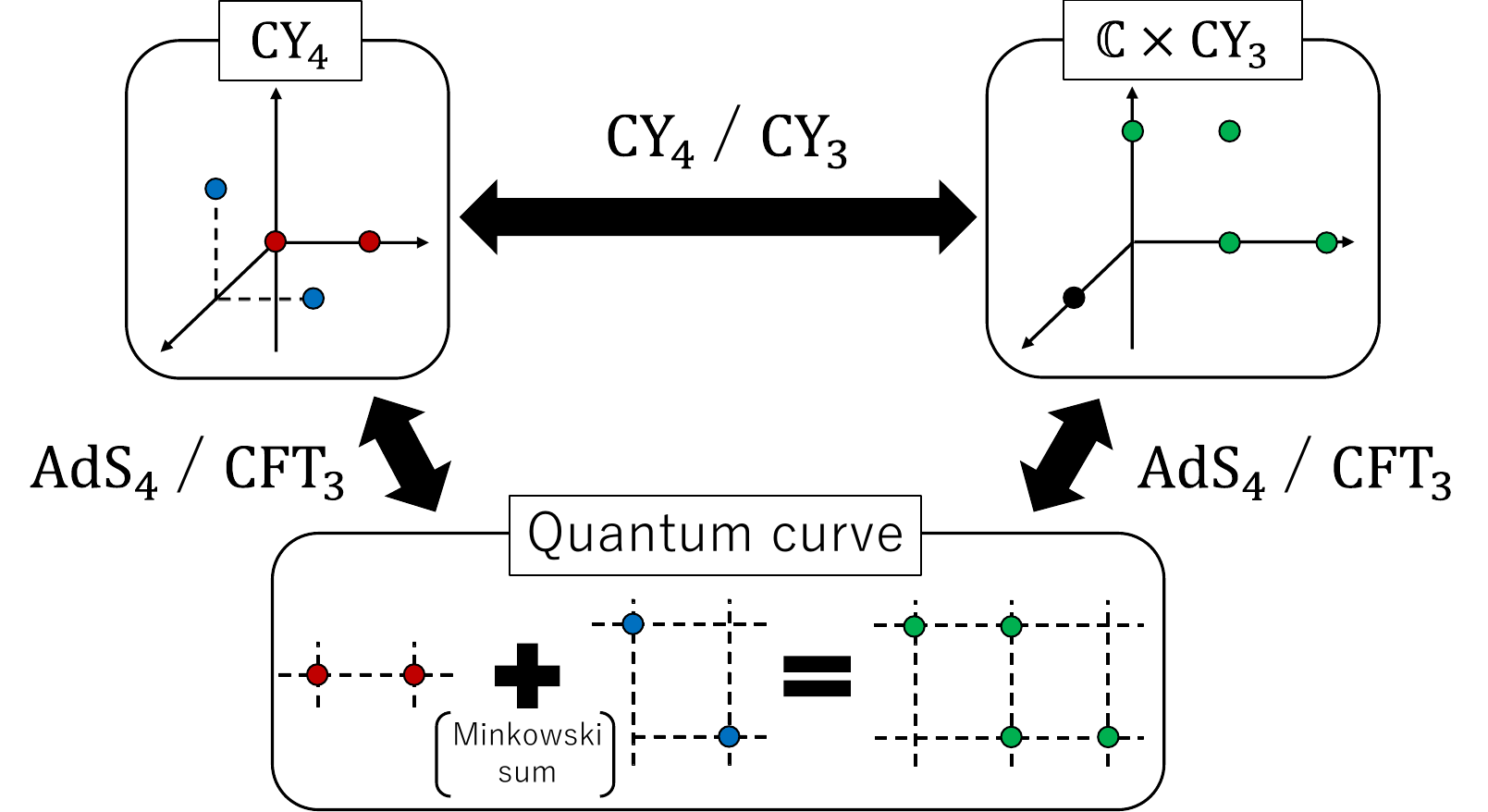}
\par\end{centering}
\caption{
An example of the conjectural CY$_{4}$/CY$_{3}$ correspondence. The Minkowski sum of the two Newton polygons in the curve connects two Calabi-Yau manifolds given by their respective toric diagrams (upper left and upper right).
}\label{fig:CY43-ABJM}
\end{figure}

In studying the AdS$_4$/CFT$_3$ correspondence through the computation of the functions $C$ and $B$ appearing in \eqref{eq:S3pf}, we uncover a non-trivial identification between brane configurations and Calabi–Yau geometries. Combining this observation with the quantum curve description—most notably through the use of the Wigner transform and the Minkowski sum—we identify an emergent structure that links pairs of toric diagrams in a precise manner. A representative example, corresponding to the ABJM theory at level $k=1$, is shown in figure \ref{fig:CY43-ABJM}.

These observations motivate a conjectural correspondence between toric Calabi–Yau manifolds of the form $\mathrm{CY}_4 \leftrightarrow \mathbb{C}\times\mathrm{CY}_3$, structured by the Minkowski sum of toric diagrams and accompanied by a non-trivial mapping between the equivariant parameters $\e$ of the CY$_4$ geometry and $\tilde\e$ of the $\mathbb{C}\times\mathrm{CY}_3$ geometry. From the field-theoretic perspective, this correspondence translates into a map between supersymmetric deformation parameters of distinct brane web configurations and their associated three-dimensional theories, which need not admit a Lagrangian description.

At the level of the three-sphere partition function, this correspondence manifests itself through the relations
\be
\label{eq:correspondence1}
	C_{\mathrm{CY}_4} (\e) = C_{\mathbb{C} \times \mathrm{CY}_3} (\tilde \e)\ ,
\ee
as well as
\be
\label{eq:correspondence2}
	B_{\mathrm{CY}_4} (\e) - B_{\mathbb{C} \times \mathrm{CY}_3} (\tilde \e) = const\ ,
\ee
where the explicit mapping between $\e$ and $\tilde\e$ (both subject to \eqref{eq:CY43-Rcond}) will be presented in section \ref{sec:correspondence}, after introducing the relevant geometric framework. In that section, we also clarify the physical interpretation of the constant appearing on the right-hand side of \eqref{eq:correspondence2}.

We test this proposal geometrically across a range of representative examples and consistently find precise agreement with both \eqref{eq:correspondence1} and \eqref{eq:correspondence2}. This matching provides strong evidence that the three-sphere partition functions of the paired theories are related at the perturbative level, and it suggests that an exact non-perturbative correspondence may also hold. More broadly, these results point toward a deeper equivalence between the BPS sectors of the two theories. This expectation is further supported by the conjecture of \cite{Cassia:2025aus, Cassia:2025jkr}, which proposes that equivariant volumes govern a wider class of supersymmetric observables, including the superconformal and topologically twisted indices and their spindle generalizations \cite{Inglese:2023wky, Colombo:2024mts}.

\subsection*{Plan of the paper}
The paper is structured as follows.
In section \ref{sec:Review} we provide an extensive review of the main ingredients relevant to our work, as well as the intricate connections between them. While none of this material is new per se, we present it in a way that is tailored to and motivated by our main results. In particular, we describe the brane setup and the resulting holographic picture in terms of 3d quiver theories, and we discuss in detail the two principal computational tools used throughout the paper: the quantum curve description of the quivers and the equivariant volume of the transverse geometry.

In section \ref{sec:Examples} we present our main new results. These concern the round-sphere partition function at finite $N$,
more specifically its perturbative Airy coefficients $C$ and $B$, for a number of new examples: M2-branes on $\mathbb{C} \times \cC$ (the conifold), the cone over the Sasakian space $Q^{1,1,1}$, and $\mathbb{C} \times \mathrm{SPP}$ (the suspended pinch point). 
In each case, we compute these coefficients independently using the field-theory quantum curve and the geometric prescription, and find exact agreement under the stated parameter constraints.

We then turn, in section \ref{sec:correspondence}, to a more careful formulation of the proposed equivariant CY$_4$/CY$_3$ correspondence. This proposal is further supported by a series of detailed examples of dual pairs, which illustrate the scope and robustness of the correspondence.

Finally, in section \ref{sec:Discussion}, we conclude with a broader discussion of the relation between the present work and the TS/ST and AdS/CFT correspondences. Several technical appendices are included, providing both established and novel supplementary material that supports various parts of our analysis and is referenced at the appropriate points in the main text.

The appendices are organized as follows. In appendix \ref{sec:BCtoQC} we collect the information needed to compute quantum curves from the $S^3$ partition function. In appendix \ref{sec:Poly-Web} we summarize the relevant properties of polygons and webs, with particular emphasis on those used in this paper. In appendix \ref{sec:WT} we review the Wigner transform. In appendix \ref{app:An equivariant duality} we present several observations on non-geometric phases.

\section{From brane webs to quantum curves and equivariant geometry}\label{sec:Review}

\begin{figure}[ht]
\begin{tikzpicture}[scale=1, every node/.style={scale=.9}]

	\draw[fill = white,thick, rounded corners, draw = black] (-1.9,-.75) rectangle (0.7,.75);
	\node at (-0.6,0.4){\large D3-branes};
	\node at (-0.57,0){\large on $(p,q)$ };
\node at (-0.6,-0.4){$5$-brane webs};
	
	\draw[fill = white,thick, rounded corners, draw = black] (2.8,-0.75) rectangle (5.4,.75);
\node at (4.1,0.4){\large M2-branes};
\node at (4.1,0){\large on toric};
\node at (4.1,-0.4){\large CY$_4$ singularity};

	\draw[fill = white,thick, rounded corners, draw = black] (6.5,-3.75) rectangle (9.2,-2.25);
	\node at (7.85,-2.6){Matrix model/};
	\node at (7.85,-3){Fermi gas/};
	\node at (7.85,-3.4){Quantum curve};

	\draw[fill = white, thick, rounded corners, draw = black] (7.85,-.75) rectangle (10.15,.75);
	\node at (9,0.4){equivariant};
	\node at (9,-0){top. string};
	\node at (9,-0.4){on toric CY$_4$};

	\draw[<->,>=stealth, black, ultra thick] (.8,0) -- (2.7,0);
    \node at (1.7,0.3){T-duality};
	\node at (1.7,-0.3){and uplift};

	\draw[->,>=stealth, black, ultra thick] (3.5,-3) -- (6.4,-3);
	\node at (4.9,-3.2){on $S^3$};
    \node at (4.9,-2.7){localization};

	\draw[<->,>=stealth, black, ultra thick] (9.25,-2.8) -- (10.8,-2);
	\node at (10.25,-2.8){TS/ST};

	\draw[fill = white,  thick, rounded corners, draw = black] (0.2,-3.75) rectangle (3.4,-2.4);
	\node at (1.8,-2.8){mirror dual pair of};
	\node at (1.8,-3.3){3d $\cN =4$ SCFTs};

	\draw[fill = white,thick, rounded corners, draw = black] (10.85,-2.55) rectangle (12.95,-1.05);
	\node at (11.9,-1.4){(equivariant)};
	\node at (11.9,-1.8){top. string};
	\node at (11.9,-2.2){on toric CY$_3$};

	\draw[<->,>=stealth, red, ultra thick] (10.25,0.) -- (11.5,-1);

	\draw[->,>=stealth, black, ultra thick] (3.2,-0.9) -- (2.1,-2.1);
	\draw[->,>=stealth, black, ultra thick] (-0.5,-0.9) -- (0.9,-2.1);
	\node at (1.5,-1.5){decoupling};
	\node at (1.5,-1.9){limits};

	\draw[<->,>=stealth, black, ultra thick] (8.9,-0.8) -- (7.7,-2.2);	
    \node at (7.35,-1.45){holography};

	\draw[->,>=stealth, black, ultra thick] (5.45,0) -- (7.75,0);
	\node at (6.6,0.3){bulk limit};
    \node at (6.6,-0.3){(geometry)};				

	\node at (11.25,-0.3){\color{red} here};

    \draw[red, dashed, line width=0.5pt] (9.65,-1.55) circle (3.25cm);

	\end{tikzpicture}
	\caption{Schematic diagram of the duality web we explore. Apart from the red arrow, which we discuss here for the first time, we establish the holographic relations within the dashed red circle for several new non-trivial examples of dual pairs.}
	\label{fig:1}
\end{figure}
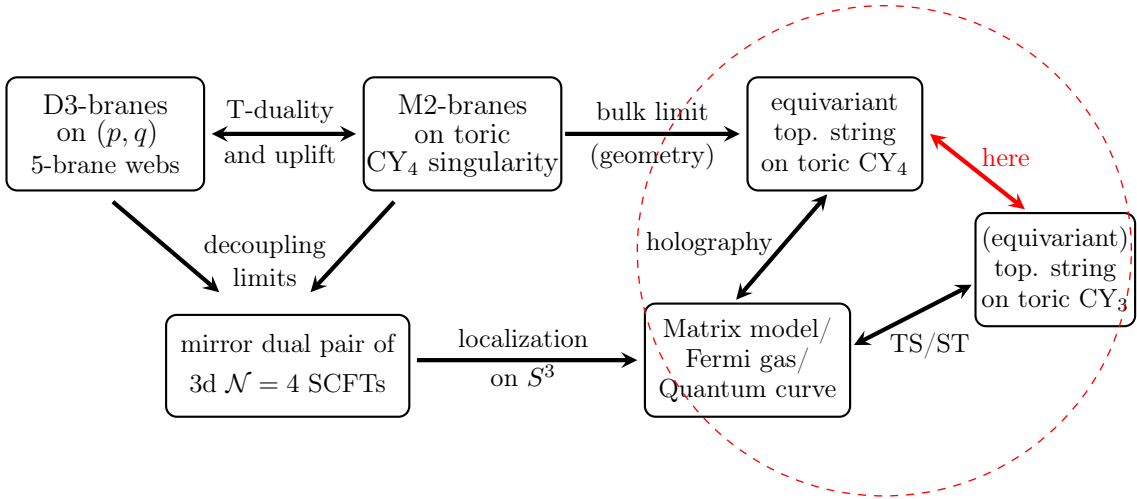

In this section we introduce the main setup and methodology underlying our analysis, which is also summarized schematically in figure \ref{fig:1}. Our aim is to present the essential ideas in a clear and logically self-contained manner, while keeping the exposition as rigorous as possible. More technical details and auxiliary derivations, which may obscure the main line of reasoning, are deferred to the appendices.

This section serves three purposes. First, we review the brane constructions and the associated quantum-curve description of the $S^3$ partition function, culminating in the main formulas \eqref{eq:VolForm}-\eqref{eq:VolC-K12}. Second, we summarize the geometric interpretation of the Airy-function data in terms of equivariant volumes of toric CY$_4$ manifolds, leading to the results \eqref{eq:Cgeometry}-\eqref{eq:Bgeometry}. Third, we prepare the ground for the CY$_4$/CY$_3$ correspondence introduced in Section 4 by highlighting the structural features that will be essential later.

\subsection{Brane setup and quiver theory}
\label{subsec:BC}
As announced previously, we are interested in M2-branes probing various geometries. Instead of working directly with the M-theory description, one can obtain equivalent brane configurations in type IIB string theory by performing a dimensional reduction followed by a $T$-duality. The resulting configurations are of the type originally introduced by Hanany and Witten \cite{Hanany:1996ie}, namely D3-branes suspended on $\left(p,q\right)$ webs. In this picture, M2-branes map to D3-branes extended along the directions $0126$, with the 6-direction compactified, while the geometries probed by the M2-branes are replaced by 5-branes in type IIB \cite{Hanany:1996ie, Aharony:1997ju, Bergman:1999na,Aharony:2008ug, Jensen:2009xh, Cremonesi:2010ae}.

Let us focus on the NS5-branes, D5-branes, and their bound states, the $\left(p,q\right)$5-branes. Here, the $\left(1,0\right)$5-brane corresponds to the NS5-brane, while the $\left(0,1\right)$5-brane corresponds to the D5-brane, with $p$ and $q$ coprime integers. We consider two types of $\left(p,q\right)$5-branes, distinguished by their orientations. In our conventions, a $\left(p,q\right)$5-brane extends along the directions $01234\left[5,9\right]_{\theta}$, while a $\left(p,q\right)$5$'$-brane extends along $012\left[3,7\right]_{\varphi}\left[4,8\right]_{\varphi}\left[5,9\right]_{\theta}$ with $\varphi \neq 0$. To preserve at least 3d $\mathcal{N}=2$ supersymmetry, the angle $\theta$ must satisfy $\tan\theta = q/p$. A summary of the brane configurations is provided in table \ref{tab:Brane}. In the following, we will mainly work with the $\left(p,q\right)$5-brane.

An important feature that allows the construction of various geometries is that 5-branes can form $\left(p,q\right)$ webs. All $\left(p,q\right)$5-branes share the directions 01234 as well as the 59-plane. Two such 5-brane bound states can meet at a junction and combine into a single web, provided that the charges are conserved at the junction. Since the angle $\theta$ of a $\left(p,q\right)$5-brane must always satisfy $\tan\theta = q/p$, the forces are automatically balanced at each junction, ensuring the stability of the web.

\begin{table}
\begin{centering}
\begin{tabular}{|c|c|c|c|c|c|c|c|c|}
\hline 
 & 012 & 3 & 4 & 5 & 6 & 7 & 8 & 9\tabularnewline
\hline 
\hline 
D3 & $\bigcirc$ &  &  &  & $\bigcirc$ &  &  & \tabularnewline
\hline 
$\left(p,q\right)$5 & $\bigcirc$ & $\bigcirc$ & $\bigcirc$ & $/_{\theta}^{59}$ &  &  &  & $/_{\theta}^{59}$\tabularnewline
\hline 
$\left(p,q\right)$5$'$ & $\bigcirc$ & $/_{\varphi}^{37}$ & $/_{\varphi}^{48}$ & $/_{\theta}^{59}$ &  & $/_{\varphi}^{37}$ & $/_{\varphi}^{48}$ & $/_{\theta}^{59}$\tabularnewline
\hline 
\end{tabular}
\par\end{centering}
\caption{The brane setup for 3d ${\cal N}=2$ Chern-Simons theories. The direction 6 is periodic. $/_{\theta}^{xy}$ means that a 5-brane has an angle $\theta$ in the $xy$ plane, where $\tan\theta=q/p$. For example, an NS5-brane extends in the directions 012345 while a D5-brane extends in the directions 012349. }\label{tab:Brane}
\end{table}

\begin{figure}
\begin{centering}
\includegraphics[scale=0.6]{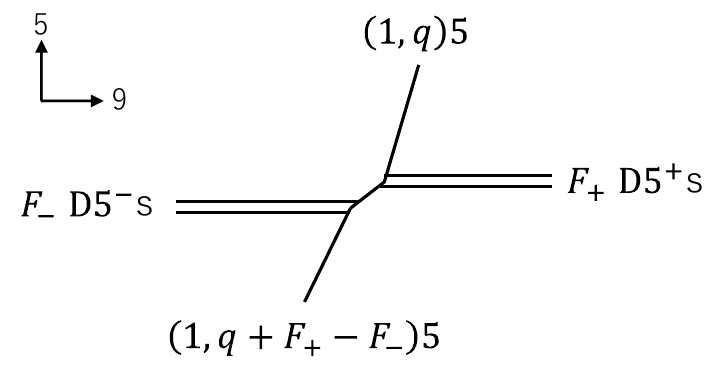}
\par\end{centering}
\caption{
A $\left(p,q\right)$ web which leads to a Lagrangian theory.
Here $\mathrm{D5}^{\pm}$ denotes a D5-brane extending in the $\pm$9 direction.
}\label{fig:BW-gen}
\end{figure}
$\left(p,q\right)$ webs of the type depicted in figure \ref{fig:BW-gen} play a particularly important role. The brane configuration consisting solely of these $\left(p,q\right)$ webs, together with isolated D5-branes, gives rise to Lagrangian theories. We represent the $\left(p,q\right)$ web in figure \ref{fig:BW-gen} and the isolated D5-branes as
\begin{equation}
\left[\left(1,q\right)+F_{-}\mathrm{D5}^{-}+F_{+}\mathrm{D5}^{+}\right],\quad\left[\left(0,1\right)\right]\ .
\label{eq:BW-gen}
\end{equation}
Further, we use the notation for a whole brane configuration 
\begin{equation}
\mathsf{w}_{1}-\mathsf{w}_{2}-\cdots\mathsf{w}_{R}-_{\mathrm{p}}\ ,
\label{eq:BC-gen}
\end{equation}
where $\mathsf{w}_{r}$ are the brane webs, with subscript $\mathrm{p}$ emphasizing that direction 6 is periodic.

In this work we focus on the following two types of brane configurations:\footnote{
Comparing with the sign-based notation in \eqref{eq:BW-gen}, the
flavored SYM web has $F_{+}=F_{2}$ and $F_{-}=F_{1}$. For the first
and second webs in \eqref{eq:fABJM-BC}, the corresponding pairs are
$(F_{+},F_{-})=(\tilde F_{1},F_{1})$ and
$(F_{2},\tilde F_{2})$, respectively.
}
\begin{equation}
\text{flavored SYM: }
\left[\left(1,q\right)+F_{1}\mathrm{D5}^{-}+F_{2}\mathrm{D5}^{+}\right]-F_{3}\left[\left(0,1\right)\right]-_{\mathrm{p}}\ ,
\label{eq:fSYM-BC}
\end{equation}
and
\begin{equation}
\text{flavored }\ccz\text{: }
\left[\left(1,q_1\right)+F_{1}\mathrm{D5}^{-}+\tilde{F}_{1}\mathrm{D5}^{+}\right]-\left[\left(1,q_2\right)+\tilde{F}_{2}\mathrm{D5}^{-}+F_{2}\mathrm{D5}^{+}\right]-_{\mathrm{p}}\ .
\label{eq:fABJM-BC}
\end{equation}
As will be explained immediately below, the labels show their worldvolume theories.
We will also use two types of brane configurations that are related to \eqref{eq:fABJM-BC}.
The first brane configuration is
\begin{equation}
\text{ flavored ABJM: }
\left[\left(1,q_1\right)+F_{1}\mathrm{D5}^{-}+\tilde{F}_{1}\mathrm{D5}^{+}\right]-\left[\left(1,q_2\right)+\tilde{F}_{2}\mathrm{D5}^{-}+F_{2}\mathrm{D5}^{+}\right]'-_{\mathrm{p}}\ ,
\label{eq:fABJM-BC2}
\end{equation}
which differs from \eqref{eq:fABJM-BC} by the second web being 5' instead of 5. 
The second brane configuration is
\begin{equation}
\text{flavored }\cczp\text{: }
\left[\left(1,q_{1}\right)+\left(1,q_{2}\right)+\left(F_{1}+\tilde{F}_{2}\right)\mathrm{D5}^{-}+\left(\tilde{F}_{1}+F_{2}\right)\mathrm{D5}^{+}\right]-_{\mathrm{p}}\ ,
\label{eq:fABJM-BC0}
\end{equation}
which is obtained by combining the two $\left(p,q\right)$ webs in \eqref{eq:fABJM-BC}.
Although we label this brane configuration the flavored $\cczp$ theory, it gives rise to a non-Lagrangian 3d theory.

The above setup allows us to describe the brane dynamics in the two standard limits: a superconformal field theory living on the suspended D3-branes, and a bulk gravitational theory exhibiting an AdS$_4$ factor. We first discuss the 3d field theories.

\subsubsection*{3d theories}\label{subsec:3dThs}

The worldvolume theories of the D3-branes become 3d supersymmetric Chern-Simons (CS) theories. A D3-segment stretched between two $\left(1,q_{i}\right)$5-branes ($i=1,2$) contributes a $\mathrm{U}\left(N\right)$ factor to the gauge group. In terms of $\mathcal{N}=2$ supersymmetry, this corresponds to a vector multiplet and an adjoint hypermultiplet. The CS level $k$ is given by $k=p_{1}-p_{2}$. An isolated D5-brane, on the other hand, contributes $\mathcal{N}=4$ fundamental matter, which decomposes into an $\mathcal{N}=2$ fundamental multiplet and an $\mathcal{N}=2$ anti-fundamental multiplet.

The $\left(p,q\right)$ webs give rise to various quiver theories. In particular, consider adding D5-branes to a $\left(p,q\right)$5-brane. Unlike isolated D5-branes, D5-branes placed on top of a $\left(p,q\right)$5-brane introduce $\mathcal{N}=4$ fundamental matter to both $\mathrm{U}\left(N\right)$ factors due to flavor doubling \cite{Brodie:1997sz,Brunner:1998jr}. When a D5-brane forms a junction, it extends only along $\pm x^{9}$. In this case, this “half” D5-brane contributes an $\mathcal{N}=2$ fundamental matter to one $\mathrm{U}\left(N\right)$ factor and an $\mathcal{N}=2$ anti-fundamental to the other, while also reproducing $\pm 1/2$ CS levels.

\begin{figure}
\begin{centering}
\includegraphics[scale=0.5]{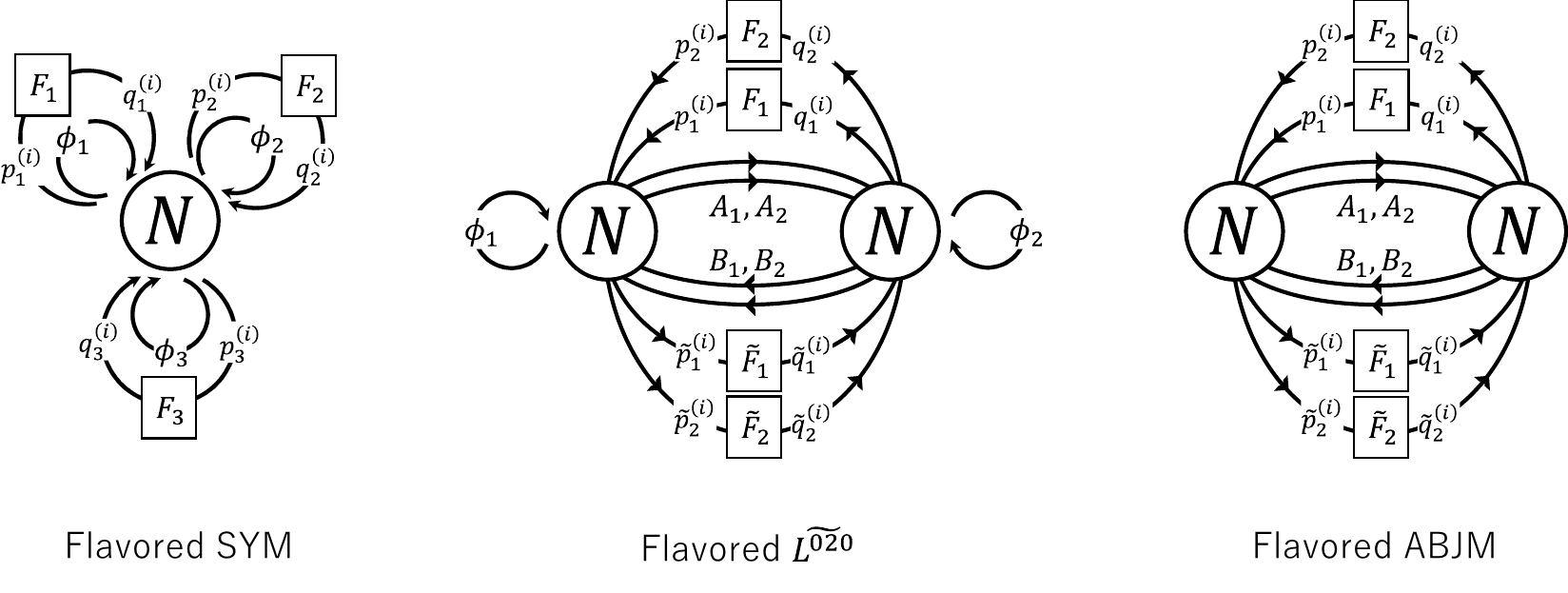}
\par\end{centering}
\caption{The $\mathcal{N}=2$ quiver diagrams. 
Left: the flavored SYM theory, which is the worldvolume theory of \eqref{eq:fSYM-BC}. 
Center: the flavored $\ccz$ theory, which is the worldvolume theory of \eqref{eq:fABJM-BC}.
Right: the flavored ABJM theory, which is the worldvolume theory of \eqref{eq:fABJM-BC2}.}
\label{fig:SYM_fABJM_Quiv}
\end{figure}

Let us consider the two examples corresponding to \eqref{eq:fSYM-BC} and \eqref{eq:fABJM-BC}. 
The brane configuration \eqref{eq:fSYM-BC} gives rise to the flavored version of the $\mathcal{N}=8$ Super-Yang-Mills (SYM) theory, while \eqref{eq:fABJM-BC} corresponds to the flavored $\ccz$ theory. 
The quiver diagrams for these theories are shown in figure \ref{fig:SYM_fABJM_Quiv}. 
In particular, the flavored $\ccz$ theory is the flavored version of the modified $L^{020}$ theory \cite{Berline1982classes,Hanany:2008cd,Hanany:2008fj}.
The flavored $\ccz$ theory features CS levels $\pm k$, where
\begin{equation}
k=q+\left(F_{1}-\tilde{F_{1}}\right)+\left(F_{2}-\tilde{F_{2}}\right)\ .\label{eq:CS-fABJM}
\end{equation}
Note that $L^{020}$ is equivalent to $\mathbb{C}^{2}/\mathbb{Z}_{2}\times\mathbb{C}$.
The unflavored $\ccz$ theory corresponds to an $\mathcal{N}=2$ version of ABJM theory.
In this sense, the $\ccz$ theory and the flavored ABJM theory can be regarded as different generalizations of ABJM theory from the perspective of the gauge theory. More colloquially, both generalizations may be referred to as ``flavored ABJM theories,'' as we did in the abstract for simplicity.

The superpotential of these theories can be decomposed into two parts, $W_{1}$ and $W_{2}$. $W_{1}$ is the cubic superpotential involving the adjoint and bi-fundamental fields, while $W_{2}$ is the cubic superpotential involving the bi-fundamental and fundamental fields. For the flavored SYM theory,
\begin{subequations}
\label{eq:SYM-Ws}
\begin{align}
W_{1}^{\left(\mathrm{fSYM}\right)} & \sim\mathrm{tr}\left[\phi_{1}\left[\phi_{2},\phi_{3}\right]\right]\ ,\label{eq:SYM-W1}\\
W_{2}^{\left(\mathrm{fSYM}\right)} & \sim\mathrm{tr}\left[
\sum_{i=1}^{F_{1}}p_{1}^{\left(i\right)}\phi_{1}q_{1}^{\left(i\right)}
+\sum_{i=1}^{F_{2}}p_{2}^{\left(i\right)}\phi_{2}q_{2}^{\left(i\right)}
+\sum_{i=1}^{F_{3}}p_{3}^{\left(i\right)}\phi_{3}q_{3}^{\left(i\right)}
\right]\ .\label{eq:SYM-W2}
\end{align}
\end{subequations}
For the flavored $\ccz$ theory,
\begin{subequations}
\label{eq:FABJM-Ws}
\begin{align}
W_{1}^{\fccz} & \sim\mathrm{tr}\left[\phi_{1}\left(B_{1}A_{1}-B_{2}A_{2}\right)-\phi_{2}\left(A_{1}B_{1}-A_{2}B_{2}\right)\right]\ ,
\label{eq:fABJM-W1}\\
W_{2}^{\fccz} & \sim\mathrm{tr}\left[
\sum_{i=1}^{F_{1}}p_{1}^{\left(i\right)}A_{1}q_{1}^{\left(i\right)}
-\sum_{i=1}^{\tilde{F}_{1}}\tilde{p}_{1}^{\left(i\right)}B_{1}\tilde{q}_{1}^{\left(i\right)}
-\sum_{i=1}^{F_{2}}p_{2}^{\left(i\right)}A_{2}q_{2}^{\left(i\right)}+\sum_{i=1}^{\tilde{F}_{2}}\tilde{p}_{2}^{\left(i\right)}B_{2}\tilde{q}_{2}^{\left(i\right)}
\right]\ .
\label{eq:fABJM-W2}
\end{align}
\end{subequations}
The worldvolume theory of the brane configuration \eqref{eq:fABJM-BC2} is the flavored version of the ABJM theory.
This is similar to the flavored $\ccz$ theory, with the difference that the adjoint matters are massive and integrated out in the infrared. 
As a result, the superpotential takes the form
\begin{equation}
W_{1}^{\left(\mathrm{fABJM}\right)}\sim\mathrm{tr}\left[\epsilon^{ij}\epsilon^{k\ell}A_{i}B_{k}A_{j}B_{\ell}\right]\ .
\label{eq:fABJM2-W1}
\end{equation}
This is the superpotential for the ordinary ABJM theory. 

Since the superpotential has R-charge two, this imposes constraints on the R-charges of the fields. 
In this paper, we assume that the adjoint fields arising from the D3-brane segments have R-charge one.
This constraint will play an important role in the $\mathrm{CY}_4$/$\mathrm{CY}_3$ correspondence.
Technically, this is a crucial assumption for deriving the quantum curve, which will be discussed below. 
For simplicity, we also assume that the fundamental fields share a common R-charge.
The monopole operators may also carry the bare R-charges $\Delta_{\mathrm{m}}^{(r)}$, one for each gauge node. 
We assume that these bare monopole R-charges are the same at all gauge nodes.

For the flavored SYM theory, the superpotential \eqref{eq:SYM-Ws}, together with the assumption above, implies the following constraints:
\begin{align}
\Delta_{\phi_{1}}+\Delta_{\phi_{2}} & =1\ ,\quad\Delta_{\phi_{3}}=1\ ,\nonumber \\
\Delta_{p_{i}^{\left(j\right)}}=\Delta_{q_{i}^{\left(j\right)}} & 
=1-\frac{1}{2}\Delta_{\phi_{i}}\ ,\quad
\left(i=1,2,3,\quad j=1,\ldots,F_{i}\right)\ .
\label{eq:SYM-Rcond}
\end{align}
For the flavored $\ccz$ theory, the superpotential \eqref{eq:FABJM-Ws}, together with the assumption above, implies the following constraints:
\begin{align}
 & \Delta_{\phi_{i}}=1\ ,\quad\Delta_{A_{i}}+\Delta_{B_{i}}=1\ ,\quad\left(i=1,2\right)\ ,\nonumber \\
 & \Delta_{p_{i}^{\left(j\right)}}=\Delta_{q_{i}^{\left(j\right)}}
 =1-\frac{1}{2}\Delta_{A_{i}}\ ,\quad
 \left(i=1,2,\quad j=1,\ldots,F_{i}\right)\ ,\nonumber \\
 & \Delta_{\tilde{p}_{i}^{\left(j\right)}}=\Delta_{\tilde{q}_{i}^{\left(j\right)}}
 =1-\frac{1}{2}\Delta_{B_{i}}\ ,\quad
 \left(i=1,2,\quad j=1,\ldots,\tilde{F}_{i}\right)\ .
 \label{eq:fABJM-Rcond}
\end{align}
For later convenience, we also define alternative variables respecting the above constraints for the flavored SYM theory,
\begin{equation}
D:=\Delta_{\phi_{1}}-\frac{1}{2}
=-\left(\Delta_{\phi_{2}}-\frac{1}{2}\right)\ ,
\quad D_{\mathrm{m}}:=\Delta_{\mathrm{m}}\ ,
\label{eq:SYM-RD}
\end{equation}
and for the flavored $\ccz$ theory,
\begin{align}
&D_{1}:=\Delta_{A_{1}}-\frac{1}{2}
=-\left(\Delta_{B_{1}}-\frac{1}{2}\right)\ ,
\quad
D_{2}:=\Delta_{B_{2}}-\frac{1}{2}
=-\left(\Delta_{A_{2}}-\frac{1}{2}\right)\ ,
\nonumber \\
&D_{\mathrm{m}}
:=2\Delta_{\mathrm{m}}^{(1)}=2\Delta_{\mathrm{m}}^{(2)}\ .
\label{eq:fABJM-RD}
\end{align}
The bare monopole R-charge is related to the monopole R-charges as \cite{Jafferis:2011zi}
\begin{equation}
D_{\mathrm{m}}=\frac{1}{2}\left(\Delta_{T}-\Delta_{\tilde{T}}\right)\ .
\label{eq:m-T-R}
\end{equation}

We assume that $D$, $D_i$ and $D_{\mathrm{m}}$ are small enough that the partition function does not diverge.
The explicit constraint depends on the models, but as we will comment in section \ref{subsec:QCtoLargeN}, the allowed region can be intuitively explained in terms of the Fermi surface.

\subsection{The quantum curve at large \texorpdfstring{$N$}{N}}\label{subsec:QC-LargeN}

In this section we analyze the $S^{3}$ partition function from the field-theory perspective. In particular, we demonstrate that, in the large-$N$ limit, the partition function reduces to the Airy form \eqref{eq:S3pf}, and we outline a systematic strategy for extracting the coefficients $C$ and $B$ appearing in this expression.

Our approach proceeds in two steps. The first step is to make use of the result of \cite{Kubo:2025jxi}, with a minor generalization appropriate for the present setup. In that work, it was shown that the partition function associated with the brane configuration takes the form
\begin{equation}
Z=\frac{1}{N!}\int\prod_{a=1}^{N}d\mu_{a}\det\left(\left[\braket{\mu_{a}|\hat{\mathcal{O}}\left(\hat{x},\hat{p}\right)^{-1}|\mu_{b}}\right]_{a,b}^{N\times N}\right)\ ,
\label{eq:FGF-QC}
\end{equation}
where $\left[e_{a,b}\right]_{a,b}^{N\times N}$ denotes an $N\times N$ matrix whose $\left(a,b\right)$ element is $e_{a,b}$. The operator $\hat{\mathcal{O}}$ is the \emph{quantum curve} of the form~\footnote{
The quantum curve \eqref{eq:QCform} in our context often takes the form of \begin{equation}
\hat{\mathcal{O}}\left(\hat{x},\hat{p}\right)
=\sum_{\left(\mathfrak{m},\mathfrak{n}\right)\in\mathrm{NP}}c_{\mathfrak{m},\mathfrak{n}}e^{\left(\mathfrak{m}+\mathfrak{m}_\mathrm{c}\right)\hat{x}+\left(\mathfrak{n}+\mathfrak{n}_\mathrm{c}\right)\hat{p}}\ .
\end{equation}
$\mathfrak{m}_\mathrm{c},\mathfrak{n}_\mathrm{c}$ are constants, which will be related to the R-charge deformations.
Namely, we frequently shift the subscripts of $c$ and the coefficients of $(x, p)$ by constants associated with R-charge deformations; this causes no ambiguity.
}
\begin{equation}
\hat{\mathcal{O}}\left(\hat{x},\hat{p}\right)
=\sum_{\left(\mathfrak{m},\mathfrak{n}\right)\in\mathrm{NP}}c_{\mathfrak{m},\mathfrak{n}}e^{\mathfrak{m}\hat{x}+\mathfrak{n}\hat{p}}\ ,
\label{eq:QCform}
\end{equation}
where for any two $\mathfrak{m}$ or two $\mathfrak{n}$ the difference must be an integer, and $\mathrm{NP}$ denotes the Newton polygon, see appendix \ref{subsec:QC-NP}.
In this formalism, all information about the partition function (and equivalently about the underlying brane configuration) is encoded in the quantum curve $\hat{\mathcal{O}}$.
The second step of our analysis is to examine the behavior of the quantum curve in the small-$\hbar$ expansion. A key observation is that, for the purpose of determining the coefficients $C$ and $B$ in the Airy function, only the leading and subleading terms in this expansion are needed. This part of the analysis closely follows the method developed in \cite{Marino:2011eh}.

We begin by clarifying the structure of \eqref{eq:FGF-QC} in the context of our setup. The structure of the calculation closely parallels that of \cite{Kubo:2025jxi}.~\footnote{Although \eqref{eq:FGF-QC} was already derived in \cite{Kubo:2025jxi} together with the explicit form of the quantum curve $\hat{\mathcal{O}}$, their analysis fixed the R-charges of the bi-fundamental matter fields and did not incorporate the bare monopole R-charges. In \cite{Kubo:2025jxi}, the expression \eqref{eq:FGF-QC} was obtained analytically for the corresponding Lagrangian theory. Since in the present work we focus on the class of Lagrangian theories represented in figure \ref{fig:SYM_fABJM_Quiv}, it is natural to expect that \eqref{eq:FGF-QC} can likewise be derived analytically for arbitrary R-charge assignments of the bi-fundamental multiplets. In what follows, we show that this expectation is indeed correct.} The expression \eqref{eq:FGF-QC} conveys two important messages. The first is that the partition function can be written in the form
\begin{equation}
Z=\frac{1}{N!}\int\prod_{a=1}^{N}d\mu_{a}\det\left(\left[\braket{\mu_{a}|\hat{\rho}\left(\hat{x},\hat{p}\right)|\mu_{b}}\right]_{a,b}^{N\times N}\right)\ .
\end{equation}
Since \eqref{eq:FGF-QC} can be interpreted as the partition function of an ideal Fermi gas, this framework is referred to as the Fermi gas formalism \cite{Marino:2011eh}. The second message conveyed by \eqref{eq:FGF-QC} is that the inverse of the one-particle density matrix $\hat{\rho}$ takes precisely the form of the quantum curve \eqref{eq:QCform}. We will show that this continues to hold even after turning on non-trivial R-charges for the bi-fundamental matter multiplets.

To establish the first message, we make use of two powerful tools: supersymmetric localization \cite{Pestun:2007rz,Kapustin:2009kz} (see also \cite{Willett:2016adv} for a review) and the Fermi gas formalism \cite{Marino:2011eh}. 
Supersymmetric localization reduces the $S^{3}$ path integral to a matrix model, whose detailed derivation is reviewed in appendix \ref{subsec:Localization}. 
The essential point for our purposes is that the integrands of the relevant matrix models can be expressed in terms of the matrix factors defined in \eqref{eq:1qD5-MF} and \eqref{eq:D5-MF}. 
With these ingredients, we can write the matrix models corresponding to the theories depicted in figure \ref{fig:SYM_fABJM_Quiv}. 
For the flavored SYM theory (equivalently, for the brane configuration \eqref{eq:fSYM-BC}),
\begin{equation}
Z_{q,(F_1,F_2,F_3)}^{\mathrm{fSYM}}\left(D,D_{\mathrm{m}}\right)  =\frac{1}{N!}\int\prod_{a=1}^{N}d\mu_{a}e^{D_{\mathrm{m}}\sum_{a=1}^{N}\mu_{a}}\left(\mathcal{Z}^{\left(0,1\right)}\left(\mu\right)\right)^{F_{3}}{\cal Z}_{F_{2},F_{1}}^{\left(1,q\right)}\left(D;\mu,\mu\right)\ .
\label{eq:MM-SYM}
\end{equation}
For the flavored $\ccz$ theory (or equivalently, for the brane configuration \eqref{eq:fABJM-BC} or \eqref{eq:fABJM-BC2}),
\begin{align}
 & Z_{
\left(q_1,F_{1},\tilde{F}_{1}\right),
\left(q_2,F_{2},\tilde{F}_{2}\right)
}^{\fccz}\left(D_{1},D_{2},D_{\mathrm{m}}\right)\nonumber \\
 & =\frac{1}{\left(N!\right)^{2}}\int\prod_{a=1}^{N}d\mu_{a}d\nu_{a}e^{\frac{1}{2}D_{\mathrm{m}}\sum_{a=1}^{N}\mu_{a}}{\cal Z}_{F_{2},\tilde{F}_{2}}^{\left(1,q_{2}\right)}\left(D_{2};\mu,\nu\right)e^{\frac{1}{2}D_{\mathrm{m}}\sum_{a=1}^{N}\nu_{a}}{\cal Z}_{\tilde{F}_{1},F_{1}}^{\left(1,q_{1}\right)}\left(D_{1};\nu,\mu\right)\ .
 \label{eq:MM-fABJM}
\end{align}
Note that we already used the R-charge parametrization in \eqref{eq:SYM-RD} and \eqref{eq:fABJM-RD}.

We then apply the Fermi gas formalism to obtain the structure \eqref{eq:FGF-QC}. 
The full derivation is presented in appendix \ref{subsec:FGF}, where we also show explicitly that the inverse of the density matrix indeed reproduces the quantum curve, thereby establishing the second message. 
The key observation is that the quantum curves relevant for our theories can be expressed as products of the elementary quantum curves defined in \eqref{eq:QC-1q} and \eqref{eq:QC-01}. 
In the main text we summarize the resulting expressions for the theories depicted in figure \ref{fig:SYM_fABJM_Quiv}. 
For the flavored SYM theory, the corresponding quantum curve takes the form
\begin{equation}
\hat{\mathcal{O}}_{q,(F_1,F_2,F_3)}^{\mathrm{fSYM}}\left(D,D_{\mathrm{m}}\right)=e^{\frac{1}{2}D_{\mathrm{m}}\hat{x}}\hat{\mathcal{O}}_{F_{2},F_{1}}^{\left(1,q\right)}\left(D\right)e^{\frac{1}{2}D_{\mathrm{m}}\hat{x}}\left(\hat{\mathcal{O}}^{\left(0,1\right)}\right)^{F_{3}}\ ,
\label{eq:SYM-QC}
\end{equation}
while for the flavored $\ccz$ theory,
\begin{equation}
\hat{\mathcal{O}}_{
\left(q_1,F_{1},\tilde{F}_{1}\right),
\left(q_2,F_{2},\tilde{F}_{2}\right)
}^{\fccz}\left(D_{i},D_{\mathrm{m}}\right)
=e^{\frac{1}{4}D_{\mathrm{m}}\hat{x}}\hat{\mathcal{O}}_{\tilde{F}_{1},F_{1}}^{\left(1,q_{1}\right)}\left(D_{1}\right)e^{\frac{1}{2}D_{\mathrm{m}}\hat{x}}\hat{\mathcal{O}}_{F_{2},\tilde{F}_{2}}^{\left(1,q_{2}\right)}\left(D_{2}\right)e^{\frac{1}{4}D_{\mathrm{m}}\hat{x}}\ .
\label{eq:fABJM-QC}
\end{equation}
The salient point is that the $S^{3}$ partition functions of these theories are indeed captured by the general expression \eqref{eq:FGF-QC}, with the operator $\hat{\mathcal{O}}$ replaced by the corresponding quantum curves derived above. In this formalism, the fundamental commutation relation is
\begin{equation}
\left[\hat{x},\hat{p}\right]=2\pi i\ .
\label{eq:ComRel}
\end{equation}

Let us finally explain an important aspect of the relation between the $\left(p,q\right)$ webs and the quantum curves. It was pointed out in \cite{Kubo:2025jxi} that the Newton polygons of the quantum curves are dual to the $\left(p,q\right)$ webs (in a sense explained in appendix \ref{sec:Poly-Web}).~\footnote{
Note that the described correspondence also follows logically as an extension of the TS/ST correspondence due to the connection with the underlying Calabi-Yau mirror curve, \cite{Grassi:2014cla, Grassi:2014zfa,Codesido:2015dia}. 
This is, however, not established for the arbitrary R-charges we consider here.
}
Namely, for a general brane configuration \eqref{eq:BC-gen}, the quantum curve is expected to be given as
\begin{equation}
\hat{\mathcal{O}}\left(\hat{x},\hat{p}\right)=\prod_{i=1}^{R}\hat{\mathcal{O}}^{\mathsf{w}_{i}}\left(\hat{x},\hat{p}\right)\ ,
\label{eq:QC-Gen}
\end{equation}
where $\hat{\mathcal{O}}^{\mathsf{w}_{i}}$ has the Newton polygon dual to $\mathsf{w}_{i}$. 
One can indeed see that the Newton polygons of $\hat{\mathcal{O}}_{F_{+},F_{-}}^{\left(1,q\right)}$ in \eqref{eq:QC-1q} and $\hat{\mathcal{O}}^{\left(0,1\right)}$ in \eqref{eq:QC-01} are dual to the $\left(p,q\right)$ webs in \eqref{eq:BW-gen}. Note that the quantum curve is specified by three key components: the Newton polygon, the total exponents of $e^{x}$ and $e^{p}$, and the coefficients associated with each term.~\footnote{The coefficients were also conjectured to be determined via the web deformations in \cite{Kubo:2025jxi}. Since in this paper we do not consider the web deformations, which are the source of the FI deformations and the mass deformations of the fundamental matters, the coefficients are trivial in the expression \eqref{eq:QC-Gen} (such as \eqref{eq:QC-1q} and \eqref{eq:QC-01}). The quantum curve can still have non-trivial coefficients in the expression \eqref{eq:QCform} through \eqref{eq:CBHform}. On the other hand, the conjecture does not fix the total exponents of $e^{x}$ and $e^{p}$, which play an important role since the R-charge deformations affect the exponents (such as \eqref{eq:SYM-QC} and \eqref{eq:fABJM-QC}).}

\subsubsection*{Modified grand potential}

We have obtained explicit quantum-curve expressions for the partition functions of both the flavored SYM and flavored $\ccz$ theories. 
In general, once the Fermi-gas representation \eqref{eq:FGF-QC} is available, the coefficients $C$ and $B$ in the Airy-function form \eqref{eq:S3pf} can be computed systematically \cite{Marino:2011eh}. 
When the operator $\hat{\mathcal{O}}$ appearing in \eqref{eq:FGF-QC} assumes the structure of a quantum curve, as in \eqref{eq:QCform}, the computation is significantly streamlined and the resulting expressions admit a clear pictorial interpretation. 
In what follows, we illustrate this strategy.

We begin by passing to the grand canonical ensemble, introducing $\mu$ as the chemical potential conjugate to the gauge group rank $N$:
\begin{equation}
\Xi\left(\mu\right):=1+\sum_{N=1}^{\infty}Z\left(N\right)e^{\mu N}\ .
\label{eq:GPF-Def}
\end{equation}
For the Fermi-gas representation \eqref{eq:FGF-QC}, the grand partition function can be written as the spectral determinant associated with the quantum curve,
\begin{equation}
\Xi\left(\mu\right)=\mathrm{Det}\left(1+e^{\mu}\hat{\mathcal{O}}\left(\hat{x},\hat{p}\right)^{-1}\right)\ ,
\label{eq:GPF-FGF}
\end{equation}
keeping the general commutation relation $\left[\hat{x},\hat{p}\right]=i\hbar$. The modified grand potential $J\left(\mu\right)$ is then defined via
\begin{equation}
\Xi\left(\mu\right)=\sum_{n=-\infty}^{\infty}e^{J\left(\mu+2\pi in\right)}\ ,
\label{eq:GP-Def}
\end{equation}
such that it respects the periodicity $\mu\rightarrow\mu+2\pi i$ in \eqref{eq:GPF-Def} and cancels the oscillation in the non-perturbative part \cite{Hatsuda:2012dt}. 
Then the canonical partition function is recovered through the simple relation
\begin{equation}
Z\left(N\right)=\frac{1}{2\pi i}\int_{-i\infty}^{+i\infty}d\mu\exp\left[J\left(\mu\right)-\mu N\right]\ .
\label{eq:GP-PF}
\end{equation}
Since we are interested only in the perturbative contribution, and since large $N$ corresponds to large $\mu$, the modified grand potential for \eqref{eq:GPF-FGF} can be expressed simply as~\footnote{
In this paper we ignore the distinction between $O\left(e^{-\Lambda}\right)$, $O\left(e^{-c\Lambda}\right)$, and $O\left(\Lambda e^{-c\Lambda}\right)$ with a constant $c$ for large $\Lambda$, as none of these terms contribute to the perturbative part.
}
\begin{equation}
J\left(\mu\right)
=\mathrm{Tr}\log\left(
1+e^{\mu}\hat{\mathcal{O}}\left(\hat{x},\hat{p}\right)^{-1}
\right)+O\left(e^{-\mu}\right)\ .
\end{equation}
To compute the large $\mu$ expansion, we define the Hamiltonian
\begin{equation}
\hat{H}\left(\hat{x},\hat{p}\right):=\log\hat{\mathcal{O}}\left(\hat{x},\hat{p}\right)\ ,
\end{equation}
and introduce the distribution operator at zero temperature
\begin{equation}
\hat{\mathsf{N}}\left(\hat{x},\hat{p};E\right)=\theta\left(E-\hat{H}\left(\hat{x},\hat{p}\right)\right)\ ,
\label{eq:DO-Def}
\end{equation}
where $\theta\left(x\right)$ is the Heaviside step function. By using the number of eigenstates of $\hat{H}$ with $E_{n}\leq E$ defined by
\begin{equation}
n\left(E\right):={\rm Tr}\ \hat{\mathsf{N}}\left(\hat{x},\hat{p};E\right)\ ,
\label{eq:NE}
\end{equation}
we find
\begin{equation}
J\left(\mu\right)=\int_{0}^{\infty}dE\frac{dn\left(E\right)}{dE}\log\left(1+e^{\mu-E}\right)=\int_{0}^{\infty}dE\frac{n\left(E\right)}{1+e^{E-\mu}}\ .
\label{eq:GP-DO}
\end{equation}
In the last equality, we assumed $n\left(0\right)=0$ and $\lim_{E\rightarrow\infty}n\left(E\right)e^{-E}=0$, both of which are correct in our setup. 

Hence, a key result follows: the large $N$ partition function can be determined once we know the behavior of $n\left(E\right)$ at large $E$. In the following, we will show that the large-$E$ expansion of $n\left(E\right)$ takes the form
\begin{equation}
n\left(E\right)=CE^{2}+n_{0}+n_{\mathrm{np}}\left(E\right)\ .
\label{eq:NE-Gen}
\end{equation}
Assuming for the moment that this is correct, from \eqref{eq:GP-DO}, we find
\begin{equation}
J\left(\mu\right)=\frac{C}{3}\mu^{3}+B\mu+A+O\left(e^{-\mu}\right)\ ,
\label{eq:GP-NE}
\end{equation}
where
\begin{align}
B & =n_{0}+\frac{\pi^{2}}{3}C\ .
\label{eq:B-n0}
\end{align}
Here, we used the formula
\begin{equation}
\int_{0}^{\infty}dE\frac{E^{a}}{1+e^{E-\mu}}=-\Gamma\left(a+1\right)\mathrm{Li}_{a+1}\left(-e^{\mu}\right),
\end{equation}
and the asymptotics of the Polylogarithm $\mathrm{Li}_{a}(z)$.
Upon a Laplace transform, \eqref{eq:GP-PF}, one can check that $C$ and $B$ are the same as those in \eqref{eq:S3pf}.

\subsubsection{Computing the number of states}\label{subsec:QCtoLargeN}
To compute $n\left(E\right)$ at large $E$, we introduce the Wigner transform. By using \eqref{eq:WT-Taylor}, we find that the Wigner transform of $\hat{\mathsf{N}}\left(\hat{x},\hat{p};E\right)$ is
\begin{equation}
\mathsf{N}\left(x,p;E\right)_{\mathrm{W}}=\theta\left(E-H_{\mathrm{W}}\left(x,p\right)\right)+\sum_{r=2}^{\infty}\frac{1}{r!}\mathcal{G}_{r}\delta^{\left(r-1\right)}\left(E-H_{\mathrm{W}}\left(x,p\right)\right)\ ,
\end{equation}
where
\begin{equation}
\mathcal{G}_{r}
=\left[\left(\hat{H}-H_{\mathrm{W}}\left(x,p\right)\right)^{r}\right]_{\mathrm{W}}\ .
\label{eq:Gr-Def}
\end{equation}
Then, thanks to \eqref{eq:Tr-Wig-Form}, the number of eigenstates $n\left(E\right)$ is given as
\begin{equation}
n\left(E\right)
=\frac{\mathrm{Vol}\left(E\right)}{2\pi\hbar}
+\sum_{r=2}^{\infty}\frac{1}{r!}\int\frac{dx\, d p}{2\pi\hbar}\mathcal{G}_{r}\delta^{\left(r-1\right)}\left(E-H_{\mathrm{W}}\left(x,p\right)\right)\ ,
\end{equation}
where
\begin{equation}
\mathrm{Vol}\left(E\right):=\int_{H_{\mathrm{W}}\left(x,p\right)\leq E} dx\, d p\ .
\label{eq:Vol-Def}
\end{equation}
Note that the boundary of the region of this integration,
\begin{equation}
H_{\mathrm{W}}\left(x,p\right)=E\ ,
\label{eq:FS-Gen}
\end{equation}
is the Fermi surface. As discussed in \cite{Marino:2011eh}, the contribution from $\mathcal{G}_{r}$ only gives corrections of order $O\left(e^{-E}\right)$. Therefore, when we focus on the perturbative part, the number of eigenstates $n\left(E\right)$ is simply given by the volume of the phase space,
\begin{equation}
n\left(E\right)=\frac{\mathrm{Vol}\left(E\right)}{2\pi\hbar}+O\left(e^{-E}\right)\ .
\label{eq:NE-Vol}
\end{equation}
In the following, we outline the computation of the volume $\mathrm{Vol}\left(E\right)$ at large $E$. While the detailed results depend on the specific examples, a universal feature emerges: assuming the curve has the form \eqref{eq:QCform}, we have $\mathrm{Vol}\left(E\right)=O\left(E^{2}\right)$ for large $E$ \cite{Marino:2011eh}. Moreover, in all examples considered here, the structure of the large-$E$ volume takes the form~\footnote{
Although an $O(E^1)$ term can also appear in general, it can be canceled by introducing an appropriate phase factor of the form $\exp\left(iN\theta\right)$ in the definition of the matrix model.
Interestingly, this term vanishes automatically in all examples considered in this paper.
\label{E1}}
\begin{equation}
\mathrm{Vol}\left(E\right)=v_{2}E^{2}+v_{0}+O\left(e^{-E}\right)\ .
\label{eq:Vol-Gen}
\end{equation}
From this result, the coefficients $C$ and $B$ in the Airy form \eqref{eq:S3pf} can finally be computed using \eqref{eq:NE-Vol} and \eqref{eq:NE-Gen}-\eqref{eq:B-n0} as
\begin{equation}
C=\frac{v_{2}}{2\pi\hbar}\ ,\quad B=\frac{v_{0}}{2\pi\hbar}+\frac{\pi v_{2}}{6\hbar}\ .
\label{eq:Vol-CB}
\end{equation}

To compute $v_{2}$ and $v_{0}$, we need to evaluate $H_{\mathrm{W}}\left(x,p\right)$. A crucial observation is that the small-$\hbar$ expansion of the eigenvalue density corresponds to an expansion in $\left(\hbar\frac{d}{dE}\right)^{2}$ \cite{Marino:2011eh}. Since we require only terms up to $O\left(E^{0}\right)$ at large $E$ (see \eqref{eq:Vol-Gen}), it suffices to expand $H_{\mathrm{W}}\left(x,p\right)$ in small $\hbar$ up to $O\left(\hbar^{2}\right)$. Using the Taylor expansion \eqref{eq:WT-Taylor} together with \eqref{eq:F0123r} and the property \eqref{eq:QC-W}, one finds
\begin{equation}
H_{\mathrm{W}}\left(x,p\right)
=H_{\mathrm{W}}^{\left(0\right)}\left(x,p\right)
+\hbar^{2}H_{\mathrm{W}}^{\left(1\right)}\left(x,p\right)
+O\left(\hbar^{4}\right)\ ,
\label{eq:HW-WKB}
\end{equation}
where
\begin{subequations}
\label{eq:HW-OW}
\begin{align}
H_{\mathrm{W}}^{\left(0\right)}\left(x,p\right) & 
=\log\mathcal{O}_{\mathrm{W}}\ ,\label{eq:HW0-OW}\\
H_{\mathrm{W}}^{\left(1\right)}\left(x,p\right) & 
=\frac{1}{8}\mathcal{O}_{\mathrm{W}}^{-2}\left[\frac{\partial^{2}\mathcal{O}_{{\rm W}}}{\partial x^{2}}\frac{\partial^{2}\mathcal{O}_{{\rm W}}}{\partial p^{2}}-\left(\frac{\partial^{2}\mathcal{O}_{{\rm W}}}{\partial x\partial p}\right)^{2}\right] \\
 & \quad-\frac{1}{12}\mathcal{O}_{\mathrm{W}}^{-3}\left[\left(\frac{\partial\mathcal{O}_{{\rm W}}}{\partial x}\right)^{2}\frac{\partial^{2}\mathcal{O}_{{\rm W}}}{\partial p^{2}}+\left(\frac{\partial\mathcal{O}_{{\rm W}}}{\partial p}\right)^{2}\frac{\partial^{2}\mathcal{O}_{{\rm W}}}{\partial x^{2}}-2\frac{\partial\mathcal{O}_{{\rm W}}}{\partial x}\frac{\partial\mathcal{O}_{{\rm W}}}{\partial p}\frac{\partial^{2}\mathcal{O}_{{\rm W}}}{\partial x\partial p}\right]\ .\nonumber
 \label{eq:HW1-OW}
\end{align}
\end{subequations}

The volume \eqref{eq:Vol-Def} at large $E$ can be evaluated using \eqref{eq:HW-WKB}. Note that the Wigner transform of the quantum curve \eqref{eq:QCform} coincides with its classical counterpart, as in \eqref{eq:QC-W}. It then follows straightforwardly that the Fermi surface \eqref{eq:FS-Gen} for $H_{\mathrm{W}}^{\left(0\right)}\left(x,p\right)$, defined by $H_{\mathrm{W}}^{\left(0\right)}\left(x,p\right)=E$, asymptotically approaches a polygon bounded by the equations
\begin{equation}
\mathfrak{m}x+\mathfrak{n}p=E-\log c_{\mathfrak{m},\mathfrak{n}}\ ,\label{eq:FS-Poly-Gen}
\end{equation}
where the term $c_{\mathfrak{m},\mathfrak{n}}e^{\mathfrak{m}x+\mathfrak{n}p}$ corresponds to a vertex of the Newton polygon of the curve $\mathcal{O}_{\mathrm{W}}\left(x,p\right)$
(providing the polygonal approximation of $H_{\mathrm{W}}^{\left(0\right)}\left(x,p\right)$).
An example is shown in figure \ref{fig:FermiSurface}.
We denote the volume enclosed by this polygon as $\mathrm{Vol}_{\mathrm{Poly}}$. The full volume can then be naturally decomposed into the following two contributions
\begin{equation}
\mathrm{Vol}\left(E\right)=\mathrm{Vol}_{\mathrm{Poly}}\left(E\right)+\dvol\ .
\label{eq:Vol-Poly_Corr}
\end{equation}
Notice that $\mathrm{Vol}_{\mathrm{Poly}}$ includes both terms of the order $E^{2}$ and $E^{0}$ for general $c_{\mathfrak{m},\mathfrak{n}}$, while $\dvol$ contains only a term of order $E^{0}$.
The strategy for computing the volume is to first evaluate the polygon part $\mathrm{Vol}_{\mathrm{Poly}}$, and then compute $\dvol$ as the correction.
In the following, we discuss generalities for the computation of the polygon part and the corrections.

The volume of the polygon part $\mathrm{Vol}_{\mathrm{Poly}}$ can be computed as follows.
Let $c_{\mathfrak{m}_{j},\mathfrak{n}_{j}}e^{\mathfrak{m}_{j}x+\mathfrak{n}_{j}p}$ denote the monomial associated with the j-th vertex of the Newton polygon of the classical curve, with the vertices ordered counterclockwise.
Here, $j$ labels the vertices.
The polygon is surrounded by the equations \eqref{eq:FS-Poly-Gen} with $\left(\mathfrak{m}_{j},\mathfrak{n}_{j}\right)$.
After a short calculation, one finds that the volume is given as
\begin{align}
\label{eq:VolForm}
 & \qquad  \mathrm{Vol}_{\mathrm{Poly}}\left(E\right) =  \\
 & \sum_{j=1}^{s}\left\{ \frac{\left(\mathfrak{m}_{j-1}\mathfrak{n}_{j+1}-\mathfrak{m}_{j+1}\mathfrak{n}_{j-1}\right)\left(E-\log c_{\mathfrak{m}_{j},\mathfrak{n}_{j}}\right)^{2}}{2\left(\mathfrak{m}_{j-1}\mathfrak{n}_{j}-\mathfrak{m}_{j}\mathfrak{n}_{j-1}\right)\left(\mathfrak{m}_{j}\mathfrak{n}_{j+1}-\mathfrak{m}_{j+1}\mathfrak{n}_{j}\right)}-\frac{\left(E-\log c_{\mathfrak{m}_{j},\mathfrak{n}_{j}}\right)\left(E-\log c_{\mathfrak{m}_{j+1},\mathfrak{n}_{j+1}}\right)}{\mathfrak{m}_{j}\mathfrak{n}_{j+1}-\mathfrak{m}_{j+1}\mathfrak{n}_{j}}\right\}\ , \nonumber
\end{align}
where $s$ is the number of the vertices and $\left(\mathfrak{m}_{0},\mathfrak{n}_{0}\right)=\left(\mathfrak{m}_{s},\mathfrak{n}_{s}\right)$, $\left(\mathfrak{m}_{s+1},\mathfrak{n}_{s+1}\right)=\left(\mathfrak{m}_{1},\mathfrak{n}_{1}\right)$. 
Here, although we initially assumed $c_{\mathfrak{m}_{i},\mathfrak{n}_{i}}$ to be real, the coefficients $c_{\mathfrak{m}_{i},\mathfrak{n}_{i}}$ appearing in this paper are often complex. 
By invoking analytic continuation, we continue to use \eqref{eq:VolForm} in this context.~\footnote{
When the coefficients \(c_{m,n}\) become complex, the quantum curve is generically non-Hermitian. In this case, the argument leading to \eqref{eq:VolForm}, which relies on a Hermitian spectral problem, cannot be applied straightforwardly \cite{Marino:2011eh}. Thus, the analytic continuation of \eqref{eq:VolForm} requires an additional justification. We expect that such a justification may be achieved along the lines of \cite{Grassi:2017qee,Anderson:2018gjo,Emery:2019znd}. Remark that, as already mentioned in footnote \ref{E1}, in all examples considered in this paper, assuming this analytic continuation leads to a nontrivial cancellation of the term linear in $E$, and the resulting free energy is real. We thank Nikolay Bobev, Pieter-Jan De Smet, Junho Hong and Xuao Zhang for drawing our attention to this subtlety.
\label{NonHermit}}
Note that, while the effect of $c_{m,n}$ in \eqref{eq:VolForm} is often incorporated into the $\dvol$ term in previous literature, we employ the form of \eqref{eq:VolForm} because the structure remains a polygon even when these effects are included.

For the derivation of this formula, we assume that the R-charges are small enough so that the partition function does not diverge.
In terms of the Fermi surface, this condition is equivalent to requiring that the volume of the polygon-approximated Fermi surface, bounded by the equations \eqref{eq:FS-Poly-Gen}, is finite.
Equivalently, in terms of the quantum curve, we assume that the Newton polygon in the $\left(\mathfrak{m},\mathfrak{n}\right)$-plane contains the point $\left(\mathfrak{m},\mathfrak{n}\right)=(0,0)$ in its strict interior.
This assumption ensures that the denominator in \eqref{eq:VolForm} does not vanish.

\subsubsection*{Leading corrections}
An important observation is that the corrections arise from the vertices of the polygon-approximated Fermi surface; see figure \ref{fig:FermiSurface} as an example.
Let us focus on a single vertex and compute the correction to the volume from this region, $\dvol_{j}$. Here, $j$ labels the vertices, and $\dvol$ in \eqref{eq:Vol-Poly_Corr} is then given by
\begin{equation}
\dvol=\sum_{j=1}^{s}\dvol_{j}\ .
\end{equation}
At large $E$, after an appropriate canonical transformation,~\footnote{
In terms of operators, the canonical transformation is the change of variables which keeps the commutation relation invariant.
This includes, for example, the shift $x\rightarrow x+f(p)+c$ or $p\rightarrow p+f(x)+c$, where $f(\cdot)$ is a function and $c$ is a constant.
} the relevant terms in $\mathcal{O}_{{\rm W}}\left(x,p\right)$ can always be as
\begin{equation}
\left.\mathcal{O}_{\mathrm{W}}\left(x,p\right)\right|_{j\text{-th vertex}}=e^{\mathfrak{m}x}\prod_{\ell=1}^{K}\left(c_{\ell}e^{-\frac{\mathfrak{n}}{2}p}+c_{\ell}^{-1}e^{\frac{\mathfrak{n}}{2}p}\right)\ ,
\label{eq:OW-Canonical}
\end{equation}
with $\prod_{\ell=1}^{K}c_{\ell}=1$. Here $\mathfrak{m}>0$, $\mathfrak{n}>0$ and $K\geq1$.
Note that the volume $\dvol_{j}$ depends only on $\mathfrak{m}\mathfrak{n}$ due to the canonical transformation which keeps this expression.
For all the examples in this paper $K$ is less than three, and thus in the following we consider the cases $K=1,2$.
The results are
\begin{equation}
\displaystyle
\dvol_{j}=\begin{cases}
\displaystyle
-\frac{\pi^{2}}{6\mathfrak{m}\mathfrak{n}}-\frac{\mathfrak{m}\mathfrak{n}}{24}\hbar^{2}\ , & (K=1)\\
\displaystyle
-\frac{1}{\mathfrak{m}\mathfrak{n}}\left[\frac{\pi^{2}}{3}+\left(2\log c_{1}\right)^{2}\right]-\frac{\mathfrak{m}\mathfrak{n}}{12}\hbar^{2}\ , & (K=2)
\end{cases}
\ ,
\label{eq:VolC-K12}
\end{equation}
where we have suppressed the subscript $j$ on $\mathfrak{m}, \mathfrak{n}, c_1$ at the given vertex. 
These results follow from the calculation below.
Throughout this calculation, we first take $c_\ell$ to be real and then analytically continue the result to complex $c_\ell$; see footnote \ref{NonHermit}.
\begin{itemize}
\item
When $K=1$,
\begin{equation}
\left.\mathcal{O}_{\mathrm{W}}\left(x,p\right)\right|_{j\text{-th vertex}}
=e^{\mathfrak{m}x}\left(e^{-\frac{\mathfrak{n}}{2}p}+e^{\frac{\mathfrak{n}}{2}p}\right)\ .
\label{eq:OW-K1}
\end{equation}
Remark that for a curve
\begin{equation}
d_{1}e^{\mathfrak{m}_{1}x+\mathfrak{n}_{1}p}+d_{2}e^{\mathfrak{m}_{2}x+\mathfrak{n}_{2}p}\ ,
\end{equation}
after an appropriate canonical transformation, \eqref{eq:OW-K1} can be obtained with
\begin{equation}
\mathfrak{m}\mathfrak{n}=\left|\mathfrak{m}_{1}\mathfrak{n}_{2}-\mathfrak{m}_{2}\mathfrak{n}_{1}\right|\ .
\label{eq:OW-K1-mn}
\end{equation}
In this region, the relevant part of the polygon considered above—namely, the one defined by \eqref{eq:FS-Poly-Gen}—is given by
\begin{equation}
\mathfrak{m}x+\frac{\mathfrak{n}}{2}\left|p\right|=E\ .
\end{equation}
Thus the boundary of the polygon is defined as
\begin{equation}
x_{j}^{\mathrm{Poly}}\left(p\right)=\frac{1}{\mathfrak{m}}\left(E-\frac{\mathfrak{n}}{2}\left|p\right|\right)\ .
\end{equation}
We then calculate the corrections, which are concentrated around $\left(x,p\right)=\left(E/\mathfrak{m},0\right)$.
After a short calculation, one finds that the curve \eqref{eq:HW-WKB} for \eqref{eq:OW-K1} becomes
\begin{align}
\left.H_{\mathrm{W}}^{\left(0\right)}\left(x,p\right)\right|_{j\text{-th vertex}} & 
=\mathfrak{m}x+\log\left(e^{-\frac{\mathfrak{n}}{2}p}+e^{\frac{\mathfrak{n}}{2}p}\right)\ ,\nonumber \\
\left.H_{\mathrm{W}}^{\left(1\right)}\left(x,p\right)\right|_{j\text{-th vertex}} & 
=\frac{\mathfrak{m}^{2}\mathfrak{n}^{2}}{24}\frac{1}{\left(e^{-\frac{\mathfrak{n}}{2}p}+e^{\frac{\mathfrak{n}}{2}p}\right)^{2}}\ .
\end{align}
Hence, the corrected Fermi surface \eqref{eq:FS-Gen} is given by
\begin{equation}
x_{j}^{\mathrm{Corr}}\left(p\right)
=\frac{1}{\mathfrak{m}}\left(E-\log\left(e^{-\frac{\mathfrak{n}}{2}p}+e^{\frac{\mathfrak{n}}{2}p}\right)\right)
-\frac{\mathfrak{m}\mathfrak{n}^{2}\hbar^{2}}{24}\frac{1}{\left(e^{-\frac{\mathfrak{n}}{2}p}+e^{\frac{\mathfrak{n}}{2}p}\right)^{2}}\ .
\end{equation}
Thus, the correction for the volume from this region becomes
\begin{align}
\dvol_{j} & 
\approx\int_{-\infty}^{\infty}dp\left(x_{j}^{\mathrm{Corr}}\left(p\right)-x_{j}^{\mathrm{Poly}}\left(p\right)\right)\nonumber \\
 & =-\frac{1}{\mathfrak{m}}\left[\int_{-\infty}^{0}dp\log\left(1+e^{\mathfrak{n}p}\right)+\int_{0}^{\infty}dp\log\left(e^{-\mathfrak{n}p}+1\right)\right]\nonumber \\
 & \quad-\frac{\mathfrak{m}\mathfrak{n}^{2}\hbar^{2}}{24}\int_{-\infty}^{\infty}dp\frac{1}{\left(e^{-\frac{\mathfrak{n}}{2}p}+e^{\frac{\mathfrak{n}}{2}p}\right)^{2}}\ .
\end{align}
Here we extend the region of the integration to $\pm\infty$ since the difference is $O\left(e^{-E}\right)$. By using
\begin{equation}
\int_{0}^{\infty}dp\log\left[ce^{-p}+1\right]
=-\mathrm{Li}_2(-c)\ ,\quad
\int_{-\infty}^{\infty}dp\frac{1}{\left(e^{-\frac{1}{2}p}+e^{\frac{1}{2}p}\right)^{2}}=1\ ,
\label{eq:Int-Form1}
\end{equation}
we finally obtain \eqref{eq:VolC-K12}.

\item 
When $K=2$,
\begin{equation}
\left.\mathcal{O}_{\mathrm{W}}\left(x,p\right)\right|_{j\text{-th vertex}}=e^{\mathfrak{m}x}\left(c_{1}e^{-\frac{\mathfrak{n}}{2}p}+c_{1}^{-1}e^{\frac{\mathfrak{n}}{2}p}\right)\left(c_{1}^{-1}e^{-\frac{\mathfrak{n}}{2}p}+c_{1}e^{\frac{\mathfrak{n}}{2}p}\right)\ .
\label{eq:OW-K2}
\end{equation}
Remark that for a specific curve
\begin{equation}
e^{\mathfrak{m}_{1}x+\mathfrak{n}_{1}p}\left(d_{1}e^{-\mathfrak{m}_{2}x-\mathfrak{n}_{2}p}+\left(d_{1}+d_{2}\right)+d_{2}e^{\mathfrak{m}_{2}x+\mathfrak{n}_{2}p}\right),
\end{equation}
after an appropriate canonical transformation, \eqref{eq:OW-K2} can be obtained with
\begin{equation}
\mathfrak{m}\mathfrak{n}=\left|\mathfrak{m}_{1}\mathfrak{n}_{2}-\mathfrak{m}_{2}\mathfrak{n}_{1}\right|,\quad 
c_{1}=\left(\frac{d_{1}}{d_{2}}\right)^{\frac{1}{4}}\ .
\label{eq:OW-K2-mn}
\end{equation}
In this region, the curve of the polygon is defined as
\begin{equation}
x_{j}^{\mathrm{Poly}}\left(p\right)=\frac{1}{\mathfrak{m}}\left(E-\mathfrak{n}\left|p\right|\right)\ .
\end{equation}
On the other hand, after a short calculation, one finds
\begin{align}
\left.H_{\mathrm{W}}^{\left(0\right)}\left(x,p\right)\right|_{j\text{-th vertex}} & 
=\mathfrak{m}x+\log\left(c_{1}e^{-\frac{\mathfrak{n}}{2}p}+c_{1}^{-1}e^{\frac{\mathfrak{n}}{2}p}\right)+\log\left(c_{1}^{-1}e^{-\frac{\mathfrak{n}}{2}p}+c_{1}e^{\frac{\mathfrak{n}}{2}p}\right)\ ,\nonumber \\
\left.H_{\mathrm{W}}^{\left(1\right)}\left(x,p\right)\right|_{j\text{-th vertex}} & 
=\frac{\mathfrak{m}^{2}\mathfrak{n}^{2}}{24}
\frac{\left(c_{1}^{-2}+c_{1}^{2}\right)\left(e^{-\mathfrak{n}p}+e^{\mathfrak{n}p}\right)+4}{\left(c_{1}e^{-\frac{\mathfrak{n}}{2}p}+c_{1}^{-1}e^{\frac{\mathfrak{n}}{2}p}\right)^{2}\left(c_{1}^{-1}e^{-\frac{\mathfrak{n}}{2}p}+c_{1}e^{\frac{\mathfrak{n}}{2}p}\right)^{2}}\ .
\end{align}
Hence the corrected Fermi surface \eqref{eq:FS-Gen} is given by
\begin{align}
x_{j}^{\mathrm{Corr}}\left(p\right) & =\frac{1}{\mathfrak{m}}\left(E-\log\left(c_{1}e^{-\frac{\mathfrak{n}}{2}p}+c_{1}^{-1}e^{\frac{\mathfrak{n}}{2}p}\right)-\log\left(c_{1}^{-1}e^{-\frac{\mathfrak{n}}{2}p}+c_{1}e^{\frac{\mathfrak{n}}{2}p}\right)\right)\nonumber \\
 & \quad-\frac{\mathfrak{m}\mathfrak{n}^{2}\hbar^{2}}{24}
 \frac{\left(c_{1}^{-2}+c_{1}^{2}\right)\left(e^{-\mathfrak{n}p}+e^{\mathfrak{n}p}\right)+4}{\left(c_{1}e^{-\frac{\mathfrak{n}}{2}p}+c_{1}^{-1}e^{\frac{\mathfrak{n}}{2}p}\right)^{2}\left(c_{1}^{-1}e^{-\frac{\mathfrak{n}}{2}p}+c_{1}e^{\frac{\mathfrak{n}}{2}p}\right)^{2}}\ .
\end{align}
$\dvol_{j}$ is given by $\int_{-\infty}^{\infty}dp\left(x_{j}^{\mathrm{Corr}}\left(p\right)-x_{j}^{\mathrm{Poly}}\left(p\right)\right)$ as in the $K=1$ case, and by using \eqref{eq:Int-Form1} and
\begin{align}
&\mathrm{Li}_{2}\left(-c\right)+\mathrm{Li}_{2}\left(-c^{-1}\right)
=2\pi^{2}B_{2}\left(\frac{\log c}{2\pi i}+\frac{1}{2}\right)
=-\frac{\left(\log c\right)^{2}}{2}-\frac{\pi^{2}}{6}\ ,\nonumber \\
&\int_{-\infty}^{\infty}dp
\frac{\left(c^{-1}+c\right)\left(e^{-p}+e^{p}\right)+4}{\left(e^{-\frac{1}{2}p}+c^{-1}e^{\frac{1}{2}p}\right)^{2}\left(e^{-\frac{1}{2}p}+ce^{\frac{1}{2}p}\right)^{2}} =2\ ,
\label{eq:Int-Form2}
\end{align}
we finally obtain \eqref{eq:VolC-K12}.

\end{itemize}

In summary, the large-$N$ Airy-function behavior follows from two inputs: (i) the quadratic growth of the phase-space volume at large energy, and (ii) the fact that higher-order quantum corrections only affect exponentially suppressed terms. This universality underlies the geometric interpretation discussed next.

\subsection{Toric CY geometry and equivariant intersection numbers}\label{subsec:TCY-Loc}

Next, we review the geometric perspective. By performing a $T$-dualization and an uplift to M-theory, the D3-branes become M2-branes probing a toric $\mathrm{CY}_{4}$, which corresponds to the cone over a Sasaki-Einstein seven-fold. This describes the geometric moduli space of the (abelian version of the) above 3d theories \cite{Martelli:2008si,Benini:2009qs}. 

In order to be slightly more general, and in view of the following calculations, we begin with smooth resolutions, denoted by $X$, of the cone over a $(2d-1)$-dimensional Sasakian manifold $L$ (with $d=4$ in the present case). Imposing the toric Sasaki condition on $L$ ensures that the cone over it (generally singular) is a non-compact Calabi–Yau manifold of complex dimension $d$. The resolution $X$ is taken to be smooth and to preserve the Calabi–Yau condition. Owing to the toric structure, $X$ can be realized as a Kähler quotient of $\mathbb{C}^n$ by a torus $U(1)^r$, with $d = n - r$, so that $X = \mathbb{C}^n//U(1)^r$.

A larger torus $U(1)^n$ acts diagonally on $\mathbb{C}^n$, and the quotient is described by a matrix of integer charges $Q^a_i$ (also known as GLSM charges) providing the embedding of $U(1)^r$ inside $U(1)^n$, where $i = 1,\dots,n$ and $a = 1,\dots,r$. The CY condition for a vanishing first Chern class in the toric case can be translated into the following condition on the charges:
\be
\label{eq:CYcondonQ}
	\sum_i Q^a_i = 0\ , \qquad \forall a\ .
\ee
We can then consider the $U(1)^n$-equivariant cohomology and correspondingly upgrade the symplectic form $\omega$ to be closed under the equivariant differential ${\rm d} + \epsilon_i\, \iota_{u^i}$, where $u^i$ generate the $U(1)^n$ action. Additionally, we parametrize $\omega = \omega_\lambda$ in terms of redundant Kähler parameters $\lambda^i$, which serve as formal variables overparametrizing the physical Kähler moduli $t^a$, conjugate to all of the isometries of the prequotient $\BC^n$.

Next, it is useful to also introduce the map $v: \mathbb{R}^n \rightarrow \mathbb{R}^d$ as the cokernel of $Q$,
\be
\sum_i v^i_\alpha Q^a_i = 0\ , \qquad \forall a, \alpha\ ,
\ee
with $\alpha = 1, \dots, d = n - r$. We refer to the vectors $v^i = \{ (v^i)_\alpha := v^i_\alpha \}$ as toric fan vectors, which a priori have the (complex) dimension of the manifold, $d$. On the other hand, note that the CY condition on the charges $Q$, \eqref{eq:CYcondonQ}, simply states that there is always one among the $n$-dimensional vectors $v_\alpha = \{(v_\alpha)^i := v^i_\alpha \}$, conventionally chosen as $v_1$, consisting entirely of unit entries, $v_1 = (1, \dots, 1)$. This allows us to effectively reduce the dimension of the toric fan vectors $v^i$,
\be
v^i = (1, v^i_2, \dots, v^i_d) =: (1, \tilde v^i_{\alpha-1})\ , \qquad \forall i\ ,
\ee
where the effective toric fan vectors $\tilde v$ are $(d-1)$-dimensional. 
This standard representation is used to depict threefolds as a collection of $n$ points (each point representing a vector $\tilde v^i$) on a planar diagram and fourfolds as a collection of points on a three-dimensional diagram, as long as the CY condition is obeyed. This explains the origin of the toric diagrams we use extensively in this work.

\subsubsection{Equivariant volume}
\label{subsec:equiVol}
We now review the definition of equivariant intersection numbers based on \cite{Cassia:2025aus, Cassia:2025jkr} and the closely related construction of \cite{Martelli:2023oqk,Colombo:2023fhu}. For a more detailed derivation of the main formulae below we refer the reader to the original references.

The general toric setup described above allows for the definition of the following quantity, using the equivariant completion $\omega_\lambda - \epsilon_i H^i$:
\be
	\BV_X (\lambda, \epsilon) := \int_X \mathe^{\omega_\lambda - \epsilon_i H^i}\ ,
\ee
which we refer to as the equivariant volume.~\footnote{Different definitions of the equivariant volume appear in the literature, depending on whether the parametrization is in terms of $(t, \epsilon)$ or $(\lambda, \epsilon)$. We discuss only the latter quantity, which is directly the generating function for the equivariant intersection numbers used below.} A key property of $\BV_X$ is that it serves as a generating function for the equivariant intersection numbers $D_{i_1 \dots i_k}$:
\be
\label{eq:equiinters}
D_{i_1 \dots i_k} := \frac{\partial^k}{\partial \lambda^{i_1} \dots \partial \lambda^{i_k}}\, \BV_X (\lambda, \epsilon) \Big|_{\lambda = 0}\ .
\ee
This property is essential for our purposes, as it allows one to define the equivariant generalization of the perturbative part of the topological string partition function on $X$, denoted $F^\text{pert.}_X(\lambda, \epsilon)$ in \cite{Cassia:2025aus, Cassia:2025jkr}.

Using further the toric fan vectors defined previously, one can prove, cf.\ \cite{Cassia:2025aus}, that in fact the function $\BV_X (\lambda, \epsilon)$ does not independently depend on all $\epsilon_i$ parameters, but only on the special combinations
\be
\label{eq:nufrome}
	\nu_\alpha (\epsilon) := \sum_i v_\alpha^i \epsilon_i\ ,
\ee
which can be regarded as equivariant parameters for the faithfully acting torus symmetry on $X$.~\footnote{We should warn the reader that the present notation follows \cite{Cassia:2025aus} and differs from the one in \cite{Martelli:2023oqk} in a potentially confusing way, $\epsilon^\text{here} = \bar \epsilon^\text{there}$ and $\nu^\text{here} = \epsilon^\text{there}$.} This means that there exists a lift of $\BV_X (\lambda, \epsilon)$ to a formally different function, $\tilde \BV_X (\lambda, \nu)$:
\be
\label{eq:usingnu}
	 \tilde \BV_X (\lambda, \nu) \Big|_{\nu_\alpha = v^i_\alpha \epsilon_i} = \BV_X (\lambda, \epsilon)\ ,
\ee
where we assumed summation over the repeated up-down index $i$. By an abuse of notation, which we believe not to be confusing for the present purposes, we are henceforth going to drop the tilde and use the notation $\BV_X (\lambda, \nu)$, since the difference does not affect $\lambda$-derivatives and therefore the equivariant intersection numbers in \eqref{eq:equiinters}.

Given the central role of the equivariant volume in what follows, it is useful—beyond direct evaluation from the definition above—to review some more practical computational methods. Naturally, these rely on equivariant localization and the associated fixed-point theorem.

One such approach employs equivariant localization in the spirit of Jeffrey–Kirwan (JK), rewriting the equivariant volume as a residue integral over the equivariant extensions of the original Chern roots $\phi_a$,
\be
x_i := \e_i + \sum_a\, \phi_a Q^a_i\ ,
\ee
so that
\be
\label{eq:evol-JK}
\BV_X(\lam,\e) = \oint_\text{JK} \prod_{a=1}^r\frac{\mathd\phi_a}{2\pi\mathi} \frac{\mathe^{x_i\lam^i}}{\prod_{i=1}^n x_i}\ .
\ee
This expression makes explicit use of the GLSM charge matrix $Q$. The choice of integration contour depends on the chamber associated with the Kähler parameters, commonly referred to as a \emph{phase} of the symplectic quotient. Importantly, not all such chambers correspond to \emph{geometric} phases—i.e.\ regions where $X$ is a smooth resolution of the singular cone. While this distinction is important to keep in mind, it is worth emphasizing that there is no inherent reason to restrict the analysis to geometric phases at the outset, since the general formalism remains valid regardless of such specialization.

The JK-residue formula above can be implemented explicitly for any toric manifold with a known charge matrix; see \cite{Cassia:2025aus} for a range of examples that we incorporate into our analysis below. That said, we should note that identifying the relevant chamber becomes increasingly cumbersome as the parameters $n$ and $r$ grow, even for fixed complex dimension $d = n - r$.

An alternative method, which makes more direct use of the toric fan vectors and becomes particularly simple in the case $d = 3$, is the fixed-point formula discussed in detail in \cite[Section 2.3]{Martelli:2023oqk}. The formulation presented below, which involves a triangulation of the toric diagram with each point (or vertex) on the toric diagram being necessarily at a tip, is specific to the geometric phases discussed earlier. A generalization to non-geometric phases instead requires triangulations with internal points on the sides of the triangles, see also appendix \ref{app:An equivariant duality}.

Let us now specialize to the case of threefolds, where the triangulation procedure can be applied straightforwardly by inspection of the toric diagram. Any such diagram can be decomposed into one (or several) sets of triangles by drawing internal lines between lattice points, with each point only allowed at a tip of a triangle. Each triangle is then labeled by its defining triplet of points, $(\tilde v^{i_1}, \tilde v^{i_2}, \tilde v^{i_3})$, for which we introduce the label $A = (i_1, i_2, i_3)$, ranging over all triangles in the chosen triangulation. Since each vector $v^i$ (including now the first unit element) is a three-dimensional vector, we define:
\be
d_A := |{\rm det} (v^{i_1}, v^{i_2}, v^{i_3})|\ .
\ee
Additionally, for each triangle, we associate a set of auxiliary vectors $u_A^{i_m}$, defined (up to normalization) as the cross product of the two vectors in the triangle that exclude $v^{i_m}$. The overall normalization and sign ambiguity are fixed by requiring:
\be
u_A^{i_m} \cdot v^{i_n} = d_A\, \delta_{m n}\ .
\ee
The fixed-point formula then evaluates the equivariant volume in terms of the effective parameters $\nu = (\nu_1, \nu_2, \nu_3)$, as a sum over all triangles $A$—which correspond to the fixed points of the torus action in the given phase:
\be
\label{eq:evol-FP}
\BV_X(\lam,\nu) =\sum_A \frac{(d_A)^2\, \mathe^{\tfrac{1}{d_A} \sum_m \lam^{i_m} (\nu \cdot u_A^{i_m})}}{\prod_m (\nu \cdot u_A^{i_m})}\ .
\ee

Finally, we note that the fixed-point formula presented in \cite[Section 2.3]{Martelli:2023oqk} admits a straightforward generalization to Calabi–Yau fourfolds, where triangles are naturally replaced by tetrahedra. Accordingly, the denominator of \eqref{eq:evol-FP} becomes quartic. In the explicit computations that follow, we restrict attention to geometric phases only, namely to triangulations in which every vertex of the toric diagram appears as a tip of a tetrahedron. As in the threefold case, the choice of triangulation can be unambiguously reconstructed from the final expression for the equivariant volume, since each exponential term is accompanied by the specific set of $\lam$-parameters associated with the corresponding vertices.

\subsubsection*{Example: the conifold \texorpdfstring{$\cC$}{C}}

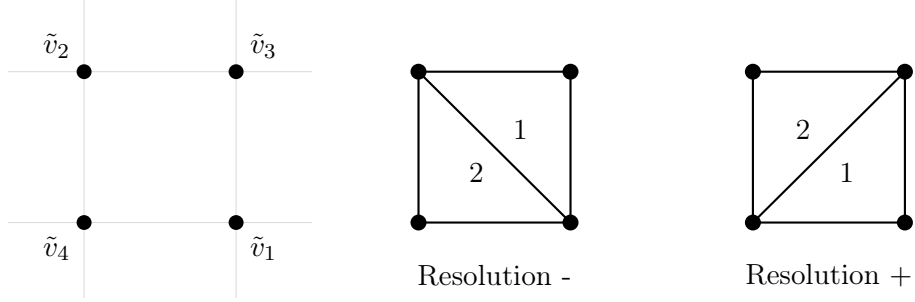
\begin{figure}
\centering
\begin{tikzpicture}[scale=2]
\begin{scope}[shift={(-2.2,0)}]
  \draw[step=1cm,gray!30,very thin] (-0.5,-0.5) grid (1.5,1.5);

  \coordinate (v1) at (0,0);
  \coordinate (v2) at (1,0);
  \coordinate (v3) at (0,1);
  \coordinate (v4) at (1,1);

  \foreach \p in {v1,v2,v3,v4}{
    \fill (\p) circle (1.4pt);
  }

  \node[below left=2pt of v1]  {$\tilde v_4$};
  \node[below right=2pt of v2] {$\tilde v_1$};
  \node[above left=2pt of v3]  {$\tilde v_2$};
  \node[above right=2pt of v4] {$\tilde v_3$};
\end{scope}

\begin{scope}
  \coordinate (v1) at (0,0);
  \coordinate (v2) at (1,0);
  \coordinate (v3) at (0,1);
  \coordinate (v4) at (1,1);
  \draw[thick] (v1) -- (v2) -- (v4) -- (v3) -- cycle;
  \draw[thick] (v2) -- (v3);
  \foreach \p in {v1,v2,v3,v4}{\fill (\p) circle (1.5pt);}
  \node at ($(v2)!0.43!(v1)!0.33!(v3)$) {2};
  \node at ($(v2)!0.43!(v4)!0.33!(v3)$) {1};
  \node at (0.5,-0.35) {Resolution -};
\end{scope}

\begin{scope}[shift={(2.2,0)}]
  \coordinate (v1) at (0,0);
  \coordinate (v2) at (1,0);
  \coordinate (v3) at (0,1);
  \coordinate (v4) at (1,1);
  \draw[thick] (v1) -- (v2) -- (v4) -- (v3) -- cycle;
  \draw[thick] (v1) -- (v4);
  \foreach \p in {v1,v2,v3,v4}{\fill (\p) circle (1.5pt);}
  \node at ($(v1)!0.43!(v2)!0.33!(v4)$) {1};
  \node at ($(v1)!0.43!(v3)!0.33!(v4)$) {2};
  \node at (0.5,-0.35) {Resolution $+$};
\end{scope}

\end{tikzpicture}
\caption{The toric diagram of the conifold, together with its two possible resolutions that correspond to two different triangulations.}
\label{fig:conifold}
\end{figure}

Since the equivalence of the above formulas is not immediately evident, let us consider the simplest non-trivial example: the resolved conifold, $\cC$. This space can be viewed as a smooth resolution of the cone over the Sasakian manifold $T^{1,1}$, and is defined via the GLSM charge vector
\be
	Q = (1, 1, -1, -1)\ , 
\ee
or alternatively via the toric fan
\be
 v =
 \begin{pmatrix}
 1 & 1 & 0 \\
 1 & 0 & 1 \\
 1 & 1 & 1 \\
 1 & 0 & 0
 \end{pmatrix}\ ,
\ee
such that $v^1 = (1, 1, 0)$, $v^2 = (1, 0, 1)$, etc., and the toric fan can be depicted in figure \ref{fig:conifold}.

Now, using \eqref{eq:nufrome}, we find
\be
\label{eq:conifnufrome}
	 \nu_1 = \e_1+\e_2+\e_3+\e_4\ , \quad \nu_2 = \e_1+\e_3\ , \quad \nu_3 = \e_2+\e_3\ ,
\ee
which will allow us to compare the two calculations of the equivariant volume described above.

Starting from the JK-residue formula, \eqref{eq:evol-JK}, 
\be
	 \BV_\cC (\lam,\e) = \oint_\text{JK} \prod_{a=1}^r\frac{\mathd\phi}{2\pi\mathi} \frac{\mathe^{\lam^1 (\e_1+\phi) + \lam^2 (\e_2+\phi) +\lam^3 (\e_3 - \phi) +\lam^4 (\e_4 - \phi)}}{(\e_1 + \phi) (\e_2 + \phi) (\e_3 - \phi) (\e_4 - \phi)}\ ,
\ee
we find that in this case there are only two chambers (corresponding to the two signs of the single K\"ahler parameter), see \cite{Cassia:2022lfj}. First, picking the positive sign, we need to pick the poles for $\phi = -\e_1$ and $\phi = -\e_2$, resulting in
\be
\label{eq:conifJKplus}
	 \BV^+_\cC (\lam, \e) = \frac1{(\e_2-\e_1)}\, \left(
 \frac{\mathe^{\lam^2 (\e_2-\e_1)+\lam^3 (\e_3+\e_1) +\lam^4 (\e_4+\e_1)}}
 {(\e_3+\e_1) (\e_4+\e_1)}
 - \frac{\mathe^{\lam^1 (\e_1-\e_2)+\lam^3 (\e_3+\e_2) +\lam^4 (\e_4+\e_2)}}
 {(\e_3+\e_2) (\e_4+\e_2)} \right)\ .
\ee
Similarly, for the other chamber (corresponding to the negative sign), we pick the poles $\phi = \e_3$ and $\phi = \e_4$, such that
\be
\label{eq:conifJKminus}
	 \BV^-_\cC (\lam, \e) = \frac1{(\e_4-\e_3)}\, \left(
 \frac{\mathe^{\lam^1 (\e_1+\e_3)+\lam^2 (\e_2+\e_3) +\lam^4 (\e_4-\e_3)}}
 {(\e_1+\e_3) (\e_2+\e_3)}
 - \frac{\mathe^{\lam^1 (\e_1+\e_4)+\lam^2 (\e_2+\e_4) +\lam^3 (\e_3-\e_4)}}
 {(\e_1+\e_4) (\e_2+\e_4)} \right)\ .
\ee

Utilizing instead the triangulation procedure, it is obvious that there are precisely two ways of splitting the toric diagram in this case (which is a square) in two triangles: along the diagonal $(1-2)$, or along the diagonal $(3-4)$. In the first triangulation, which would correspond to chamber $-$ above, we have the two triangles: $A_-=1 = (123)$ and $A_-=2 = (124)$. Then, we can readily calculate
\be
	\begin{split}
		d_1 &= 1\ , \qquad u^1_{1} = (1, 0, -1)\ , \quad u^2_1 = (1, -1, 0)\ , \quad u^3_1 = (-1,1,1)\ , \\
	d_2 &= 1\ , \qquad u^1_2 = (0, 1, 0)\ , \quad u^2_2 = (0, 0, 1)\ , \quad u^4_2 = (1,-1,-1)\ . \\
	\end{split}
\ee
We can then readily evaluate \eqref{eq:evol-FP} as
\be
\label{eq:conifFPminus}
	\BV^-_\cC (\lam, \nu) = \frac1{(\nu_1-\nu_2-\nu_3)}\, \left(
 \frac{\mathe^{\lam^1 \nu_2+\lam^2 \nu_3 +\lam^4 (\nu_1-\nu_2-\nu_3)}}
 { \nu_2 \nu_3}
 - \frac{\mathe^{\lam^1 (\nu_1 - \nu_3)+\lam^2 (\nu_1-\nu_2) +\lam^3 (\nu_2+\nu_3-\nu_1)}}
 {(\nu_1 - \nu_2) (\nu_1 - \nu_3)} \right)\ .
\ee
A simple inspection reveals that \eqref{eq:conifJKminus} and \eqref{eq:conifFPminus} coincide upon the relation \eqref{eq:conifnufrome}, as expected.

Similarly, the other triangulation corresponds to chamber $+$ in the JK prescription, with the two triangles $A_+=1 = (134)$ and $A_+=2 = (234)$. We find
\be
	\begin{split}
		d_1 &= 1\ , \qquad u^1_{1} = (0, 1, -1)\ , \quad u^3_1 = (0, 0, 1)\ , \quad u^4_1 = (1,-1,0)\ , \\
	d_2 &= 1\ , \qquad u^2_2 = (0, -1, 1)\ , \quad u^3_2 = (0, 1, 0)\ , \quad u^4_2 = (1,0,-1)\ , \\
	\end{split}
\ee
and correspondingly evaluate \eqref{eq:evol-FP} as
\be
\label{eq:conifFPplus}
	\BV^+_\cC (\lam, \nu) = \frac1{(\nu_3-\nu_2)}\, \left(
 \frac{\mathe^{\lam^2 (\nu_3-\nu_2)+\lam^3 \nu_2 +\lam^4 (\nu_1-\nu_3)}}
 { \nu_2 (\nu_1-\nu_3)}
 - \frac{\mathe^{\lam^1 (\nu_2 - \nu_3)+\lam^3 \nu_3 +\lam^4 (\nu_1-\nu_2)}}
 {(\nu_2- \nu_3) \nu_3} \right)\ .
\ee
It is again clear that \eqref{eq:conifJKplus} and \eqref{eq:conifFPplus} coincide using \eqref{eq:conifnufrome}, which concludes our explicit check on the consistency of the general formulae.

\subsubsection{Geometric prediction for the \texorpdfstring{$S^3$}{S3} partition function}\label{subsec:Holo-PF}
Assuming the equivariant volume has been computed using the methods described above, we arrive at the main insight and proposal of \cite{Cassia:2025aus, Cassia:2025jkr}: a relation between equivariant topological strings on a toric fourfold (conical resolution) $X$, and M2-brane partition functions (with M2-branes at the tip of the cone). In the present context, this implies that the perturbative part of the $S^3$ partition function of M2-brane theories is holographically predicted to be related to $F^\text{pert.}_X(\lambda, \epsilon)$. Consequently, we can express \eqref{eq:S3pf} entirely in terms of $\BV_X(\lambda, \epsilon)$ and its derivatives, which relate to the equivariant intersection numbers of $X$. Explicitly, the round sphere partition function was predicted in \cite{Cassia:2025jkr} to be,~\footnote{We have taken the round sphere limit of the general formula presented in \cite{Cassia:2025jkr}, for a squashing parameter $b=1$.}
\be
\label{eq:centralresultgeometry}
	Z_{S^3}^\text{pert} (\Delta, N) \simeq {\rm Ai} \Big[ \left( C_{X} (\Delta) \right)^{-1/3} \left(N - \frac{\chi(X)}{24}- \frac{ c_2 (\Delta) -2\, c_3 (\Delta)}{24}\right) \Big]\ ,
\ee
after the identification 
\be
\label{eq:e-Delta}
	\e_i = \Delta_i\ ,
\ee
between the equivariant parameters and the field theory mesonic deformations,~\footnote{Note that, for the holographic purpose of matching with the dual sphere partition function, \cite{Cassia:2025jkr} implemented the so-called mesonic twist; see also \cite{Hosseini:2019ddy}. Geometrically, this procedure corresponds to a blowdown of the two-cycles of the manifold $X$, and it plays a crucial role in enabling the identification of the equivariant parameters $\epsilon$ with the dual R-charges $\Delta$. This additional condition appears necessary to achieve agreement with all field-theoretic computations presented in this work. For a detailed analysis of the large $N$ behavior in the presence of additional baryonic symmetries, see \cite{Hosseini:2025mgf}.} under the supersymmetric constraint
\be
\label{eq:Deltaconstr}
	\sum_i \Delta_i = 2\ .
\ee
The quantities above, depending explicitly on the toric manifold $X$, can be expressed via the equivariant volume $\BV_X$ as follows:~\footnote{We have rescaled the definition of $C_X$ here with respect to \cite{Cassia:2025jkr} in order to fit the answer to the standard field theory conventions used in \eqref{eq:S3pf}.}
\be
\label{eq:Cgeometry}
	C_X (\e) := \frac{1}{8 \pi^2} \BV_X (\lambda = 0, \epsilon)\ ,
\ee
while the equivariant Chern numbers~\footnote{Note the somewhat unconventional notation we use here, defining the equivariant Chern numbers $c_p (\e) := \int_X c_p^\mathbb{T}$, with the subscript $\mathbb{T}$ denoting the equivariant upgrade of the ordinary Chern class. Precisely due to equivariance, the Chern numbers are in general not vanishing for any $p \leq d$.} are given by
\be
\label{eq:shorthandc2andc3}
\begin{split}	
	c_2 (\e) :=  \sum_{i < j} \frac{\partial^2 \BV_X (\lam,  \e)}{\partial \lam^i \partial \lam^j} & \Bigg|_{\lam = 0}\ , \qquad c_3 (\e) := \sum_{i < j < k} \frac{\partial^3 \BV_X (\lam,  \e)}{\partial \lam^i \partial \lam^j \partial \lam^k} \Bigg|_{\lam = 0}\ , \\
	\chi (X) := c_4 &= c_4 (\e) = \sum_{i < j < k < l} \frac{\partial^4 \BV_X (\lam,  \e)}{\partial \lam^i \partial \lam^j \partial \lam^k \partial \lam^l} \Bigg|_{\lam = 0}\ .
\end{split}
\ee
Note that in the last equation we used that $X$ is a fourfold. The Euler character, being the top Chern number, is always independent of the equivariant parameters $\e$ even if they appear in intermediate calculations on the right-hand side above.

Upon a direct comparison between the general form of the $S^3$ partition function, \eqref{eq:S3pf}, and \eqref{eq:centralresultgeometry}, apart from the function $C_X$ directly defined above, we can also identify the function $B_X$ incorporating the perturbative corrections,
\be
\label{eq:Bgeometry}
	B_X (\e) := \frac1{24}\, \left( c_2 (\e) -2\, c_3 (\e) + \chi (X)  \right)\ .
\ee

Note that, as in the discussion around \eqref{eq:usingnu}, we could bypass the redundant parametrization in terms of the $\epsilon_i$ parameters and their field-theoretic counterparts $\Delta_i$, and instead express all results directly using the effective parameters $\nu_\alpha$. In this formulation, the supersymmetric constraint \eqref{eq:Deltaconstr} simplifies to
\be
\nu_1 = 2\ ,
\ee
with $\nu_2, \dots, \nu_d$ remaining completely unconstrained. In what follows, we mainly use the parametrization in terms of the redundant $\e$ parameters, or holographically in terms of $\Delta$ parameters, due to the additional R-charge constraints, \eqref{eq:CY43-Rcond} that are readily applicable.

Lastly, let us emphasize that the proposed form of the $S^3$ partition function in \eqref{eq:centralresultgeometry} applies to M2-brane constructions in the absence of fractional brane charges—i.e.\ when there is no discrete torsion; see \cite{Aharony:2008gk}. The presence of discrete torsion can particularly affect the subleading terms encoded in $B_X(\epsilon)$, as discussed in \cite{Bergman:2009zh}.

\subsection{Holographic duality}\label{subsec:Holography}
 
In this section, we review the holographic dual pairs derived from the brane configurations. We focus particularly on the configurations \eqref{eq:fSYM-BC} and \eqref{eq:fABJM-BC}, which play a central role in our analysis. Additionally, the configurations \eqref{eq:fABJM-BC2} and \eqref{eq:fABJM-BC0} will also be relevant.

On the field theory side, the quiver diagrams can be directly inferred from the brane setups. 
We have previously illustrated the quiver diagrams associated with these three brane configurations: \eqref{eq:fSYM-BC} and \eqref{eq:fABJM-BC} correspond to the flavored SYM and flavored $\ccz$ theories, respectively, while Eq.~\eqref{eq:fABJM-BC2} corresponds to a quiver theory without adjoint matter.
See figure \ref{fig:SYM_fABJM_Quiv}.

\begin{figure}
\begin{centering}
\includegraphics[scale=0.7]{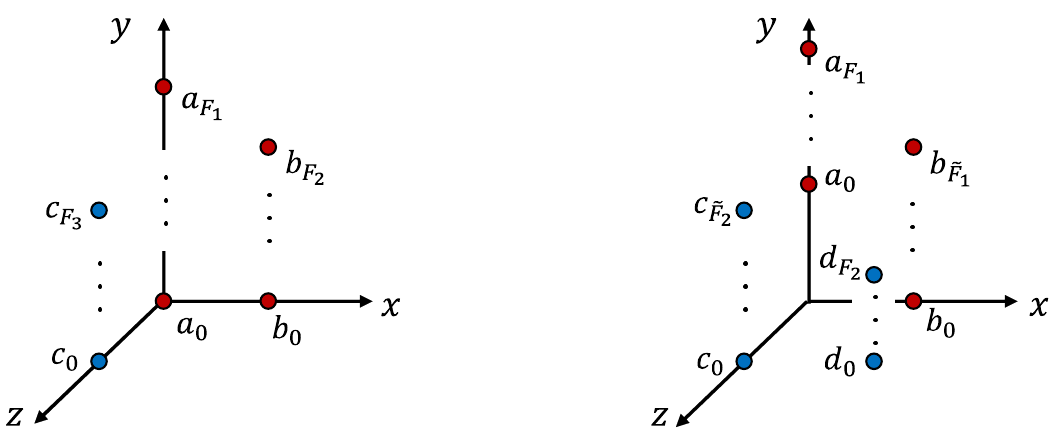}
\par\end{centering}
\caption{The 3d toric diagrams corresponding to \eqref{eq:fSYM-BC} (left) and \eqref{eq:fABJM-BC2} (right). The 2d toric diagrams at $z=0$ (red) and $z=1$ (blue) are dual to the corresponding $\left(p,q\right)$ webs. The perfect matching variables shown in the figure will be explained in due course. (Remark that the subscripts do not refer to the values of the $z$-coordinate.)}
\label{fig:newfig}
\end{figure}

\begin{figure}
\begin{centering}
\includegraphics[scale=0.7]{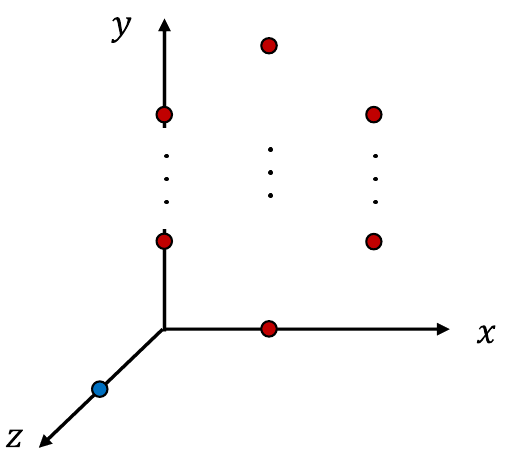}
\par\end{centering}
\caption{The 3d toric diagrams corresponding to the brane configurations \eqref{eq:fABJM-BC} and \eqref{eq:fABJM-BC0}.
There is a 2d toric diagram in the $z=0$ plane (red) which is dual to the $\left(p,q\right)$ web in \eqref{eq:fABJM-BC0} and a single point at $(x,y,z)=(0,0,1)$ (blue) corresponding to the adjoint matters.}\label{fig:fABJM1_TD}
\end{figure}

From the bulk perspective, a toric $\mathrm{CY}_4$ manifold can be obtained via T-duality from the type IIB brane setup, followed by uplifting to M-theory. 
In this process, D3-branes transform into M2-branes probing the toric $\mathrm{CY}_4$ geometry. 
This geometry can also be identified as the moduli space of (the abelianized version of) the corresponding three-dimensional theories. 
Numerous examples of toric $\mathrm{CY}_4$ manifolds have been classified in the literature \cite{Benini:2009qs,Cremonesi:2010ae,Amariti:2012tj}. 
For the brane configurations \eqref{eq:fSYM-BC} and \eqref{eq:fABJM-BC2}, the toric diagrams of the corresponding $\mathrm{CY}_4$s are shown in figure \ref{fig:newfig}. 
The toric diagram associated with the brane configuration \eqref{eq:fABJM-BC} has also been analyzed in \cite{Benini:2009qs,Cremonesi:2010ae,Amariti:2012tj} and corresponds to the direct product case $\mathbb{C}\times \mathrm{CY}_3$; see figure \ref{fig:fABJM1_TD}.


Our aim in this paper is to test, at the level of the perturbative Airy coefficients $C$ and $B$, the equality of the partition functions predicted by holography, focusing particularly on the brane configurations \eqref{eq:fSYM-BC} and \eqref{eq:fABJM-BC}. 
The dual theory associated with the brane configuration \eqref{eq:fSYM-BC} is straightforward: the flavored SYM theory depicted on the left side of figure \ref{fig:SYM_fABJM_Quiv} is dual to the topological string theory on the toric $\mathrm{CY}_4$, whose toric diagram is shown on the left side of figure \ref{fig:newfig} \cite{Benini:2009qs}. 
Here the 2d toric diagrams in the $z=0,1$ planes are dual (in a sense explained in appendix \ref{sec:Poly-Web}) to the $(p,q)$ webs $\left[\left(1,q\right)+F_{1}\mathrm{D5}^{-}+F_{2}\mathrm{D5}^{+}\right]$ and $F_{3}\left[\left(0,1\right)\right]$ in \eqref{eq:fSYM-BC}, respectively.
In specific examples, we show that the field-theory and geometric expressions for $C$ and $B$ coincide.

In contrast, the dual theories arising from the brane configuration \eqref{eq:fABJM-BC} are less straightforward. 
Notably, the matrix model for the flavored $\ccz$ theory nearly coincides with that of the ABJM theory once the R-charge condition $\Delta_{\phi_{i}}=1$ 
is imposed as \eqref{eq:fABJM-Rcond} for the flavored $\ccz$ theory.
This condition ensures the cancellation of the adjoint matter contributions, as discussed around \eqref{eq:DS-Prop}.
The sole difference remaining lies in the R-charge constraints for the bifundamental matter, governed by the marginality of the superpotential. 
It is evident that the constraint derived from \eqref{eq:fABJM2-W1} is weaker than the one from \eqref{eq:fABJM-W1} with $\Delta_{\phi_{i}}=1$. 
In other words, the $S^{3}$ partition function of the flavored $\ccz$ theory with \eqref{eq:fABJM-Rcond} can be obtained directly from the expression associated with the brane configuration \eqref{eq:fABJM-BC2} by further imposing the R-charge condition 
\begin{equation}
\Delta_{A_{i}}+\Delta_{B_{i}}=1\ ,
\label{eq:RAB-cond}
\end{equation}
for the ordinary ABJM theory.
Since the flavored ABJM theory is dual to the $\mathrm{CY}_{4}$ on the right-hand side of figure \ref{fig:newfig} via the brane configuration \eqref{eq:fABJM-BC2}, this reasoning leads to the expected relation between the flavored $\ccz$ theory and the toric $\mathrm{CY}_{4}$ depicted in that figure. 
Here the 2d toric diagrams in the $z=0,1$ planes are dual to the first and second $(p,q)$ webs in \eqref{eq:fABJM-BC}, respectively.
As shown in section \ref{sec:Examples}, once the R-charge condition is imposed, the two descriptions yield the same Airy coefficients $C$ and $B$.

Let us now comment on the apparent tension with the fact that the moduli space of the 3d theories shown on the right of figure \ref{fig:SYM_fABJM_Quiv} is known to be captured by the toric diagram in figure \ref{fig:fABJM1_TD} \cite{Benini:2009qs,Amariti:2012tj}. Although this may seem contradictory at first sight, there is actually no inconsistency. Rather, it signals the $\mathrm{CY}_{4}/\mathrm{CY}_{3}$ correspondence introduced in the introduction. In the present context, the relevant $\mathrm{CY}_{4}$ is the geometry on the right of figure \ref{fig:newfig}, whereas the associated $\mathrm{CY}_{3}$ is the 2d toric diagram at $z=0$ in figure \ref{fig:fABJM1_TD}. We will analyze this correspondence in detail in section \ref{sec:correspondence}. For now, we simply emphasize that this observation is fully compatible with existing results.

In the preceding discussion, the toric diagram of figure \ref{fig:fABJM1_TD} naturally enters. Since we argue that this geometry is not associated with the worldvolume theory of \eqref{eq:fABJM-BC}, it is reasonable to ask which brane configuration it does correspond to. A useful observation is that, in general—and at least for brane configurations built solely from the $\left(p,q\right)$ webs listed in table \ref{tab:Brane}—the ordering of the $\left(p,q\right)$ webs along the $x^{6}$ direction does not affect the dual toric geometry. This includes situations in which several $\left(p,q\right)$ webs sit at the same value of $x^{6}$ and merge into a single large $\left(p,q\right)$ web. Under the identification \eqref{eq:centralresultgeometry}, we find that the toric diagram in figure \ref{fig:fABJM1_TD} is dual to the brane configuration in which the two $\left(p,q\right)$ webs are placed at the same $x^{6}$ position, as in \eqref{eq:fABJM-BC0}.~\footnote{More generally, we expect that toric diagrams of this type correspond to brane configurations containing a single $\left(p,q\right)$ web, with that web dual to the 2d toric diagram itself.} When combined with the quantum curve computation, this identification again leads to the $\mathrm{CY}_{4}/\mathrm{CY}_{3}$ correspondence in section \ref{sec:correspondence}.~\footnote{Note that, in order to confirm this identification, it is essential to analyze not only $C$ but also $B$. This is because the $C$ associated with \eqref{eq:fABJM-BC} and \eqref{eq:fABJM-BC0} coincides, whereas $B$ does not. In other words, $C$ is insensitive to the ordering or relative positions of the $\left(p,q\right)$ webs along the $x^{6}$ direction, while $B$ is, in general, sensitive to such details.}

To summarize, in studying the holographic correspondence using the $S^{3}$ partition function \eqref{eq:S3pf}, we have identified three relevant types of dual pairs. 
First, the flavored SYM theory is dual to the left toric diagram in figure \ref{fig:newfig}. 
Second, the flavored $\ccz$ theory is dual to the right toric diagrams in figure \ref{fig:newfig} once the R-charge condition is imposed. 
Third, the worldvolume theory of the brane configuration \eqref{eq:fABJM-BC0} is dual to the toric diagram shown in figure~\ref{fig:fABJM1_TD}. 
Below, we focus on the specific holographically dual identifications concerning the $S^3$ partition function; in particular, we review the R-charge correspondence for these models.

\subsubsection{R-charge identification}\label{subsec:Holo-R}
We have described both the CFT and AdS theories and their mutual correspondence. In this section, we review the R-charge identification relevant for the holographic duality in our setup. As explained in section \ref{subsec:Holo-PF}, we have already matched the equivariant parameters $\epsilon_{i}$ with the field-theory parameters (basic R-charges) $\Delta_{i}$ via $\epsilon_{i}=\Delta_{i}$.
Each parameter $\Delta_{i}$ is assigned to the corresponding toric vertex $v^{i}$.
In what follows, we clarify how the basic R-charges $\Delta_{i}$ are related to the R-charges $\Delta_{X}$ of the matter fields.
Our discussion will focus in particular on the flavored SYM and flavored ABJM theories.

A key ingredient in this analysis is the notion of perfect matchings \cite{Benini:2009qs,Jafferis:2011zi}. Originally introduced in the context of 4d gauge theories, perfect matchings provide a map between parameters of the 4d quiver gauge theory and points in the toric diagram of the Calabi–Yau threefold.
This construction was later extended to 3d theories, where the toric diagram is uplifted to the three-dimensional diagram of the toric $\mathrm{CY}_{4}$. In our conventions, this uplift proceeds along the $y$-direction. 

Concretely, the relation between matter fields and monopole operators on the one hand, and the perfect matchings shown in figure \ref{fig:newfig} on the other, is given as follows.
For the flavored SYM theory, the relation is given by \cite{Jafferis:2011zi}:
\begin{align}
&\phi_{1} =\prod_{i=0}^{F_{1}}a_{i}\ ,\quad 
\phi_{2}=\prod_{i=0}^{F_{2}}b_{i}\ ,\quad 
\phi_{3}=\prod_{i=0}^{F_{3}}c_{i}\ ,\nonumber \\
&T =\left(\prod_{i=0}^{F_{1}}a_{i}^{F_{1}-i}\right)\left(\prod_{i=0}^{F_{2}}b_{i}^{F_{2}-i}\right)\left(\prod_{i=0}^{F_{3}}c_{i}^{F_{3}-i}\right)\ ,\quad
\tilde{T} =\left(\prod_{i=0}^{F_{1}}a_{i}^{i}\right)\left(\prod_{i=0}^{F_{2}}b_{i}^{i}\right)\left(\prod_{i=0}^{F_{3}}c_{i}^{i}\right)\ .
\label{eq:F-PM-fSYM}
\end{align}
For the flavored ABJM theory, the relation is \cite{Benini:2009qs}
\begin{align}
A_{1} & =\prod_{i=0}^{F_{1}}a_{i}\ ,\quad
B_{1}=\prod_{i=0}^{\tilde{F}_{1}}b_{i}\ ,\quad 
B_{2}=\prod_{i=0}^{\tilde{F}_{2}}c_{i}\ ,\quad 
A_{2}=\prod_{i=0}^{F_{2}}d_{i}\ ,\nonumber \\
T & =\left(\prod_{i=0}^{F_{1}}a_{i}^{F_{1}-i}\right)\left(\prod_{i=0}^{\tilde{F}_{1}}b_{i}^{\tilde{F}_{1}-i}\right)\left(\prod_{i=0}^{\tilde{F}_{2}}c_{i}^{\tilde{F}_{2}-i}\right)\left(\prod_{i=0}^{F_{2}}d_{i}^{F_{2}-i}\right)\ ,\nonumber \\
\tilde{T} & =\left(\prod_{i=0}^{F_{1}}a_{i}^{i}\right)\left(\prod_{i=0}^{\tilde{F}_{1}}b_{i}^{i}\right)\left(\prod_{i=0}^{\tilde{F}_{2}}c_{i}^{i}\right)\left(\prod_{i=0}^{F_{2}}d_{i}^{i}\right)\ .
\label{eq:F-PM-fABJM}
\end{align}
The above relations imply, for instance, that the field-theory operator $\det\left(\phi_{1}\right)$ of the flavored SYM corresponds to an M5-brane wrapping the entire $a_{i}$ tower \cite{Jafferis:2011zi}.
The scaling dimension of such operators is determined by the volume of the cycle wrapped by the corresponding M5-brane.

Crucially for our purposes here, the perfect matchings allow us to extract the relations between the basic R-charges and the R-charges of the matter fields and monopole operators.
In particular, when a field $X$ is related to a set of perfect matchings $t_i$, the corresponding relation takes the form
\begin{equation}
X=\prod_{i\in P_{X}}t_{i}^{\ell_{i}}\ .
\end{equation}
Then, since $t_{i}$ is associated with $v^{i}$,~\footnote{In general, the correspondence between the vertices and the perfect matchings is not one-to-one.
However, it is so for the SYM and fABJM theories, as illustrated in figure \ref{fig:newfig}. Therefore, in these cases one can directly identify each perfect matching $t_{i}$ with the corresponding vertex $v^{i}$.} one can read off the corresponding R-charge identification as follows:
\begin{equation}
\Delta_{X}=\sum_{i\in P_{X}}\ell_{i}\Delta_{i}\ .
\label{eq:F-PM-R}
\end{equation}

\section{\texorpdfstring{$S^3$}{S3} partition functions}\label{sec:Examples}

In this section we test the AdS/CFT correspondence for various examples. 
On the field theory side, we use the quantum curve technique reviewed and developed in section \ref{subsec:QC-LargeN}, while on the toric geometry side, we calculate the equivariant volume reviewed in section \ref{subsec:TCY-Loc}.
We then check the correspondence under the R-charge identification discussed in section \ref{subsec:Holography}.
More specifically, we verify the equivalence of the Airy coefficients $C$ and $B$ on both sides of the AdS/CFT correspondence.
The first example described in section \ref{subsec:C4} is useful for understanding the computation in the other examples.

\subsection{\texorpdfstring{$\mathbb{C}^{4}$}{C4} geometry}\label{subsec:C4}

In this section we study the $\mathbb{C}^{4}$ geometry, which corresponds to the cone over the seven-sphere.
Although this example has already been studied in the literature \cite{Benini:2009qs,Jafferis:2011zi,Cassia:2025jkr}, we revisit it here to illustrate the main concepts in section \ref{sec:Review}. It also serves as a prime example in the equivariant correspondence we discuss in the next section.

We consider two brane configurations which realize this toric diagram
\begin{subequations}
\label{eq:BC-C4}
\begin{align}
 & \left[\left(1,0\right)\right]-\left[\left(0,1\right)\right]-_{\mathrm{p}}\ ,
 \label{eq:BC-S001}\\
 & \left[\left(1,1\right)+\mathrm{D5}^{-}\right]-_{\mathrm{p}}\ .
 \label{eq:BC-S010}
\end{align}
\end{subequations}

\subsubsection*{CFT: Quantum curve result}

In this section we compute the large $N$ partition function on the field theory side using the quantum curve expression.

\begin{itemize}
\item Configuration $a$.\\ 
We start with the first brane configuration \eqref{eq:BC-S001}. 
The associated quantum curve is given by \eqref{eq:SYM-QC} with $q=0$, $F_{1}=F_{2}=0$, $F_{3}=1$. 
Concretely,
\begin{equation}
\hat{\mathcal{O}}_{0,\left(0,0,1\right)}^{\mathrm{fSYM}}\left(\hat{x},\hat{p};D,D_{\mathrm{m}}\right)
=e^{\frac{1}{2}D_{\mathrm{m}}\hat{x}}\left(e^{\left(-\frac{1}{2}+D\right)\hat{p}}+e^{\left(\frac{1}{2}+D\right)\hat{p}}\right)e^{\frac{1}{2}D_{\mathrm{m}}\hat{x}}\left(e^{-\frac{1}{2}\hat{x}}+e^{\frac{1}{2}\hat{x}}\right)\ .
\end{equation}
Note that the Dirac constant is $\hbar=2\pi$. 
Using \eqref{eq:OpSim} and \eqref{eq:CBHform}, the quantum curve can be written as
\begin{equation}
\hat{\mathcal{O}}_{0,\left(0,0,1\right)}^{\mathrm{fSYM}}\left(\hat{x},\hat{p};D,D_{\mathrm{m}}\right)
=\left(\begin{array}{cc}
+c_{\frac{1}{2},-\frac{1}{2}}e^{\left(\frac{1}{2}+D_{\mathrm{m}}\right)\hat{x}+\left(-\frac{1}{2}+D\right)\hat{p}} & +c_{\frac{1}{2},\frac{1}{2}}e^{\left(\frac{1}{2}+D_{\mathrm{m}}\right)\hat{x}+\left(\frac{1}{2}+D\right)\hat{p}}\\
+c_{-\frac{1}{2},-\frac{1}{2}}e^{\left(-\frac{1}{2}+D_{\mathrm{m}}\right)\hat{x}+\left(-\frac{1}{2}+D\right)\hat{p}} & +c_{-\frac{1}{2},\frac{1}{2}}e^{\left(-\frac{1}{2}+D_{\mathrm{m}}\right)\hat{x}+\left(\frac{1}{2}+D\right)\hat{p}}
\end{array}\right)\ ,
\label{eq:QC-S001}
\end{equation}
where
\begin{align}
 & c_{\frac{1}{2},-\frac{1}{2}}=e^{-\frac{1}{2}\left(-\frac{1}{2}+D\right)\pi i}\ ,\quad c_{\frac{1}{2},\frac{1}{2}}=e^{-\frac{1}{2}\left(\frac{1}{2}+D\right)\pi i}\ ,\nonumber \\
 & c_{-\frac{1}{2},-\frac{1}{2}}=e^{\frac{1}{2}\left(-\frac{1}{2}+D\right)\pi i}\ ,\quad c_{-\frac{1}{2},\frac{1}{2}}=e^{\frac{1}{2}\left(\frac{1}{2}+D\right)\pi i}\ .
 \label{eq:cmn-S001}
\end{align}
Note that we often express the quantum curves in a manner that follows the Newton polygon like \eqref{eq:QC-S001}. 
In the expression \eqref{eq:QC-S001}, we can use the property \eqref{eq:QC-W}, and thus the Wigner transform of the quantum curve is 
\begin{equation}
\left(\mathcal{O}_{0,\left(0,0,1\right)}^{\mathrm{fSYM}}\right)_{\mathrm{W}}\left(x,p;D,D_{\mathrm{m}}\right)
=e^{D_{\mathrm{m}}x+Dp}\left(\begin{array}{cc}
+c_{\frac{1}{2},-\frac{1}{2}}e^{\frac{1}{2}x-\frac{1}{2}p} & +c_{\frac{1}{2},\frac{1}{2}}e^{\frac{1}{2}x+\frac{1}{2}p}\\
+c_{-\frac{1}{2},-\frac{1}{2}}e^{-\frac{1}{2}x-\frac{1}{2}p} & +c_{-\frac{1}{2},\frac{1}{2}}e^{-\frac{1}{2}x+\frac{1}{2}p}
\end{array}\right)\ .
\label{eq:OW-S001}
\end{equation}
We then follow the analysis discussed in section \ref{subsec:QCtoLargeN} with this expression, which can be numerically checked, cf.\ figure \ref{fig:FermiSurface}. 
\begin{figure}
\begin{centering}
\includegraphics[scale=0.55]{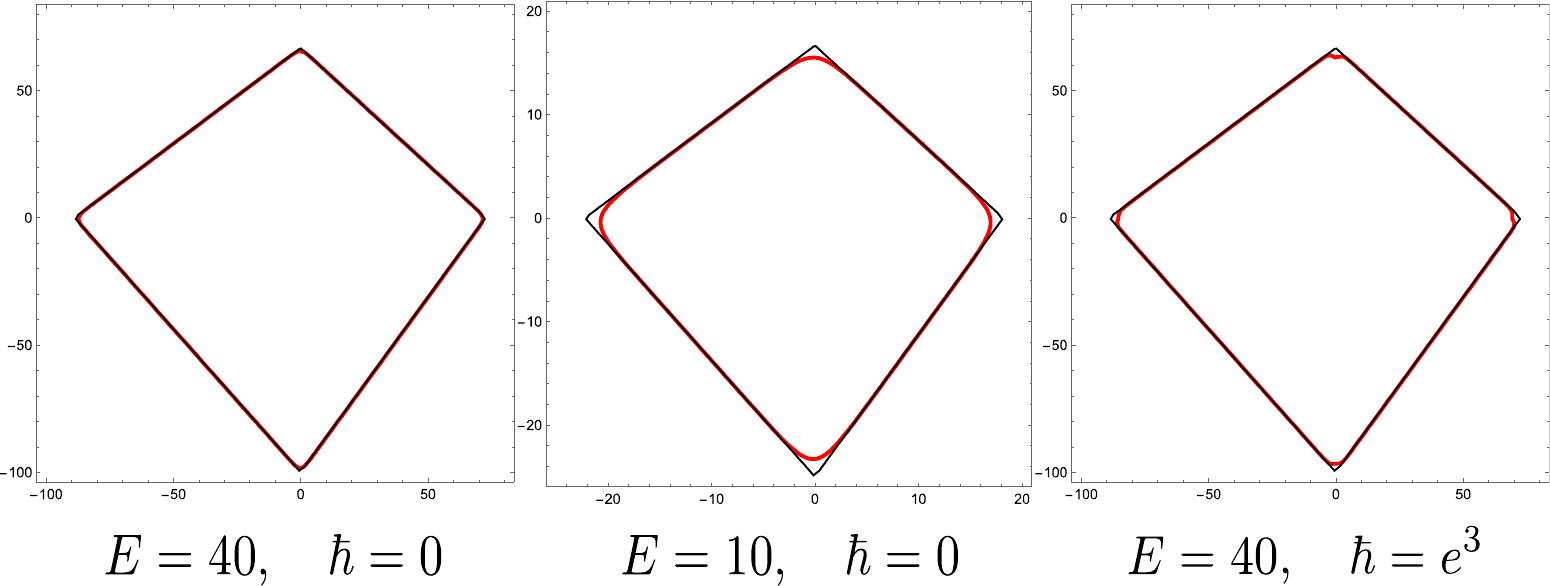}
\par\end{centering}
\caption{
The Fermi surface associated with \eqref{eq:QC-S001} for various $E$ and $\hbar$.
Here $D_{\mathrm{m}}=0.05$, $D=0.1$. 
Black lines are the exact polygon $\left(\pm\frac{1}{2}+D_{\mathrm{m}}\right)x+\left(\pm\frac{1}{2}+D\right)p=E$. 
Red curves represent $H_{\mathrm{W}}^{\left(0\right)}+\hbar^{2}H_{\mathrm{W}}^{\left(1\right)}=E$ with $(E,\hbar)$ shown below the panels.
At large $E$ the Fermi surface can be approximated by the polygon (left) and the corrections from $O\left(e^{-E}\right)$ and $H_{\mathrm{W}}^{\left(1\right)}$ are localized near the vertices of the polygon (center and right).
}
\label{fig:FermiSurface}
\end{figure}
We first compute the volume for the polygon part \eqref{eq:VolForm}. For the curve \eqref{eq:OW-S001} each term sits at a vertex, hence we need to consider all the terms. The volume is
\begin{equation}
\mathrm{Vol}_{\mathrm{Poly}}^{\left(0\right)}\left(E\right)=\frac{8}{\left(1-4D^{2}\right)\left(1-4D_{\mathrm{m}}^{2}\right)}E^{2}+\frac{1}{2}\pi^{2}\ .
\label{eq:VolP-S001}
\end{equation}
We then compute the corrections, which come from the four regions:
\begin{equation}
\mathrm{I}_{\pm}:\left(x,p\right)\approx\left(\pm2E,0\right),\quad\mathrm{II}_{\pm}:\left(x,p\right)\approx\left(0,\pm 2E\right)\ .
\end{equation}
Hence
\begin{equation}
\dvol=\sum_{\sigma=\pm}\left(\dvol_{\mathrm{I}_{\sigma}}+\dvol_{\mathrm{II}_{\sigma}}\right)\ .
\end{equation}
First, the relevant terms of the curve in region $\mathrm{I}_{\pm}$ are
\begin{equation}
\left.\left(\mathcal{O}_{0,\left(0,0,1\right)}^{\mathrm{fSYM}}\right)_{\mathrm{W}}\left(x,p;D,D_{\mathrm{m}}\right)\right|_{\mathrm{I}_{\pm}}
= e^{D_{\mathrm{m}}x+Dp}\left(c_{\pm\frac{1}{2},-\frac{1}{2}}e^{\pm\frac{1}{2}x-\frac{1}{2}p}+c_{\pm\frac{1}{2},\frac{1}{2}}e^{\pm\frac{1}{2}x+\frac{1}{2}p}\right)\ .
\end{equation}
Hence, by using \eqref{eq:VolC-K12} with \eqref{eq:OW-K1-mn}, we find
\begin{equation}
\dvol_{\mathrm{I}_{\pm}}
=-\frac{\pi^{2}}{6}\left(\frac{1}{\left(\frac{1}{2}\pm D_{\mathrm{m}}\right)}+\frac{1}{2}\pm D_{\mathrm{m}}\right)\ .
\label{eq:CorrI-S001}
\end{equation}
Similarly, after an appropriate canonical transformation, the relevant
terms in region $\mathrm{II}_{\pm}$ take the form
\begin{equation}
\left.\left(\mathcal{O}_{0,\left(0,0,1\right)}^{\mathrm{fSYM}}\right)_{\mathrm{W}}\left(x,p;D,D_{\mathrm{m}}\right)\right|_{\mathrm{II}_{\pm}}
= e^{\left(\frac{1}{2}\pm D\right)x}\left(e^{-\frac{1}{2}p}+e^{\frac{1}{2}p}\right)\ ,
\end{equation}
and hence
\begin{equation}
\dvol_{\mathrm{II}_{\pm}}=-\frac{\pi^{2}}{6}\left(\frac{1}{\left(\frac{1}{2}\pm D\right)}+\frac{1}{2}\pm D\right)\ .
\label{eq:CorrII-S001}
\end{equation}
Therefore, the total volume is
\begin{equation}
\mathrm{Vol}\left(E\right)=\frac{8}{\left(1-4D^{2}\right)\left(1-4D_{\mathrm{m}}^{2}\right)}E^{2}+\left(\frac{1}{6}-\frac{4\left(1-2D^{2}-2D_{\mathrm{m}}^{2}\right)}{3\left(1-4D^{2}\right)\left(1-4D_{\mathrm{m}}^{2}\right)}\right)\pi^{2}+O\left(e^{-E}\right)\ .
\end{equation}
Thus we arrive at
\begin{align}
C_{0,\left(0,0,1\right)}^{\mathrm{fSYM}} & =\frac{2}{\left(1-4D^{2}\right)\left(1-4D_{\mathrm{m}}^{2}\right)\pi^{2}},\nonumber \\
B_{0,\left(0,0,1\right)}^{\mathrm{fSYM}} & =\frac{1+2D^{2}+2D_{\mathrm{m}}^{2}}{3\left(1-4D^{2}\right)\left(1-4D_{\mathrm{m}}^{2}\right)}+\frac{1}{24}\ .
\label{eq:CB-S001}
\end{align}

\item Configuration $b$.\\
Next, we consider the second brane configuration \eqref{eq:BC-S010}. 
The associated quantum curve is given by \eqref{eq:SYM-QC} with $q=1$, $F_{1}=1$, $F_{2}=F_{3}=0$. 
The Wigner transform of the associated quantum curve is given by
\begin{equation}
\left(\mathcal{O}_{1,\left(1,0,0\right)}^{\mathrm{fSYM}}\right)_{\mathrm{W}}\left(x,p;D,D_{\mathrm{m}}\right)
=e^{\left(\frac{1}{4}-\frac{1}{2}D+D_{\mathrm{m}}\right)x+Dp}\left(\begin{array}{cc}
+e^{\frac{1}{2}x-\frac{1}{2}p} & +0\\
+e^{-\frac{1}{2}x-\frac{1}{2}p} & +e^{-\frac{1}{2}x+\frac{1}{2}p}
\end{array}\right)\ .
\label{eq:OW-S010}
\end{equation}
We first compute the volume for the polygon part. For the curve \eqref{eq:OW-S010} again each term sits at a vertex and we need to consider all three terms. The volume is
\begin{equation}
\mathrm{Vol}_{\mathrm{Poly}}\left(E\right)=\frac{16}{\left(1-2D\right)\left(\left(1+2D\right)^{2}-16D_{\mathrm{m}}^{2}\right)}E^{2}\ .
\label{eq:VolP-S010}
\end{equation}
We then compute the corrections, which come from the three regions:
\begin{equation}
\mathrm{I}:\left(x,p\right)\approx\left(E,E\right),\quad\mathrm{II}:\left(x,p\right)\approx\left(0,-2E\right),\quad\mathrm{III}:\left(x,p\right)\approx\left(-2E,0\right)\ .
\end{equation}
The curve in these regions can be written as
\begin{align}
\left.\left(\mathcal{O}_{1,\left(1,0,0\right)}^{\mathrm{fSYM}}\right)_{\mathrm{W}}\left(x,p;D,D_{\mathrm{m}}\right)\right|_{\mathrm{I}} & 
= e^{\alpha x+\beta p}\left(e^{\frac{1}{2}x-\frac{1}{2}p}+e^{-\frac{1}{2}x+\frac{1}{2}p}\right)\ ,\nonumber \\
\left.\left(\mathcal{O}_{1,\left(1,0,0\right)}^{\mathrm{fSYM}}\right)_{\mathrm{W}}\left(x,p;D,D_{\mathrm{m}}\right)\right|_{\mathrm{II}} & 
= e^{\alpha x+\beta p}\left(e^{\frac{1}{2}x-\frac{1}{2}p}+e^{-\frac{1}{2}x-\frac{1}{2}p}\right)\ ,\nonumber \\
\left.\left(\mathcal{O}_{1,\left(1,0,0\right)}^{\mathrm{fSYM}}\right)_{\mathrm{W}}\left(x,p;D,D_{\mathrm{m}}\right)\right|_{\mathrm{III}} & 
= e^{\alpha x+\beta p}\left(e^{-\frac{1}{2}x-\frac{1}{2}p}+e^{-\frac{1}{2}x+\frac{1}{2}p}\right)\ .
\end{align}
Here $\alpha=\frac14-\frac D2+D_{\mathrm m}$ and $\beta=D$.
Thus, using \eqref{eq:VolC-K12} with \eqref{eq:OW-K1-mn}, the corrections from each region can be computed as
\begin{equation}
\dvol=-\frac{\pi^{2}}{6}\left(1+\frac{2}{1-2D}+\frac{4}{1+2D-4D_{\mathrm{m}}}+\frac{4}{1+2D+4D_{\mathrm{m}}}\right)\ .
\label{eq:Corr-S010}
\end{equation}
Therefore, the total volume is the sum of \eqref{eq:VolP-S010} and \eqref{eq:Corr-S010} up to $O\left(e^{-E}\right)$,
\begin{align}
C_{1,\left(1,0,0\right)}^{\mathrm{fSYM}} & =\frac{4}{\left(1-2D\right)\left(\left(1+2D\right)^{2}-16D_{\mathrm{m}}^{2}\right)\pi^{2}}\ ,\nonumber \\
B_{1,\left(1,0,0\right)}^{\mathrm{fSYM}} & =\frac{6-D\left(\left(1-2D\right)^{2}-16D_{\mathrm{m}}^{2}\right)}{6\left(1-2D\right)\left(\left(1+2D\right)^{2}-16D_{\mathrm{m}}^{2}\right)}-\frac{1}{8}\ .
\label{eq:CB-S010}
\end{align}
\end{itemize}

\subsubsection*{AdS: Equivariant string result}

The toric fan vectors for \eqref{eq:BC-S001} and \eqref{eq:BC-S010} are given as
\begin{equation}
v^{(a)}=\left(\begin{array}{cccc}
1 & 0 & 0 & 0\\
1 & 1 & 0 & 0\\
1 & 0 & 1 & 1\\
1 & 0 & 0 & 1
\end{array}\right),\quad v^{(b)}=\left(\begin{array}{cccc}
1 & 0 & 0 & 0\\
1 & 1 & 0 & 0\\
1 & 0 & 1 & 0\\
1 & 0 & 0 & 1
\end{array}\right),
\end{equation}
respectively. They are related by the $\mathrm{SL}\left(3,\mathbb{Z}\right)$ transformation and thus give exactly the same equivariant volume. It is straightforward to check that the above toric fan vectors lead to an identically vanishing GLSM charge matrix, $Q = 0$. The JK-residue formula, \eqref{eq:evol-JK}, for the equivariant volume therefore trivially results in 
\be
    \BV = \frac{\mathe^{\e_i \lam^i}}{\prod_{i=1}^4 \e_i}\ ,
\ee
with a summation over repeated up-down indices in the exponent.

From \eqref{eq:Cgeometry}-\eqref{eq:shorthandc2andc3}, we find that
\begin{align}
 & C\left(\epsilon\right)=\frac{1}{8\pi^{2}\epsilon_{1}\epsilon_{2}\epsilon_{3}\epsilon_{4}}\ ,\qquad \chi=1\ ,\nonumber \\
 & c_{2}\left(\epsilon\right)=\sum_{i<j=1}^{4}\frac{1}{\epsilon_{i}\epsilon_{j}}\ ,\qquad 
 c_{3}\left(\epsilon\right)=\sum_{i=1}^{4}\frac{1}{\epsilon_{i}}\ .
 \label{eq:CB-C4}
\end{align}

Note that in this case the $\e_i$ variables are not redundant since the alternative parametrization in terms of $\nu$, cf.\ \eqref{eq:nufrome}, also produces four independent variables.

\subsubsection*{Holography}
We consider two holographic dualities derived by the two brane configurations in \eqref{eq:BC-C4}.

Let us start with the first brane configuration \eqref{eq:BC-S001}. 
The relation between the R-charges and the equivariant parameters can be read off from \eqref{eq:F-PM-fSYM} and \eqref{eq:F-PM-R} with \eqref{eq:e-Delta} as
\begin{equation}
\Delta_{\phi_{1}}^{(a)}=\epsilon_{1}\ ,\quad
\Delta_{\phi_{2}}^{(a)}=\epsilon_{2}\ ,\quad
\Delta_{\phi_{3}}^{(a)}=\epsilon_{3}+\epsilon_{4}\ ,\quad
\Delta_{T}^{(a)}=\epsilon_{4}\ ,\quad
\Delta_{\tilde{T}}^{(a)}=\epsilon_{3}\ .
\label{eq:R-Epara-S001}
\end{equation}
The R-charge constraints on $\Delta_{\phi_i}$ thus translate into the constraints
\be
    \e_1 + \e_2 = 1 = \e_3 + \e_4\ ,
\ee
such that the free parameters $D, D_{\mathrm{m}}$, cf.\ \eqref{eq:SYM-RD} and \eqref{eq:m-T-R}, become
\be
    D^{(a)} = \e_1 - \frac12 = \frac12 - \e_2\ , \qquad D_{\mathrm{m}}^{(a)} = \e_3 - \frac12 = \frac12 - \e_4\ .
\ee
With these identifications, one can check that \eqref{eq:CB-S001} is equivalent to \eqref{eq:CB-C4} with \eqref{eq:Bgeometry}.

Next, we consider the second brane configuration \eqref{eq:BC-S010}. 
The relation between the R-charges and the equivariant parameters can be read off from \eqref{eq:F-PM-fSYM} and \eqref{eq:F-PM-R} with \eqref{eq:e-Delta} as
\begin{equation}
\Delta_{\phi_{1}}^{(b)}=\epsilon_{1}+\epsilon_{3}\ ,\quad
\Delta_{\phi_{2}}^{(b)}=\epsilon_{2}\ ,\quad
\Delta_{\phi_{3}}^{(b)}=\epsilon_{4}\ ,\quad
\Delta_{T}^{(b)}=\epsilon_{1}\ ,\quad
\Delta_{\tilde{T}}^{(b)}=\epsilon_{3}\ .
\end{equation}
The R-charge constraints on $\Delta_{\phi_i}$ thus translate into the constraints
\be
    \e_1 + \e_2 + \e_3 = 1 = \e_4\ ,
\ee
such that the free parameters $D, D_{\mathrm{m}}$, cf.\ \eqref{eq:SYM-RD} and \eqref{eq:m-T-R}, become
\be
    D^{(b)} = \e_1 + \e_3 - \frac12 = \frac12 - \e_2\ , \qquad D^{(b)}_{\mathrm{m}} =\frac12\, \left(\e_1- \e_3\right)\ .
\ee
With these identifications one can check that \eqref{eq:CB-S010} is equivalent to \eqref{eq:CB-C4} with \eqref{eq:Bgeometry}. As expected, our results are in full agreement with the conjectures in \cite{Hristov:2022lcw} and \cite{Bobev:2022eus,Bobev:2025ltz}, as well as a number of exact localization results, see \cite{Herzog:2010hf,Fuji:2011km,Marino:2011eh,Nosaka:2015iiw,Hatsuda:2016uqa,Chester:2021gdw,Geukens:2024zmt,Kubo:2024qhq} and references therein.

\subsection{\texorpdfstring{$\mathbb{C}\times\mathcal{C}$}{C-Conifold} geometry}\label{subsec:Cfd}

In this section we study the $\mathbb{C}\times\mathcal{C}$ geometry, with $\cC$ the conifold we have discussed previously as an illustrative example. We consider three brane configurations which realize the corresponding toric diagram:
\begin{subequations}
\label{eq:BC-Cfd}
\begin{align}
 & \left[\left(1,1\right)+\mathrm{D5}^{-}\right]-\left[\left(1,0\right)\right]-_{\mathrm{p}}\ ,
 \label{eq:BC-A0100}\\
 & \left[\left(1,1\right)+\mathrm{D5}^{-}\right]-\left[\left(0,1\right)\right]-_{\mathrm{p}}\ ,
 \label{eq:BC-S011}\\
 & \left[\left(1,0\right)+\mathrm{D5}^{-}+\mathrm{D5}^{+}\right]-_{\mathrm{p}}\ .
 \label{eq:BC-S110}
\end{align}
\end{subequations}

\subsubsection*{CFT: Quantum curve result}
\begin{itemize}
\item Configuration $a$.\\
We start with the first brane configuration \eqref{eq:BC-A0100}. 
The associated quantum curve is given by \eqref{eq:fABJM-QC} with $q_{1}=1$, $q_{2}=0$, $\tilde{F}_{1}=F_{2}=\tilde{F}_{2}=0$, $F_{1}=1$. 
We also set $D_{\mathrm{m}} = 0$ without loss of generality. 
This is because $D_{\mathrm{m}}$ can be eliminated by shifting the integration variables as $(\mu, \nu) \to (\mu+c, \nu-c)$ with an appropriate constant $c$ at the level of the matrix model \eqref{eq:MM-fABJM}. 
This shift reflects the fact that individual bi-fundamental fields and monopole operators are not gauge invariant.
Consequently, there is a redundancy in their individual R-charges.
The Wigner transform of this quantum curve is given by
\begin{align}
 & \left(\mathcal{O}_{\left(1,1,0\right),\left(0,0,0\right)}^{\fccz}\right)_{\mathrm{W}}\left(x,p;D_{i}\right)\nonumber \\
 & =e^{\left(\frac{1}{4}-\frac{1}{2}D_{1}\right)x+\left(D_{1}+D_{2}\right)p}\left(\begin{array}{ccc}
+c_{\frac{1}{2},-1}e^{\frac{1}{2}x-p} & +c_{\frac{1}{2},0}e^{\frac{1}{2}x} & +0\\
+c_{-\frac{1}{2},-1}e^{-\frac{1}{2}x-p} & +c_{-\frac{1}{2},0}e^{-\frac{1}{2}x} & +c_{-\frac{1}{2},1}e^{-\frac{1}{2}x+p}
\end{array}\right)\ ,
\label{eq:OW-A0100}
\end{align}
where
\begin{align}
 & c_{\frac{1}{2},-1}=e^{-i\pi\left(\frac{1}{2}-D_{2}\right)\left(\frac{3}{4}-\frac{1}{2}D_{1}\right)}\ ,\quad c_{\frac{1}{2},0}=e^{i\pi\left(\frac{1}{2}+D_{2}\right)\left(\frac{3}{4}-\frac{1}{2}D_{1}\right)}\ ,\nonumber \\
 & c_{-\frac{1}{2},-1}=e^{i\pi\left(\frac{1}{2}-D_{2}\right)\left(\frac{1}{4}+\frac{1}{2}D_{1}\right)}\ ,\quad c_{-\frac{1}{2},0}=c_{-\frac{1}{2},-1}+c_{-\frac{1}{2},1}\ ,\quad c_{-\frac{1}{2},1}=e^{-i\pi\left(\frac{1}{2}+D_{1}\right)\left(\frac{1}{4}+\frac{1}{2}D_{2}\right)}\ .
\end{align}
The volume for the polygon part is given as
\begin{equation}
\mathrm{Vol}_{\mathrm{Poly}}\left(E\right)=\frac{8\left(7-2D_{1}\right)}{\left(3-2D_{1}\right)\left(1+2D_{1}\right)\left(1-D_{1}-D_{2}\right)\left(3+2D_{1}+4D_{2}\right)}E^{2}+\frac{7-2D_{1}}{16}\pi^{2}\ .
\label{eq:VolP-A0100}
\end{equation}
The corrections come from the four regions:
\begin{equation}
\left(x,p\right)\approx\left(+2E,0\right),\left(0,-E\right),\left(-2E,0\right),\left(+2E,+2E\right).
\end{equation}
Using \eqref{eq:VolC-K12} with \eqref{eq:OW-K1-mn} and \eqref{eq:OW-K2-mn}, the corrections from each region can be computed as
\begin{equation}
\dvol=-\frac{\pi^{2}}{16}\left(7-2D_{1}\right)-\frac{\pi^{2}}{6}\left(\frac{4}{3-2D_{1}}+\frac{1}{1-D_{1}-D_{2}}+\frac{8}{1+2D_{1}}+\frac{4}{3+2D_{1}+4D_{2}}\right)\ .
\label{eq:Corr-A0100}
\end{equation}
The total volume is the sum of \eqref{eq:VolP-A0100} and \eqref{eq:Corr-A0100}. Therefore, by using \eqref{eq:Vol-CB} and \eqref{eq:Vol-Gen}, we obtain
\begin{align}
C_{\left(1,1,0\right),\left(0,0,0\right)}^{\fccz} & =\frac{2\left(7-2D_{1}\right)}{\pi^{2}\left(3-2D_{1}\right)\left(1+2D_{1}\right)\left(1-D_{1}-D_{2}\right)\left(3+2D_{1}+4D_{2}\right)}\ ,\nonumber \\
B_{\left(1,1,0\right),\left(0,0,0\right)}^{\fccz} & =\frac{\left(7-2D_{1}\right)\left(1+12D_{1}^{2}+24D_{1}D_{2}-4D_{2}\left(1-4D_{2}\right)\right)}{24\left(3-2D_{1}\right)\left(1+2D_{1}\right)\left(1-D_{1}-D_{2}\right)\left(3+2D_{1}+4D_{2}\right)}\ .
\label{eq:CB-A0100}
\end{align}

\item Configuration $b$.\\
Next, we consider the second brane configuration \eqref{eq:BC-S011}. The associated quantum curve is given by \eqref{eq:SYM-QC} with $q=1$, $F_{1}=F_{3}=1$, $F_{2}=0$. The Wigner transform of this quantum curve is given by
\begin{equation}
\left(\mathcal{O}_{1,\left(1,0,1\right)}^{\mathrm{fSYM}}\right)_{\mathrm{W}}\left(x,p;D,D_{\mathrm{m}}\right)=e^{\left(\frac{1}{4}-\frac{1}{2}D+D_{\mathrm{m}}\right)x+Dp}\left(\begin{array}{cc}
+c_{1,-\frac{1}{2}}e^{x-\frac{1}{2}p} & +0\\
+c_{0,-\frac{1}{2}}e^{-\frac{1}{2}p} & +c_{0,\frac{1}{2}}e^{\frac{1}{2}p}\\
+c_{-1,-\frac{1}{2}}e^{-x-\frac{1}{2}p} & +c_{-1,\frac{1}{2}}e^{-x+\frac{1}{2}p}
\end{array}\right)\ ,
\label{eq:OW-S011}
\end{equation}
where
\begin{align*}
 & c_{1,-\frac{1}{2}}=e^{i\pi\left(\frac{1}{4}-\frac{1}{2}D\right)}\ ,\\
 & c_{0,-\frac{1}{2}}=c_{1,-\frac{1}{2}}+c_{-1,-\frac{1}{2}}\ ,\quad c_{0,\frac{1}{2}}=e^{-i\pi\left(\frac{1}{4}+\frac{1}{2}D\right)}\ ,\\
 & c_{-1,-\frac{1}{2}}=e^{-i\pi\left(\frac{1}{4}-\frac{1}{2}D\right)}\ ,\quad c_{-1,\frac{1}{2}}=e^{i\pi\left(\frac{1}{4}+\frac{1}{2}D\right)}\ ,
\end{align*}
The volume for the polygon part is given as
\begin{equation}
\mathrm{Vol}_{\mathrm{Poly}}\left(E\right)=\frac{16\left(3+2D\right)}{\left(1-4D^{2}\right)\left(\left(3+2D\right)^{2}-16D_{\mathrm{m}}^{2}\right)}E^{2}+\frac{3+2D}{8}\pi^{2}\ .
\label{eq:VolP-S011}
\end{equation}
The corrections come from the four regions:
\begin{equation}
\left(x,p\right)\approx\left(0,-2E\right),\left(-E,0\right),\left(0,+2E\right),\left(+2E,+2E\right).
\end{equation}
Using \eqref{eq:VolC-K12} with \eqref{eq:OW-K1-mn} and \eqref{eq:OW-K2-mn}, the corrections from each region can be computed as
\begin{equation}
\dvol=-\frac{\pi^{2}}{8}\left(3+2D\right)-\frac{\pi^{2}}{3}\left(\frac{3+2D}{1-4D^{2}}+\frac{4\left(3+2D\right)}{\left(3+2D\right)^{2}-16D_{\mathrm{m}}^{2}}\right)\ .
\label{eq:Corr-S011}
\end{equation}
The total volume is the sum of \eqref{eq:VolP-S011} and \eqref{eq:Corr-S011}. Therefore, by using \eqref{eq:Vol-CB} and \eqref{eq:Vol-Gen}, we obtain
\begin{align}
C_{1,\left(1,0,1\right)}^{\mathrm{fSYM}} & =\frac{4\left(3+2D\right)}{\left(1-4D^{2}\right)\left(\left(3+2D\right)^{2}-16D_{\mathrm{m}}^{2}\right)\pi^{2}}\ ,\nonumber \\
B_{1,\left(1,0,1\right)}^{\mathrm{fSYM}} & =\frac{\left(3+2D\right)\left(3-12\left(1-D\right)D+16D_{\mathrm{m}}^{2}\right)}{12\left(1-4D^{2}\right)\left(\left(3+2D\right)^{2}-16D_{\mathrm{m}}^{2}\right)}\ .
\label{eq:CB-S011}
\end{align}

\item Configuration $c$.\\
Finally, we consider the third brane configuration \eqref{eq:BC-S110}. The associated quantum curve is given by \eqref{eq:SYM-QC} with $q=0$, $F_{1}=F_{2}=1$, $F_{3}=0$. The Wigner transform of this quantum curve is given by
\begin{equation}
\left(\mathcal{O}_{0,\left(1,1,0\right)}^{\mathrm{fSYM}}\right)_{\mathrm{W}}\left(x,p;D,D_{\mathrm{m}}\right)=e^{D_{\mathrm{m}}x+Dp}\left(\begin{array}{cc}
+e^{\frac{1}{2}x-\frac{1}{2}p} & +e^{\frac{1}{2}x+\frac{1}{2}p}\\
+e^{-\frac{1}{2}x-\frac{1}{2}p} & +e^{-\frac{1}{2}x+\frac{1}{2}p}
\end{array}\right)\ .
\label{eq:OW-S110}
\end{equation}
The volume for the polygon part is given as
\begin{equation}
\mathrm{Vol}_{\mathrm{Poly}}\left(E\right)=\frac{8}{\left(1-4D^{2}\right)\left(1-4D_{\mathrm{m}}^{2}\right)}E^{2}\ .
\label{eq:VolP-S110}
\end{equation}
The corrections come from the four regions:
\begin{equation}
\left(x,p\right)\approx\left(\pm2E,0\right),\left(0,\pm2E\right).
\end{equation}
Using \eqref{eq:VolC-K12} with \eqref{eq:OW-K1-mn}, the corrections from each region can be computed as
\begin{equation}
\dvol=-\frac{\pi^{2}}{3}-\frac{4\left(1-2D^{2}-2D_{\mathrm{m}}^{2}\right)\pi^{2}}{3\left(1-4D^{2}\right)\left(1-4D_{\mathrm{m}}^{2}\right)}\ .
\label{eq:Corr-S110}
\end{equation}
The total volume is the sum of \eqref{eq:VolP-S110} and \eqref{eq:Corr-S110}. Therefore, by using \eqref{eq:Vol-CB} and \eqref{eq:Vol-Gen}, we obtain
\begin{align}
C_{0,\left(1,1,0\right)}^{\mathrm{fSYM}} & =\frac{2}{\left(1-4D^{2}\right)\left(1-4D_{\mathrm{m}}^{2}\right)\pi^{2}}\ ,\nonumber \\
B_{0,\left(1,1,0\right)}^{\mathrm{fSYM}} & =\frac{3+12D^{2}+12D_{\mathrm{m}}^{2}-16D^{2}D_{\mathrm{m}}^{2}}{12\left(1-4D^{2}\right)\left(1-4D_{\mathrm{m}}^{2}\right)}\ .
\label{eq:CB-S110}
\end{align}
\end{itemize}

\subsubsection*{AdS: Equivariant string result}

The toric fan vectors for \eqref{eq:BC-A0100}, \eqref{eq:BC-S011} and \eqref{eq:BC-S110} are given as
\begin{equation}
v^{(a)}=\left(\begin{array}{cccc}
1 & 0 & 0 & 0\\
1 & 1 & 0 & 1\\
1 & 1 & 0 & 0\\
1 & 0 & 0 & 1\\
1 & 0 & 1 & 0
\end{array}\right),\quad v^{(b)}=\left(\begin{array}{cccc}
1 & 0 & 0 & 0\\
1 & 0 & 1 & 1\\
1 & 0 & 0 & 1\\
1 & 0 & 1 & 0\\
1 & 1 & 0 & 0
\end{array}\right),\quad v^{(c)}=\left(\begin{array}{cccc}
1 & 0 & 0 & 0\\
1 & 1 & 1 & 0\\
1 & 0 & 1 & 0\\
1 & 1 & 0 & 0\\
1 & 0 & 0 & 1
\end{array}\right),
\label{eq:vec-Cfd}
\end{equation}
respectively. They are related by the $\mathrm{SL}\left(3,\mathbb{Z}\right)$ transformation and thus give the same equivariant volume. In this case we can write the GLSM charge vector as
\be
	Q = (1, 1, -1, -1, 0)\ , 
\ee
with the last entry corresponding to the $\mathbb{C}$ factor. We have evaluated the equivariant volume of the conifold in quite some detail above, and here we choose one of the possible answers (either one leads to the same holographic predictions), e.g.\ the positive chamber, \eqref{eq:conifJKplus},
\be
	 \BV (\lam, \e) = \frac{\mathe^{\lam^5 \e_5}}{(\e_2-\e_1)\, \e_5}\, \left(
 \frac{\mathe^{\lam^2 (\e_2-\e_1)+\lam^3 (\e_3+\e_1) +\lam^4 (\e_4+\e_1)}}
 {(\e_3+\e_1) (\e_4+\e_1)}
 - \frac{\mathe^{\lam^1 (\e_1-\e_2)+\lam^3 (\e_3+\e_2) +\lam^4 (\e_4+\e_2)}}
 {(\e_3+\e_2) (\e_4+\e_2)} \right)\ ,
\ee
where $\e_5, \lam^5$ correspond to the equivariant parameters in the direction of the $\mathbb{C}$ factor.

From \eqref{eq:Cgeometry}-\eqref{eq:shorthandc2andc3}, we find
\begin{align}
 & C\left(\epsilon\right)=\frac{\epsilon_{1}+\epsilon_{2}+\epsilon_{3}+\epsilon_{4}}{8\pi^{2}\left(\epsilon_{1}+\epsilon_{3}\right)\left(\epsilon_{2}+\epsilon_{3}\right)\left(\epsilon_{1}+\epsilon_{4}\right)\left(\epsilon_{2}+\epsilon_{4}\right)\epsilon_{5}}\ ,\qquad\chi=2\ ,\nonumber \\
 & c_{2}\left(\epsilon\right)=8\pi^{2}C\left(\epsilon\right)\left(\epsilon_{1}\epsilon_{2}+\epsilon_{3}\epsilon_{4}+\sum_{i<j=1}^{5}\epsilon_{i}\epsilon_{j}\right)\ ,\nonumber \\
 & c_{3}\left(\epsilon\right)=\frac{1}{\epsilon_{1}+\epsilon_{3}}+\frac{1}{\epsilon_{2}+\epsilon_{3}}+\frac{1}{\epsilon_{1}+\epsilon_{4}}+\frac{1}{\epsilon_{2}+\epsilon_{4}}+\frac{2}{\epsilon_{5}}\ .
 \label{eq:CB-cfd}
\end{align}
Note that instead of the redundant variables $\e_1, \dots, \e_4$, we could alternatively use the $\nu$-variables in the conifold directions, as in \eqref{eq:conifnufrome}.

\subsubsection*{Holography}
We consider three holographic dualities derived by the three brane configurations \eqref{eq:BC-Cfd}.

For configuration $a$, the relation between the R-charges and the
equivariant parameters follows from
\eqref{eq:F-PM-fABJM}, \eqref{eq:fABJM-RD}, and
\eqref{eq:m-T-R}, whereas for configurations $b$ and $c$ it follows
from \eqref{eq:F-PM-fSYM}, \eqref{eq:SYM-RD}, and
\eqref{eq:m-T-R}, with $\epsilon_i=\Delta_i$.
For \eqref{eq:BC-A0100},
\begin{equation}
\Delta_{A_{1}}^{(a)}=\epsilon_{1}+\epsilon_{5}\ ,\quad
\Delta_{B_{1}}^{(a)}=\epsilon_{3}\ ,\quad
\Delta_{B_{2}}^{(a)}=\epsilon_{4}\ ,\quad
\Delta_{A_{2}}^{(a)}=\epsilon_{2}\ ,\quad
\Delta_{T}^{(a)}=\epsilon_{1}\ ,
\quad\Delta_{\tilde{T}}^{(a)}=\epsilon_{5}\ ,
\end{equation}
under the constraints
\be
    \e_1 + \e_3 + \e_5 = 1 = \e_2+\e_4\ ,
\ee
such that
\be
    D^{(a)}_1 = \e_1 + \e_5 - \frac12 = \frac12 - \e_3\ , \qquad D_2^{(a)} =\e_4 - \frac12 = \frac12 - \e_2\ .
\ee

For \eqref{eq:BC-S011},
\begin{equation}
\Delta_{\phi_{1}}^{(b)}=\epsilon_{1}+\epsilon_{4}\ ,\quad
\Delta_{\phi_{2}}^{(b)}=\epsilon_{5}\ ,\quad
\Delta_{\phi_{3}}^{(b)}=\epsilon_{2}+\epsilon_{3}\ ,\quad
\Delta_{T}^{(b)}=\epsilon_{1}+\epsilon_{3}\ ,\quad
\Delta_{\tilde{T}}^{(b)}=\epsilon_{2}+\epsilon_{4}\ ,
\label{eq:Holo-Cfd-b}
\end{equation}
under the constraints
\be
    \e_1 + \e_4 + \e_5 = 1 = \e_2+\e_3\ ,
\ee
such that
\be
    D^{(b)} = \e_1 + \e_4 - \frac12 = \frac12 - \e_5\ , \qquad D_{\mathrm{m}}^{(b)} =\frac12\, \left(\e_1-\e_2+ \e_3-\e_4\right)\ .
\ee

For \eqref{eq:BC-S110},
\begin{equation}
\Delta_{\phi_{1}}^{(c)}=\epsilon_{1}+\epsilon_{3}\ ,\quad
\Delta_{\phi_{2}}^{(c)}=\epsilon_{2}+\epsilon_{4}\ ,\quad
\Delta_{\phi_{3}}^{(c)}=\epsilon_{5}\ ,\quad
\Delta_{T}^{(c)}=\epsilon_{1}+\epsilon_{4}\ ,\quad
\Delta_{\tilde{T}}^{(c)}=\epsilon_{2}+\epsilon_{3}\ ,
\label{eq:R-Epara-S110}
\end{equation}
under the constraints
\be
    \e_1 + \e_2 + \e_3 + \e_4 = 1 =\e_5\ ,
\ee
such that
\be
    D^{(c)} = \e_1 + \e_3 - \frac12 = \frac12 - \e_2 - \e_4\ , \qquad D_{\mathrm{m}}^{(c)} =\frac12\, \left(\e_1-\e_2- \e_3+\e_4\right)\ .
\ee

Under the above identifications, we confirmed that \eqref{eq:CB-A0100}, \eqref{eq:CB-S011} and \eqref{eq:CB-S110} are equivalent to \eqref{eq:CB-cfd} with \eqref{eq:Bgeometry}. As far as we are aware, we have demonstrated a successful holographic match at finite $N$ for the first time, generalizing the large $N$ result of \cite{Jafferis:2011zi}.

Note that a one-parameter family of the form $\epsilon_{i}=f_{i}\left(\Delta_{X},\delta\right)$ can still satisfy the above identifications. This is due to the redundancy of the equivariant parameters $\e$ used above, \cite{Martelli:2023oqk,Cassia:2025aus}. As previously remarked, we could have simply used the $\nu$-variables in \eqref{eq:conifFPplus}, which would fix the redundancy. This happens also for the remaining examples, where we keep using the redundant $\e$ variables.

We also remark that, regarding the first case \eqref{eq:BC-A0100}, one might expect an additional constraint $\epsilon_1 - \epsilon_5 = 0$ since we set $D_{\mathrm{m}} = 0$. 
However, this can be achieved without loss of generality by utilizing the redundancy $(\epsilon_1, \epsilon_5) \to (\epsilon_1 + c, \epsilon_5 - c)$. 
On the field theory side we also set $D_{\mathrm{m}} = 0$ without loss of generality by utilizing the gauge redundancy as briefly explained above \eqref{eq:OW-A0100}.
Thus, the redundancy appears on both the gauge theory and geometry sides and does not correspond to an additional physical restriction.

\subsection{\texorpdfstring{$C (Q^{1,1,1})$}{Q111} geometry}\label{subsec:Q111}

In this section we study the geometry of the resolution of the cone over the Sasaki-Einstein space $Q^{1,1,1}$, which by an abuse of notation we also label in the same way. We consider the following brane configuration, which realizes this toric diagram
\begin{equation}
\left[\left(1,1\right)+\mathrm{D5}^{-}\right]-\left[\left(1,0\right)+\mathrm{D5}^{+}\right]-_{\mathrm{p}}\ .
\label{eq:BC-A0110}
\end{equation}

\subsubsection*{CFT: Quantum curve result}

The quantum curve for \eqref{eq:BC-A0110} is given by \eqref{eq:fABJM-QC} with $q_{1}=1$, $q_{2}=0$, $F_{1}=F_{2}=1$, $\tilde{F}_{1}=\tilde{F}_{2}=0$. 
We also set $D_{1}=D_{2}=\tilde{D}/2$ exploiting the gauge redundancy discussed in section \ref{subsec:Cfd}.
The Wigner transform of this quantum curve  is then given by
\begin{align}
&\left(\mathcal{O}_{\left(1,1,0\right),\left(0,1,0\right)}^{\fccz}\right)_{\mathrm{W}}\left(x,p;\tilde D,D_{\mathrm{m}}\right) \nonumber \\
&\quad =e^{\left(-\frac{1}{2}\tilde D+D_{\mathrm{m}}\right)x+\tilde D p}\left(\begin{array}{ccc}
+c_{1,-1}e^{x-p} & +c_{1,0}e^{x} & +0\\
+c_{0,-1}e^{-p} & +c_{0,0} & +c_{0,1}e^{p}\\
+0 & +c_{-1,0}e^{-x} & +c_{-1,1}e^{-x+p}
\end{array}\right)\ ,
\label{eq:OW-A0110}
\end{align}
where
\begin{align}
 & c_{1,-1}=e^{-i\pi\left(\frac{1}{4}-\frac{1}{4}\tilde D\right)}\ ,\quad 
 c_{1,0}=e^{i\pi\left(\frac{1}{2}+\frac{1}{2}D_{\mathrm{m}}\right)}\ ,\nonumber \\
 & c_{0,-1}=e^{i\pi\left(\frac{1}{4}-\frac{1}{4}\tilde D\right)}\ ,\quad 
 c_{0,1}=e^{-i\pi\left(\frac{1}{4}+\frac{1}{4}\tilde D\right)}\ ,\nonumber \\
 & c_{-1,0}=e^{-i\pi\left(\frac{1}{2}-\frac{1}{2}D_{\mathrm{m}}\right)}\ ,\quad 
 c_{-1,1}=e^{i\pi\left(\frac{1}{4}+\frac{1}{4}\tilde D\right)}\ .
\end{align}
The volume for the polygon part is given as
\begin{equation}
\mathrm{Vol}_{\mathrm{Poly}}\left(E\right)=\frac{4\left(12-3\tilde D^{2}-4D_{\mathrm{m}}^{2}\right)}{\left(1-\tilde D^{2}\right)\left(4-\left(\tilde D-2D_{\mathrm{m}}\right)^{2}\right)\left(4-\left(\tilde D+2D_{\mathrm{m}}\right)^{2}\right)}E^{2}+\pi^{2}\ .
\label{eq:VolP-A0110}
\end{equation}
The corrections come from the six regions:
\begin{equation}
\left(x,p\right)\approx\left(\pm E,\pm E\right),\left(\pm E,0\right),\left(0,\pm E\right)\ .
\end{equation}
Using \eqref{eq:VolC-K12} with \eqref{eq:OW-K1-mn}, the corrections from each region can be computed as
\begin{equation}
\dvol=-\pi^{2}-\frac{\left(12-3\tilde D^{2}-4D_{\mathrm{m}}^{2}\right)\left(4-3\tilde D^{2}-4D_{\mathrm{m}}^{2}\right)\pi^{2}}{3\left(1-\tilde D^{2}\right)\left(4-\left(\tilde D-2D_{\mathrm{m}}\right)^{2}\right)\left(4-\left(\tilde D+2D_{\mathrm{m}}\right)^{2}\right)}\ .
\label{eq:Corr-A0110}
\end{equation}
The total volume is the sum of \eqref{eq:VolP-A0110} and \eqref{eq:Corr-A0110}. Therefore, by using \eqref{eq:Vol-CB} and \eqref{eq:Vol-Gen}, we obtain
\begin{align}
C_{\left(1,1,0\right),\left(0,1,0\right)}^{\fccz} & =\frac{12-3\tilde D^{2}-4D_{\mathrm{m}}^{2}}{\left(1-\tilde D^{2}\right)\left(4-\left(\tilde D-2D_{\mathrm{m}}\right)^{2}\right)\left(4-\left(\tilde D+2D_{\mathrm{m}}\right)^{2}\right)\pi^{2}}\ ,\nonumber \\
B_{\left(1,1,0\right),\left(0,1,0\right)}^{\fccz} & =\frac{\left(12-3\tilde D^{2}-4D_{\mathrm{m}}^{2}\right)\left(3\tilde D^{2}+4D_{\mathrm{m}}^{2}\right)}{12\left(1-\tilde D^{2}\right)\left(4-\left(\tilde D-2D_{\mathrm{m}}\right)^{2}\right)\left(4-\left(\tilde D+2D_{\mathrm{m}}\right)^{2}\right)}\ .
\label{eq:CB-A0110}
\end{align}

\subsubsection*{AdS: Equivariant string result}

The toric fan vector for \eqref{eq:BC-A0110} is given as
\be
 v =
 \begin{pmatrix}
 1 & 1 & 1 & 1 \\
 1 & 0 & 0 & 0 \\
 1 & 1 & 0 & 1 \\
 1 & 0 & 1 & 0 \\
 1 & 1 & 0 & 0 \\
 1 & 0 & 1 & 1
 \end{pmatrix}\ ,
\ee
leading to the GLSM charge matrix
\be
Q =
\begin{pmatrix}
1 & 1 & 0 & 0 & -1 & -1 \\
0 & 0 & 1 & 1 & -1 & -1
\end{pmatrix}\ ,
\ee
corresponding to the resolution of the cone over $L = Q^{1,1,1}$. In this case the equivariant volume was explicitly computed (in a particular geometric phase) in \cite{Cassia:2025aus},
\be
\ba
 \BV(\lam,\e) =&\frac{\mathe^{\lam^2 (\e_2 - \e_1) + \lam^4 (\e_4-\e_3) + \sum_{j=5}^6 \lam^j (\e_j+ \e_1+\e_3)}}{(\e_2-\e_1)(\e_4-\e_3)  \prod_{j=5}^6 (\e_j+ \e_1+\e_3)} +  \frac{\mathe^{\lam^2 (\e_2 - \e_1) + \lam^3 (\e_3-\e_4) + \sum_{j=5}^6 \lam^j (\e_j+ \e_1+\e_4)}}{(\e_2-\e_1)(\e_3-\e_4) \prod_{j=5}^6 (\e_j+ \e_1+\e_4)}\\
 &+ \frac{\mathe^{\lam^1 (\e_1 - \e_2) + \lam^4 (\e_4-\e_3) + \sum_{j=5}^6 \lam^j (\e_j+ \e_2+\e_3)}}{(\e_1-\e_2)(\e_4-\e_3) \prod_{j=5}^6 (\e_j+ \e_2+\e_3)} +  \frac{\mathe^{\lam^1 (\e_1 - \e_2) + \lam^3 (\e_3-\e_4) + \sum_{j=5}^6 \lam^j (\e_j+ \e_2+\e_4)}}{(\e_1-\e_2)(\e_3-\e_4) \prod_{j=5}^6 (\e_j+ \e_2+\e_4)} \ .
\ea
\ee

We finally find that
\begin{align}
 & C\left(\epsilon\right)=\frac{1}{8\pi^{2}\left(\epsilon_{1}-\epsilon_{2}\right)\left(\epsilon_{3}-\epsilon_{4}\right)}\left(\sum_{i=1}^{2}\sum_{j=3}^{4}\left(-1\right)^{i+j}\prod_{k=5}^{6}\frac{1}{\left(\epsilon_{i}+\epsilon_{j}+\epsilon_{k}\right)}\right)\ ,\quad\chi=4\ ,\nonumber \\
 & c_{2}\left(\epsilon\right)=
 \sum_{j=3}^{4}\sum_{k=5}^{6}\prod_{i=1}^{2}\frac{1}{\left(\epsilon_{i}+\epsilon_{j}+\epsilon_{k}\right)}
 +\sum_{i=1}^{2}\sum_{k=5}^{6}\prod_{j=3}^{4}\frac{1}{\left(\epsilon_{i}+\epsilon_{j}+\epsilon_{k}\right)}
 +\sum_{i=1}^{2}\sum_{j=3}^{4}\prod_{k=5}^{6}\frac{1}{\left(\epsilon_{i}+\epsilon_{j}+\epsilon_{k}\right)}\ ,\nonumber \\
 & c_{3}\left(\epsilon\right)=\sum_{i=1}^{2}\sum_{j=3}^{4}\sum_{k=5}^{6}\frac{1}{\epsilon_{i}+\epsilon_{j}+\epsilon_{k}}\ .\label{eq:CB-Q111}
\end{align}
Note that instead of the redundant variables $\e_1, \dots, \e_6$, we could alternatively use the $\nu$-variables,
\be
    \nu_1 = \sum_{i=1}^6 \e_i\ , \quad \nu_2 = \e_1+\e_3+\e_5\ , \quad \nu_3 = \e_1 + \e_4+\e_5\ , \quad \nu_4 = \e_1+\e_3+\e_6\ .
\ee

\subsubsection*{Holography}

We consider the holographic duality derived by the brane configuration \eqref{eq:BC-A0110}. The relation between the R-charges and the equivariant parameters is given by \eqref{eq:F-PM-fABJM} and \eqref{eq:fABJM-RD}-\eqref{eq:m-T-R} with $\epsilon_{i}=\Delta_{i}$ as
\begin{equation}
\Delta_{A_{1}}=\epsilon_{2}+\epsilon_{4}\ ,\quad
\Delta_{B_{1}}=\epsilon_{5}\ ,\quad
\Delta_{B_{2}}=\epsilon_{6}\ ,\quad
\Delta_{A_{2}}=\epsilon_{1}+\epsilon_{3}\ ,\quad
\Delta_{T}=\epsilon_{2}+\epsilon_{3}\ , \quad
\Delta_{\tilde{T}}=\epsilon_{1}+\epsilon_{4}\ ,
\end{equation}
under the constraints
\be
    \e_1 + \e_3 + \e_6 = 1 = \e_2+\e_4+\e_5\ ,
\ee
such that
\be
\begin{split}
    D_1 = \e_2 + \e_4 - \frac12 &= \frac12 - \e_5\ = \frac{\tilde D}{2} , \qquad 
    D_2 =\e_6 - \frac12 = \frac12 - \e_1 -\e_3 = \frac{\tilde D}{2}\ , \\
    D_{\mathrm{m}} &= \frac12\, \left( \e_2 + \e_3 - \e_1 - \e_4 \right)\ .
\end{split}
\ee
Note that the identification of $D_1$ and $D_2$ on the field theory side further imposes the following relations:
\be
    \e_6 = \e_2 + \e_4\ , \qquad \e_5 = \e_1 + \e_3\ ,
\ee
such that there are in total two free parameters on both sides of the correspondence
(as discussed in section \ref{subsec:Cfd}, this is not a physical restriction).
Under these identifications, we confirmed that \eqref{eq:CB-A0110} is equivalent to \eqref{eq:CB-Q111} with \eqref{eq:Bgeometry}. 
Similarly to the previous example, we have demonstrated a successful holographic match at finite $N$ for the first time, generalizing the large $N$ result of \cite{Jafferis:2011zi}. Note that our results appear to disagree with the subleading expression in \cite{Bobev:2025ltz}, which in our understanding is conjectured to hold at vanishing deformation parameters, $\tilde D = D_{\mathrm{m}} = 0$.~\footnote{Ref.\ \cite{Bobev:2025ltz} considered the more general geometry $Q^{1,1,1}/\mathbb{Z}_{N_{\mathrm{f}}}$. The corresponding brane configuration is 
\be
\left[\left(1,N_{\mathrm{f}}\right)+N_{\mathrm{f}}\, \mathrm{D5}^{-}\right]-\left[\left(1,0\right)+N_{\mathrm{f}}\, \mathrm{D5}^{+}\right]-_{\mathrm{p}}\ ,
\ee
for which the Airy coefficients can be computed. The quantum curve is  $\mathcal{O}_{\left(N_{\mathrm{f}},N_{\mathrm{f}},0\right),\left(0,N_{\mathrm{f}},0\right)}^{\fccz}$, and the computation proceeds similarly (we set $\tilde{D}=D_{\mathrm{m}}=0$ for simplicity). The main difference from the previous procedure is the need to consider a general $K$ in \eqref{eq:OW-Canonical}. However, after an appropriate canonical transformation, the relevant terms simplify to $\mathcal{O}_{\mathrm{W}}\sim e^{x}\left(e^{-\frac{1}{2}p}+e^{\frac{1}{2}p}\right)^{N_{\mathrm{f}}}$.
The subleading term of the small-$\hbar$ expansion for this curve is 
\begin{equation}
H_{\mathrm{W}}^{\left(1\right)}\left(x,p\right)=\frac{N_{\mathrm{f}}}{24}\, \left(e^{-\frac{1}{2}p}+e^{\frac{1}{2}p}\right)^{-2}.
\end{equation}
The integration can be performed with the formula \eqref{eq:Int-Form1}.
We then finally obtain
\begin{equation}
C_{\left(N_{\mathrm{f}},N_{\mathrm{f}},0\right),\left(0,N_{\mathrm{f}},0\right)}^{\fccz}
=\frac{3}{4\pi^{2}N_{\mathrm{f}}},\qquad B_{\left(N_{\mathrm{f}},N_{\mathrm{f}},0\right),\left(0,N_{\mathrm{f}},0\right)}^{\fccz}
=-\frac{N_{\mathrm{f}}^{2}-1}{12N_{\mathrm{f}}}.
\end{equation}
The disagreement in $B$ arises solely from the $O(N_{\mathrm{f}}^{-1})$ term, while the leading $O(N_{\mathrm{f}})$ term agrees with \cite{Bobev:2025ltz} (also derived in the planar limit in \cite{Geukens:2024zmt}).}

\subsection{\texorpdfstring{$\mathbb{C}\times\mathrm{SPP}$}{C-SPP} geometry}\label{subsec:SPP}

In this section we study the $\mathbb{C}\times\mathrm{SPP}$ geometry. We consider the following brane configuration, which realizes this toric diagram
\begin{equation}
\left[\left(1,1\right)+2\mathrm{D5}^{-}+\mathrm{D5}^{+}\right]-_{\mathrm{p}}\ .
\label{eq:BC-S120}
\end{equation}

\subsubsection*{CFT: Quantum curve result}

The quantum curve for \eqref{eq:BC-S120} is given by \eqref{eq:SYM-QC} with $q=1$, $F_{1}=2$, $F_{2}=1$, $F_{3}=0$. The Wigner transform of this quantum curve is given by
\begin{equation}
\left(\mathcal{O}_{1,\left(2,1,0\right)}^{\mathrm{fSYM}}\right)_{\mathrm{W}}\left(x,p;D,D_{\mathrm{m}}\right)=e^{\left(\frac{1}{4}-\frac{D}{2}+D_{\mathrm{m}}\right)x+Dp}\left(\begin{array}{cc}
+e^{x-\frac{1}{2}p} & +0\\
+2e^{-\frac{1}{2}p} & +e^{\frac{1}{2}p}\\
+e^{-x-\frac{1}{2}p} & +e^{-x+\frac{1}{2}p}
\end{array}\right)\ .
\label{eq:OW-S120}
\end{equation}
The volume for the polygon part is given as
\begin{equation}
\mathrm{Vol}_{\mathrm{Poly}}\left(E\right)=\frac{16\left(3+2D\right)}{\left(1-4D^{2}\right)\left(\left(3+2D\right)^{2}-16D_{\mathrm{m}}^{2}\right)}E^{2}\ .
\label{eq:VolP-S120}
\end{equation}
The corrections come from the four regions:
\begin{equation}
\left(x,p\right)\approx\left(0,-2E\right),\left(-E,0\right),\left(0,+2E\right),\left(+2E,+2E\right)\ .
\end{equation}
Using \eqref{eq:VolC-K12} with \eqref{eq:OW-K1-mn} and \eqref{eq:OW-K2-mn}, the corrections from each region can be computed as
\begin{equation}
\dvol=-\frac{\pi^{2}}{2}-\frac{\pi^{2}}{3}\left(\frac{3+2D}{1-4D^{2}}+\frac{4\left(3+2D\right)}{\left(3+2D\right)^{2}-16D_{\mathrm{m}}^{2}}\right)\ .
\label{eq:Corr-S120}
\end{equation}
The total volume is the sum of \eqref{eq:VolP-S120} and \eqref{eq:Corr-S120}. Therefore, by using \eqref{eq:Vol-CB} and \eqref{eq:Vol-Gen}, we obtain
\begin{align}
C_{1,\left(2,1,0\right)}^{\mathrm{fSYM}} & =\frac{4\left(3+2D\right)}{\left(1-4D^{2}\right)\left(\left(3+2D\right)^{2}-16D_{\mathrm{m}}^{2}\right)\pi^{2}}\ ,\nonumber \\
B_{1,\left(2,1,0\right)}^{\mathrm{fSYM}} & =-\frac{1}{8}+\frac{\left(3+2D\right)\left(3-12\left(1-D\right)D+16D_{\mathrm{m}}^{2}\right)}{12\left(1-4D^{2}\right)\left(\left(3+2D\right)^{2}-16D_{\mathrm{m}}^{2}\right)}\ .
\label{eq:CB-S120}
\end{align}

\subsubsection*{AdS: Equivariant string result}

The toric fan vector for \eqref{eq:BC-S120} is given as
\begin{equation}
v=\left(\begin{array}{cccc}
1 & 0 & 0 & 0\\
1 & 1 & 0 & 0\\
1 & 1 & 1 & 0\\
1 & 0 & 2 & 0\\
1 & 0 & 1 & 0\\
1 & 0 & 0 & 1
\end{array}\right), 
\label{eq:vec-SPP}
\end{equation}
with the first five vectors corresponding to the suspended pinch point ($\mathrm{SPP}$) manifold. In turn, the GLSM charge matrix is given by
\be
Q =
\begin{pmatrix}
1 & 0 & 0 & 1 & -2 & 0 \\
-1 & 1 & -1 & 0 & 1 & 0
\end{pmatrix}\ ,
\ee
with the last column corresponding to the $\mathbb{C}$ factor. The equivariant volume in the chamber $t^1>0, t^2>0$ is given by:
\be
\begin{split}
	 \BV (\lam, \e) &= \frac{\mathe^{\lam^6 \e_6}}{\e_6}\, \Bigg[
 \frac{\mathe^{\lam^1 (\e_1+\e_2-\e_4)+\lam^3 (\e_2+\e_3) +\lam^5 (2\e_4+\e_5-\e_2)}}
 {(\e_2+\e_3) (\e_1+\e_2-\e_4) (2 \e_4+\e_5-\e_2)} \\
 &- \frac{\mathe^{\lam^3 (\e_2+\e_3)-\lam^4 (\e_1+\e_2-\e_4) +\lam^5 (2 \e_1+\e_2+\e_5)}}
 {(\e_2+\e_3) (\e_1+\e_2-\e_4) (2 \e_1+\e_2+\e_5)} - \frac{\mathe^{\lam^1 (\e_1+\e_4+\e_5)+\lam^2 (\e_2-2 \e_4-\e_5) +\lam^3 (2 \e_3+2\e_4+\e_5)}}
 {(\e_1+\e_4+\e_5) (2 \e_4+\e_5-\e_2) ( \e_3+2\e_4+\e_5)} \Bigg]\ ,
 \end{split}
\ee
with $\e_6, \lam^6$ the parameters corresponding to the $\mathbb{C}$ factor.

We finally find that
\begin{align}
 & C\left(\epsilon\right)=\frac{2\epsilon_{1}+\epsilon_{2}+\epsilon_{3}+2\epsilon_{4}+2\epsilon_{5}}{8\pi^{2}\left(\epsilon_{2}+\epsilon_{3}\right)\left(2\epsilon_{1}+\epsilon_{2}+\epsilon_{5}\right)\left(\epsilon_{1}+\epsilon_{4}+\epsilon_{5}\right)\left(\epsilon_{3}+2\epsilon_{4}+\epsilon_{5}\right)\epsilon_{6}}\ ,\quad\chi=3\ ,\nonumber \\
 & c_{2}\left(\epsilon\right)=8\pi^{2}C\left(\epsilon\right)\left(\sum_{i<j=1}^{6}\epsilon_{i}\epsilon_{j}+\sum_{i=1}^{5}\epsilon_{i}\epsilon_{5}+2\epsilon_{1}\epsilon_{3}+3\epsilon_{1}\epsilon_{4}+2\epsilon_{2}\epsilon_{4}\right)\ ,\nonumber \\
 & c_{3}\left(\epsilon\right)=\frac{2}{\epsilon_{2}+\epsilon_{3}}+\frac{1}{2\epsilon_{1}+\epsilon_{2}+\epsilon_{5}}+\frac{1}{\epsilon_{1}+\epsilon_{4}+\epsilon_{5}}+\frac{1}{\epsilon_{3}+2\epsilon_{4}+\epsilon_{5}}+\frac{3}{\epsilon_{6}}\ .\label{eq:CB-SPP}
\end{align}
Note that instead of the redundant variables in the SPP directions, $\e_1, \dots, \e_5$, we could alternatively use the $\nu$-variables,
\be
    \nu_1 = \sum_{i=1}^6 \e_i\ , \quad \nu_2 = \e_2+\e_3\ , \quad \nu_3 = \e_3 + 2 \e_4+\e_5\ .
\ee

\subsubsection*{Holography}
We consider the holographic duality derived by the brane configuration \eqref{eq:BC-S120}. The relation between the R-charges and the equivariant parameters is given by \eqref{eq:F-PM-fSYM} and \eqref{eq:SYM-RD}-\eqref{eq:m-T-R} with $\epsilon_{i}=\Delta_{i}$ as
\begin{equation}
\Delta_{\phi_{1}}=\epsilon_{1}+\epsilon_{4}+\epsilon_{5}\ ,\quad
\Delta_{\phi_{2}}=\epsilon_{2}+\epsilon_{3}\ ,\quad
\Delta_{\phi_{3}}=\epsilon_{6}\ ,\quad
\Delta_{T}=2\epsilon_{1}+\epsilon_{2}+\epsilon_{5}\ ,\quad
\Delta_{\tilde{T}}=\epsilon_{3}+2\epsilon_{4}+\epsilon_{5}\ ,
\label{eq:Holo-SPP}
\end{equation}
under the constraints
\be
    \e_1 +\e_2+ \e_3+\e_4 + \e_5 = 1 = \e_6\ ,
\ee
such that
\be
    D = \e_1 +\e_4+ \e_5 - \frac12 = \frac12 - \e_2 - \e_3\ , \qquad D_{\mathrm{m}} =\frac12 \left( 2\, \e_1 + \e_2 - \e_3 - 2\, \e_4 \right)\ .
\ee
Under the above identifications, we confirmed that \eqref{eq:CB-S120} is equivalent to \eqref{eq:CB-SPP} with \eqref{eq:Bgeometry}. Just as in the previous examples, we have demonstrated a successful holographic match at finite $N$ for the first time in the literature.

\section{Equivariant CY\texorpdfstring{$_4$}{4}/CY\texorpdfstring{$_3$}{3} correspondence}
\label{sec:correspondence}

\subsection{Statement of the correspondence}

Let us now focus on defining the proposed geometric correspondence, which is strongly motivated and partially characterized by the quantum curve computations and the $S^3$ partition function analysis described above. We note at the outset that the CY$_3$ side of the correspondence is closely related to the topological string sector of the TS/ST correspondence \cite{Hatsuda:2012dt, Grassi:2014cla}, a connection that will be discussed in detail in the next section. For the moment, we set aside this additional structure and concentrate instead on a direct geometric construction in terms of toric diagrams. 
We emphasize that the procedure described below establishes a relation based on the underlying physics, rather than an exact mathematical duality, between equivariant volumes as opposed to the more straightforward construction in appendix \ref{app:An equivariant duality}.

As already sketched in the introduction (cf.\ figure \ref{fig:CY43-ABJM}), the underlying idea is conceptually simple. A toric Calabi–Yau four-fold is represented by a three-dimensional lattice of points with integer coordinates $(x,y,z)$, whereas a Calabi–Yau three-fold is described by a two-dimensional lattice with coordinates $(x,y)$. One may therefore formally regard the different $z$-slices of a CY$_4$ toric diagram as a collection of two-dimensional toric diagrams. The proposed correspondence is suggested by our explicit quantum curve computations, which relate these distinct $z$-slices through their Minkowski sum.
We will return to this point shortly.

In this work we restrict our attention to toric CY$_4$ geometries with two-layer toric diagrams. Concretely, we consider configurations in which all vertices lie on two $x$–$y$ planes, denoted $P_0$ and $P_1$, which without loss of generality correspond to the slices at $z=0$ and $z=1$, respectively:
\be
\begin{split}
	v^{\mathrm{CY}_4} &:= \{ (1, x_z, y_z, z) | (x_z, y_z) \in P_z,\, z=0,1 \} \\ 
    &= \{(1, x_0, y_0, 0)| (x_0, y_0) \in P_0\} \cup \{(1, x_1, y_1, 1)| (x_1, y_1) \in P_1\}\ ,
\end{split}
\label{eq:vCY4}
\ee
with $P_z$ the full set of points in the lattice at the given $z$ value.

We can then generate a toric CY$_3$ defined by the Minkowski sum of the lattices above
\be
	\tilde v^{\mathrm{CY}_3} 
    :=  \{(1, x_0 + x_1, y_0 + y_1, 0)|(x_0, y_0) \in P_0,\, (x_1, y_1) \in P_1 \}\ ,
\ee
corresponding to a 2d lattice at $z=0$. Note that different pairs of points on the $P_0$ and $P_1$ planes might end up generating the same point in the Minkowski sum, such that the counting of lattice points on the CY$_3$ is not straightforward. In addition, in order to formally retain the dimensionality, we add an additional $\mathbb{C}$ factor that corresponds to a single point on the $z=1$ plane,
\be
	\tilde v^{\mathbb{C} \times \mathrm{CY}_3} :=  \{(1, x_0 + x_1, y_0 + y_1, 0)| (x_0, y_0) \in P_0,\, (x_1, y_1) \in P_1 \} \cup \{(1, 0, 0, 1)\}\ .
    \label{eq:vCY3}
\ee
Due to the nature of the vertices, we can further introduce the notation of labeling the CY$_4$ lattice points by $v^{(z)}_{x, y}$, while the $\mathbb{C} \times \mathrm{CY}_3$ points by $\tilde v_{x,y}$ together with $\tilde v_1 := (1, 0, 0, 1)$.
We propose that the $\mathbb{C}\times\mathrm{CY}_3$ constructed in this way be paired with the $\mathrm{CY}_4$, in the sense of the equivariant volume and Airy-data relations stated below.

We briefly pause to describe how this arises from the field theory side. 
A key observation is the relationship between the flavored ABJM theory defined by \eqref{eq:fABJM-BC2} and the flavored $\cczp$ theory defined by \eqref{eq:fABJM-BC0}, mediated by the flavored $\ccz$ theory. 
The $\ccz$ and $\cczp$ theories are characterized by the same Newton polygon, which ensures that their Airy function coefficients $C$ are identical. 
From the quantum curve perspective, this is because the Wigner transform replaces the operators by classical variables, as in \eqref{eq:QC-W}, and the leading-order Fermi surface is bounded by the lines \eqref{eq:FS-Poly-Gen}.
Geometrically, this equivalence manifests as a relation between CY$_4$ and CY$_3$ through the Minkowski sum (see appendix \ref{subsec:Web-Poly}). Importantly, the partition functions are equivalent under the R-charge constraints \eqref{eq:fABJM-Rcond} and \eqref{eq:RAB-cond}, inducing \eqref{eq:eqParaCond} on the geometry side, to which we turn our attention.

As part of the correspondence, it is crucial to properly identify the equivariant parameters on the two sides. Since one may assign a (generally redundant) equivariant parameter $\e$ to each lattice point, or vertex, we distinguish in this section between two sets of equivariant parameters: those associated with the original CY$_4$ geometry,
\be
	\e^{(0)}_{x_0,y_0}\ , \, \forall (x_0, y_0) \in P_0\ , \qquad \e^{(1)}_{x_1,y_1}\ , \,  \forall (x_1, y_1) \in P_1\ ,
\ee
corresponding to the lattice points $v^{(0)}$ and $v^{(1)}$; and the ones associated with $\mathbb{C} \times \mathrm{CY}_3$ by
\be
	\tilde \e_{x, y}\ , \, \forall (x,y) \in (P_0+P_1)\ , \qquad \tilde \e_1\ ,
\ee
corresponding to the lattice points $\tilde v$. As emphasized in the introduction (see \eqref{eq:CY43-Rcond}), and explained from a field theory point of view as a constraint on the existence of the quantum curve description, we further restrict
\be
\label{eq:constr1CY4}
	\sum_{(x, y)\in P_0} \e^{(0)}_{x, y} = 1\ , \qquad \sum_{(x, y)\in P_1} \e^{(1)}_{x, y} = 1\ ,
\ee
on the CY$_4$ side; and
\be
\label{eq:constr2CY3}
	\sum_{(x, y)\in (P_0 + P_1)} \tilde \e_{x, y} = 1\ , \qquad \tilde \e_1 = 1\ ,
\ee
on the $\mathbb{C} \times \mathrm{CY}_3$ side.

We must then define a map between the two sets of equivariant parameters that respects the constraints discussed above. To this end, we introduce an auxiliary set of variables, $\eta_{(x_0, y_0),(x_1, y_1)}$, defined for each pair of lattice points $(x_0, y_0) \in P_0$ and $(x_1, y_1) \in P_1$. These auxiliary variables provide a convenient parametrization that relates the two collections of equivariant parameters across the correspondence,
\be
\begin{split}
 & \epsilon_{x,y}^{(0)}=\sum_{\left(x_{1},y_{1}\right)\in P_{1}} \eta_{\left(x,y\right),\left(x_{1},y_{1}\right)}\ ,\qquad \epsilon_{x,y}^{(1)}=\sum_{\left(x_{0},y_{0}\right)\in P_{0}}\eta_{\left(x_{0},y_{0}\right),\left(x,y\right)}\ , \\
 & \tilde{\epsilon}_{x,y}=\sum_{\left\{ x_{0}+x_{1}=x,\, y_{0}+y_{1}=y|\left(x_0,y_0\right)\in P_0,\, \left(x_1,y_1\right)\in P_1 \right\} }\eta_{\left(x_{0},y_{0}\right),\left(x_{1},y_{1}\right)}\ .
 \label{eq:CY43-Rcorr}
\end{split}
\ee
Given this parametrization, the central claim of the CY$_4$/CY$_3$ correspondence, already anticipated in the introduction, can now be stated as follows.
\be
	\BV_{\mathrm{CY}_4} (\lam = 0, \e (\eta)) = \BV_{\mathbb{C} \times \mathrm{CY}_3} (\tilde \lam = 0, \tilde \e (\eta))\ ,
\ee
such that, when written in terms of the leading part of the S$^3$ partition function,
\be
\label{eq:CY43-C}
	C_{\mathrm{CY}_4} (\e (\eta)) = C_{\mathbb{C} \times \mathrm{CY}_3} (\tilde \e (\eta))\ .
\ee
In addition, the relation extends beyond leading order and into the exact finite $N$ expression, cf.\ \eqref{eq:S3pf}, from the relation
\be
\label{eq:subleadingshift}
	B_{\mathrm{CY}_4} (\e (\eta)) - B_{\mathbb{C} \times \mathrm{CY}_3} (\tilde \e (\eta)) = const\ ,
\ee
where $B_X$ is defined geometrically in \eqref{eq:Bgeometry}. 

Several remarks are in order below.

First, the relation \eqref{eq:CY43-Rcorr} respects the structure of the Minkowski sum. The points on the slices $z=0,1$ of the 3d toric diagram for the CY$_4$ are paired with the points on the opposite slices at $z=1,0$, respectively, and are then mapped to points of the 2d toric diagram for the CY$_3$. The first line of \eqref{eq:CY43-Rcorr} translates this construction into the language of equivariant parameters; specifically, $\eta_{(x_0, y_0),(x_1, y_1)}$ almost directly corresponds to $\tilde{\epsilon}_{x_0+x_1, y_0+y_1}$. This correspondence, however, is not one-to-one, since multiple distinct pairs of points on the 3d toric diagram may map to the same point on the 2d toric diagram. The second line of \eqref{eq:CY43-Rcorr} accounts for this degeneracy by requiring that the corresponding equivariant parameter be given by the sum of all $\eta_{(x_0, y_0),(x_1, y_1)}$ associated with the relevant mappings.

Second, the number of auxiliary variables $\eta$ in \eqref{eq:CY43-Rcorr} is always greater than or equal to the number of independent $\e$ and $\tilde\e$ parameters. As a result, for any given example one can always employ the above parametrization, which consistently relates the two sets of equivariant parameters.

Third, we discuss the field-theoretic interpretation of the parameter correspondence given in \eqref{eq:CY43-Rcorr}. As mentioned previously, the CY$_4$/CY$_3$ correspondence manifests on the field theory side as the equality of the Airy function coefficient $C$, which can be verified through an analysis of the Newton polygons. In general, however, this does not provide an explicit mapping between the equivariant parameters. The reason is that, while equivariant parameters correspond to R-charges on the field theory side, the theory associated with a given CY$_3$ is often non-Lagrangian when starting from a generic CY$_4$. To address this issue, we consider the case of CY$_4 = \mathbb{C} \times \mathcal{C}$ and CY$_3 = \text{SPP}$ as an illustrative example. In terms of the brane configurations, this corresponds to \eqref{eq:BC-S011} and \eqref{eq:BC-S120}. This example is particularly instructive because the corresponding field theories are Lagrangian on both sides and, moreover, share the same matter content. Consequently, the equivalence of $C$ as a function of the R-charges reduces to the equality of the R-charges for each matter multiplet. Given the holographic dictionary in \eqref{eq:Holo-Cfd-b} and \eqref{eq:Holo-SPP}, the requirement that the R-charges of the matter fields coincide can be translated into the following expression:
\be
\begin{split}
&\e^{(0)}_{0,0}+\e^{(0)}_{0,1} 
=\tilde\e_{0,0}+\tilde \e_{0,1}+\tilde\e_{0,2}\ , \quad
\e^{(0)}_{1,0} 
=\tilde\e_{1,0}+\tilde \e_{1,1}\ , \quad
\e^{(1)}_{0,0}+\e^{(1)}_{0,1} 
=\tilde\e_1=1\ , \\
&\e^{(0)}_{0,0}+\e^{(1)}_{0,0} 
=2\tilde\e_{0,0}+\tilde\e_{0,1}+\tilde \e_{1,0}\ , \quad
\e^{(0)}_{0,1}+\e^{(1)}_{0,1} 
=\tilde\e_{0,1}+2\tilde\e_{0,2}+\tilde \e_{1,1} \ .
\end{split}
\ee
Here, the toric fan vectors are given by $v^{(b)}$ in \eqref{eq:vec-Cfd} and \eqref{eq:vec-SPP}, which we have used to align the notation of the equivariant parameters with that of the current section. This is consistent with \eqref{eq:CY43-Rcorr} (see also \eqref{eq:CY43-Cfd-para}). The above argument remains valid even when the field theory corresponding to CY$_4$ is a general flavored SYM theory described by the brane configuration \eqref{eq:fSYM-BC}. In this case, the flavor numbers on the CY$_4$ side are $(F_1, F_2, F_3)$, while those on the CY$_3$ side are $(F_1 + F_3, F_2 + F_3, 0)$. Thus, at least in this setup, \eqref{eq:CY43-Rcorr} translates in field theory into the equivalence of the R-charges of the matter content on both sides.

Fourth, throughout this analysis we have focused exclusively on the geometric phases (see section \ref{subsec:TCY-Loc}) of the corresponding manifolds, a choice that plays a crucial role in the evaluation of $B_{\mathrm{CY}_4}$ and $B_{\mathbb{C} \times \mathrm{CY}_3}$ appearing above. A deeper understanding of the holographic significance of this choice of chamber in the computation of the equivariant volume would be highly desirable. Restricting to the geometric phase, the exact value of the constant on the right-hand side of the above relation admits a field-theoretic interpretation in terms of the difference between the subleading contributions of the corresponding brane descriptions discussed earlier.

We conclude this main part of the discussion by emphasizing once more that, although the conjectured correspondence has been formulated in terms of the coefficients $C_X$ and $B_X$ appearing in the $S^3$ partition function, its origin is intrinsically geometric, arising from the equivariant triple intersection numbers of the two Calabi–Yau manifolds under consideration.

\subsubsection{Origin of the subleading shift}\label{subsec:CY43-B}
Let us focus further on \eqref{eq:subleadingshift}, which for the moment has an unspecified constant on the right-hand side, and try to justify the origin and precise value of the constant from the dual field theory and quantum-curve points of view.

In terms of the brane configuration, the CY$_4$/CY$_3$ correspondence is interpreted as
\begin{equation}
\mathsf{w}_{1}-\mathsf{w}_{2}-_{\mathrm{p}}\, \quad \leftrightarrow \quad \mathsf{w}_{1+2}-_{\mathrm{p}}\ .
\label{eq:BC-CY4-Gen}
\end{equation}
We then expect that the constant in \eqref{eq:subleadingshift} can be computed via the quantum curves for the brane configurations in \eqref{eq:BC-CY4-Gen} as
\begin{equation}
B_{\mathrm{CY}_{4}}\left(\epsilon (\eta)\right)
-B_{\mathbb{C}\times\mathrm{CY}_{3}}\left(\tilde \epsilon (\eta)\right) =
B^{\fccz} \Big|_{D_i=D_\mathrm{m}=0} -
B^{\fcczp}\ ,
\label{eq:CY43-B}
\end{equation}
Here the quantity $B^{\fccz} \Big|_{D_i=D_\mathrm{m}=0}$ is computed from the known quantum curves, while $B^{\fcczp}$ is obtained by fixing the total exponents to match those of $B^{\fccz} \Big|_{D_i=D_\mathrm{m}=0}$. 
We set $D_i=D_\mathrm{m}=0$ when calculating $B^{\fccz}$ because the left-hand side is independent of the equivariant parameter.
We will verify \eqref{eq:CY43-B} in explicit examples below. We expect that the analogous discussion is applicable for the SYM theory, and thus \eqref{eq:CY43-B} holds for both the toric diagrams in figure \ref{fig:newfig}.

Even more formally, without passing through the explicit knowledge of the brane construction and field theory details, we conjecture the generalized relation
\begin{equation}
B_{\mathrm{CY}_{4}}\left(\e(\eta)\right) -
B_{\mathbb{C}\times\mathrm{CY}_{3}}\left(\tilde \e(\eta)\right) = 
B^{\mathcal{T}\left[\mathrm{CY}_{4}\right]}-
B^{\mathcal{T}\left[\mathbb{C}\times\mathrm{CY}_{3}\right]}\ ,
\label{eq:CY43-B2}
\end{equation}
where $\mathcal{T}$ is the quantum curve associated with the corresponding toric diagram. The quantum curve for $\mathcal{T}[\mathrm{CY}_{4}]$ is given by the product of the curves associated with the two layers, while that for $\mathcal{T}[\mathbb{C}\times\mathrm{CY}_{3}]$ is obtained by reducing this product to the normal form \eqref{eq:QCform}, 
with ordering-induced phase factors
removed but multiplicities of coincident monomials retained.
This construction relies only on the identification between toric diagrams and Newton polygons and does not require a Lagrangian field-theoretic description.
We will also verify \eqref{eq:CY43-B2} in an explicit example below.

\subsection{Examples of dual pairs}
In this section we test the conjectural CY$_{4}$/CY$_{3}$ correspondence in the four examples summarized in table~\ref{tab:CY43-examples}.
For every pair, we directly verify the equality of the leading Airy coefficient in \eqref{eq:CY43-C} and show that the geometric difference of the subleading coefficient $B$ is reproduced by the corresponding
field-theory or quantum-curve computation, as in \eqref{eq:CY43-B} or
\eqref{eq:CY43-B2}. 
For completeness, we note that these checks provide non-trivial evidence for the proposed correspondence in the four examples considered here, while establishing it for general two-layer toric CY$_4$ geometries remains an open problem.

As emphasized in the previous subsection, we expect the CY$_4$/CY$_3$ correspondence to be intrinsic to the equivariant toric geometries, even though it was originally suggested by combining holography with the quantum-curve analysis. 
We therefore choose the examples so as to progressively reduce the field-theoretic input.
The first two pairs, already analyzed in section~\ref{sec:Examples}, admit Lagrangian descriptions on both sides; the third has a non-Lagrangian theory on the $\mathbb{C}\times\mathrm{CY}_{3}$ side; and the fourth is analyzed without an explicit field-theory description even on the $\mathrm{CY}_{4}$ side. 
In the latter two cases, the relevant quantum curves can still be constructed from the brane/toric or Newton-polygon data and reproduce the geometric $B$-shift.

\begin{table}[!ht]
\centering
\small
\renewcommand{\arraystretch}{1.25}
\begin{tabular}{|p{0.32\textwidth}|p{0.55\textwidth}|}
\hline
\raggedright\textbf{Pair}
&
\raggedright\textbf{Field-theory description}
\tabularnewline
\hline

\raggedright
$\mathbb{C}^{4}
 \leftrightarrow
 \mathbb{C}\times\cC$
&
\raggedright
Lagrangian on both sides.
\tabularnewline
\hline

\raggedright
$\mathbb{C}\times\cC
 \leftrightarrow
 \mathbb{C}\times\mathrm{SPP}$
&
\raggedright
Lagrangian on both sides.
\tabularnewline
\hline

\raggedright
$C(Q^{1,1,1})
 \leftrightarrow
 \mathbb{C}\times\mathrm{dP}_{3}$
&
\raggedright
Lagrangian on the $\mathrm{CY}_{4}$ side; no explicit field-theory description is available in our analysis on the $\mathbb{C}\times\mathrm{CY}_{3}$ side.
\tabularnewline
\hline

\raggedright
The unnamed pair in \eqref{eq:unnamedtoricvectors}
&
\raggedright
No explicit field-theory description is available in our analysis.
\tabularnewline
\hline
\end{tabular}

\caption{
Summary of the four examples used to test the conjectural equivariant
CY$_4$/CY$_3$ correspondence, classified by the availability of an
explicit field-theory description. In all four examples, we verify
$C_{\mathrm{CY}_{4}}
 =C_{\mathbb{C}\times\mathrm{CY}_{3}}$
and find that the geometric difference of the subleading coefficient
$B$ agrees with the corresponding field-theory or quantum-curve
calculation.
}
\label{tab:CY43-examples}
\end{table}

\subsubsection*{$\mathbb{C}^4 \leftrightarrow \mathbb{C}\times\cC$}
Let us first consider the case when $\mathrm{CY}_{4}=\mathbb{C}^{4}$, leading to $\mathrm{CY}_{3}=\mathcal{C}$:
\begin{equation}
v^{\mathbb{C}^{4}}=\left(\begin{array}{cccc}
1 & 0 & 0 & 0\\
1 & 1 & 0 & 0\\
1 & 0 & 1 & 1\\
1 & 0 & 0 & 1
\end{array}\right),\quad \Rightarrow \quad \tilde v^{\mathbb{C}\times\mathcal{C}}=
\left(\begin{array}{cccc}
1 & 0 & 0 & 0\\
1 & 1 & 1 & 0\\
1 & 0 & 1 & 0\\
1 & 1 & 0 & 0\\
1 & 0 & 0 & 1
\end{array}\right).
\end{equation}
The constraints on the respective equivariant parameters, \eqref{eq:constr1CY4} and \eqref{eq:constr2CY3}, are given by
\be
\begin{split}
	\e^{(0)}_{0,0} &+ \e^{(0)}_{1,0} = 1 \ , \qquad \e^{(1)}_{0,0} + \e^{(1)}_{0,1} = 1 \ , \\
	\tilde \e_1 &= 1\ , \qquad \tilde \e_{0,0} + \tilde \e_{0,1}+\tilde \e_{1,0} + \tilde \e_{1,1} = 1\ .
\end{split}
\ee
In terms of the auxiliary variables introduced in \eqref{eq:CY43-Rcorr}, we then find
\be
\begin{split}
	\e^{(0)}_{0,0} &= \eta_{(0,0),(0,0)} + \eta_{(0,0),(0,1)} = \tilde \e_{0,0} + \tilde \e_{0,1}\ , \\
	\e^{(0)}_{1,0} &= \eta_{(1,0),(0,0)} + \eta_{(1,0),(0,1)} = \tilde \e_{1,0} + \tilde \e_{1,1}\ , \\
	\e^{(1)}_{0,0} &= \eta_{(0,0),(0,0)} + \eta_{(1,0),(0,0)} = \tilde \e_{0,0} + \tilde \e_{1,0}\ , \\
\e^{(1)}_{0,1} &= \eta_{(0,0),(0,1)} + \eta_{(1,0),(0,1)} = \tilde \e_{0,1} + \tilde \e_{1,1}\ ,
\end{split}
\ee
where the latter equalities hold term by term in the given order. It is clear that the above identifications are compatible with the constraints above.

We already calculated the $C$ and $B$ coefficients for these particular examples. In the present notation for the corresponding equivariant parameters, \eqref{eq:CB-C4} and \eqref{eq:CB-cfd} lead to the equality
\be
	C_{\mathbb{C}^{4}} = \frac1{8 \pi^2\, \e^{(0)}_{0,0} \e^{(0)}_{1,0} \e^{(1)}_{0,0} \e^{(1)}_{0,1}} =\frac{\tilde \e_{0,0} + \tilde \e_{0,1} + \tilde \e_{1,0} + \tilde \e_{1,1}}{8\pi^{2}\left(\tilde\epsilon_{0,0}+\tilde\epsilon_{0,1}\right) \left(\tilde \epsilon_{1,0}+\tilde \epsilon_{1,1}\right)  \left(\tilde\epsilon_{0,0}+\tilde\epsilon_{1,0}\right)\left(\tilde \epsilon_{0,1}+\tilde \epsilon_{1,1}\right)\tilde \e_1}= C_{\mathbb{C}\times\mathcal{C}}\ ,
\ee
where we have used the identifications above. From \eqref{eq:CB-C4} and \eqref{eq:CB-cfd}, we can also check the validity of \eqref{eq:subleadingshift}:
\begin{equation}
B_{\mathbb{C}^{4}}\left(\epsilon (\eta)\right) - B_{\mathbb{C}\times\mathcal{C}}\left(\tilde \epsilon (\eta)\right) = \frac{1}{8}\ ,
\end{equation}
again under the identification between $\e$ and $\tilde \e$ and the additional constraints.
This is compatible with our field-theory calculations, where \eqref{eq:CB-S001} and \eqref{eq:CB-S110} with $D=D_{\mathrm{m}}=0$ give
\begin{equation}
B_{0,\left(0,0,1\right)}^{\mathrm{fSYM}} \Big|_{D = D_{\mathrm{m}} = 0} - B_{0,\left(1,1,0\right)}^{\mathrm{fSYM}}\Big|_{D = D_{\mathrm{m}} = 0} = \frac{1}{8}\ ,
\end{equation}
confirming \eqref{eq:CY43-B}.

\subsubsection*{$\mathbb{C}\times\cC \leftrightarrow \mathbb{C}\times\mathrm{SPP}$}
Second, we consider the case when $\mathrm{CY}_{4}=\mathbb{C}\times\mathcal{C}$ and $\mathrm{CY}_{3}=\mathrm{SPP}$ with the toric fan vectors
\begin{equation}
v^{\mathbb{C}\times\mathcal{C}}=\left(\begin{array}{cccc}
1 & 0 & 0 & 0\\
1 & 0 & 1 & 1\\
1 & 0 & 0 & 1\\
1 & 0 & 1 & 0\\
1 & 1 & 0 & 0
\end{array}\right),\quad \Rightarrow \quad \tilde v^{\mathbb{C}\times\mathrm{SPP}}=\left(\begin{array}{cccc}
1 & 0 & 0 & 0\\
1 & 1 & 0 & 0\\
1 & 1 & 1 & 0\\
1 & 0 & 2 & 0\\
1 & 0 & 1 & 0\\
1 & 0 & 0 & 1
\end{array}\right).
\end{equation}
The constraints on the respective equivariant parameters, \eqref{eq:constr1CY4} and \eqref{eq:constr2CY3}, are given by
\be
\begin{split}
	\e^{(0)}_{0,0} &+ \e^{(0)}_{1,0} + \e^{(0)}_{0,1} = 1 \ , \qquad \e^{(1)}_{0,0} + \e^{(1)}_{0,1} = 1 \ , \\
	\tilde \e_1 &= 1\ , \qquad \tilde \e_{0,0} + \tilde \e_{0,1} + \tilde \e_{0,2}+\tilde \e_{1,0} + \tilde \e_{1,1} = 1\ .
\end{split}
\ee
In terms of the auxiliary variables introduced in \eqref{eq:CY43-Rcorr}, we then find
\be
\begin{split}
	\e^{(0)}_{0,0} &= \eta_{(0,0),(0,0)} + \eta_{(0,0),(0,1)} = \tilde \e_{0,0} + (\tilde \e_{0,1} - \eta_{(0,1),(0,0)})\ , \\
	\e^{(0)}_{1,0} &= \eta_{(1,0),(0,0)} + \eta_{(1,0),(0,1)} = \tilde \e_{1,0} + \tilde \e_{1,1}\ , \\
	\e^{(0)}_{0,1} &= \eta_{(0,1),(0,0)} + \eta_{(0,1),(0,1)} = (\tilde \e_{0,1} - \eta_{(0,0),(0,1)}) + \tilde \e_{0,2}\ , \\
	\e^{(1)}_{0,0} &= \eta_{(0,0),(0,0)} + \eta_{(1,0),(0,0)} + \eta_{(0,1),(0,0)} = \tilde \e_{0,0} +  (\tilde \e_{0,1} - \eta_{(0,0),(0,1)}) + \tilde \e_{1,0}\ , \\
\e^{(1)}_{0,1} &= \eta_{(0,0),(0,1)} + \eta_{(1,0),(0,1)} + \eta_{(0,1),(0,1)} = (\tilde \e_{0,1} - \eta_{(0,1),(0,0)}) + \tilde \e_{1,1} + \tilde \e_{0,2}\ ,
\end{split}
\label{eq:CY43-Cfd-para}
\ee
where the latter equalities hold term by term in the given order, since $\tilde \e_{0,1} = \eta_{(0,0),(0,1)} + \eta_{(0,1),(0,0)}$. It can be checked that the above identifications are compatible with the constraints above.

We already calculated the $C$ and $B$ coefficients for these particular examples. In the present notation for the corresponding equivariant parameters, \eqref{eq:CB-cfd} and \eqref{eq:CB-SPP} lead to the equality
\be
\begin{split}
	C_{\mathbb{C}\times\mathcal{C}} &=\frac{ \e^{(0)}_{0,0} + \e^{(1)}_{0,1} + \e^{(1)}_{0,0} + \e^{(0)}_{0,1}}{8\pi^{2}\left(\e^{(0)}_{0,0}+\e^{(1)}_{0,0}\right) \left(\e^{(1)}_{0,1}+\e^{(1)}_{0,0}\right)  \left(\e^{(0)}_{0,0}+\e^{(0)}_{0,1}\right)\left(\e^{(1)}_{0,1}+\e^{(0)}_{0,1}\right) \e^{(0)}_{1,0}} \\
&= \frac{2\tilde \epsilon_{0,0}+\tilde \epsilon_{1,0}+\tilde\epsilon_{1,1}+2\tilde\epsilon_{0,2}+2\tilde\epsilon_{0,1}}{8\pi^{2}\left(\tilde\epsilon_{1,0}+\tilde\epsilon_{1,1}\right)\left(2\tilde\epsilon_{0,0}+\tilde\epsilon_{1,0}+\tilde\epsilon_{0,1}\right)\left(\tilde\epsilon_{0,0}+\tilde\epsilon_{0,2}+\tilde\epsilon_{0,1}\right)\left(\tilde\epsilon_{1,1}+2\tilde\epsilon_{0,2}+\tilde\epsilon_{0,1}\right) \tilde \epsilon_{1}} = C_{\mathbb{C}\times\mathrm{SPP}}\ ,
\end{split}
\ee
where we have used the identifications above. From \eqref{eq:CB-cfd} and \eqref{eq:CB-SPP}, we can also check the validity of \eqref{eq:subleadingshift}:
\begin{equation}
B_{\mathbb{C}\times\mathcal{C}}\left(\epsilon (\eta)\right) - B_{\mathbb{C}\times\mathrm{SPP}}\left(\tilde \epsilon (\eta)\right) = \frac{1}{8}\ ,
\end{equation}
again under the identification between $\e$ and $\tilde \e$ and the additional constraints.
This is compatible with our field-theory calculations, where \eqref{eq:CB-S011} and \eqref{eq:CB-S120} with $D=D_{\mathrm{m}}=0$ give
\begin{equation}
B_{1,\left(1,0,1\right)}^{\mathrm{fSYM}} \Big|_{D = D_{\mathrm{m}} = 0} - B_{1,\left(2,1,0\right)}^{\mathrm{fSYM}}\Big|_{D = D_{\mathrm{m}} = 0} = \frac{1}{8}\ ,
\end{equation}
again confirming \eqref{eq:CY43-B}.

\subsubsection*{$C(Q^{1,1,1}) \leftrightarrow \mathbb{C}\times\mathrm{dP}_{3}$}
The third example is the case when $\mathrm{CY}_{4}= C(Q^{1,1,1})$. The corresponding toric $\mathrm{CY}_{3}$ is $\mathrm{CY}_{3}=\mathrm{dP}_{3}$. The toric vectors are 
\begin{equation}
\label{eq:dP3toricvectors}
v^{Q^{1,1,1}}=\left(\begin{array}{cccc}
1 & 0 & 0 & 0\\
1 & 1 & 1 & 1\\
1 & 1 & 0 & 0\\
1 & 0 & 1 & 1\\
1 & 0 & 1 & 0\\
1 & 1 & 0 & 1
\end{array}\right),\quad \Rightarrow \quad \tilde v^{\mathbb{C}\times\mathrm{dP}_{3}}=\left(\begin{array}{cccc}
1 & 2& 1 & 0\\
1 & 1 & 2 & 0\\
1 & 0 & 2 & 0\\
1 & 0 & 1 & 0\\
1 & 1 & 0 & 0\\
1 & 2 & 0 & 0\\
1 & 1 & 1 & 0\\
1 & 0 & 0 & 1
\end{array}\right).
\end{equation}
The constraints on the respective equivariant parameters, \eqref{eq:constr1CY4} and \eqref{eq:constr2CY3}, are given by
\be
\begin{split}
	\e^{(0)}_{0,0} &+ \e^{(0)}_{1,0} + \e^{(0)}_{0,1} = 1 \ , \qquad \e^{(1)}_{1,1} + \e^{(1)}_{0,1} + \e^{(1)}_{1,0} = 1 \ , \\
	\tilde \e_1 &= 1\ , \qquad \tilde \e_{2,1} + \tilde \e_{1,2}+\tilde \e_{0,2} + \tilde \e_{0,1} + \tilde \e_{2,0}+\tilde \e_{1,0} + \tilde \e_{1,1} = 1\ .
\end{split}
\ee
In terms of the auxiliary variables introduced in \eqref{eq:CY43-Rcorr}, we then find
\be
\begin{split}
	\e^{(0)}_{0,0} &= \eta_{(0,0),(1,1)} + \eta_{(0,0),(0,1)} + \eta_{(0,0),(1,0)} = (\tilde \e_{1,1} - \eta_{(0,1),(1,0)} -\eta_{(1,0),(0,1)}) + \tilde \e_{0,1}+\tilde \e_{1,0}\ , \\
	\e^{(0)}_{1,0} &= \eta_{(1,0),(1,1)} + \eta_{(1,0),(0,1)} + \eta_{(1,0),(1,0)} = \tilde \e_{2,1} + (\tilde \e_{1,1} - \eta_{(0,1),(1,0)} -\eta_{(0,0),(1,1)}) + \tilde \e_{2,0}\ , \\
	\e^{(0)}_{0,1} &= \eta_{(0,1),(1,1)} + \eta_{(0,1),(0,1)} + \eta_{(0,1),(1,0)} = \tilde \e_{1,2} + \tilde \e_{0,2} + (\tilde \e_{1,1} - \eta_{(1,0),(0,1)} -\eta_{(0,0),(1,1)})\ , \\
	\e^{(1)}_{1,1} &= \eta_{(0,0),(1,1)} + \eta_{(1,0),(1,1)} + \eta_{(0,1),(1,1)} = (\tilde \e_{1,1} - \eta_{(0,1),(1,0)} -\eta_{(1,0),(0,1)}) + \tilde \e_{2,1} +\tilde \e_{1,2}\ , \\
\e^{(1)}_{0,1} &= \eta_{(0,0),(0,1)} + \eta_{(1,0),(0,1)} + \eta_{(0,1),(0,1)} =\tilde \e_{0,1} +  (\tilde \e_{1,1} - \eta_{(0,1),(1,0)} -\eta_{(0,0),(1,1)})+ \tilde\e_{0,2}\ , \\
\e^{(1)}_{1,0} &= \eta_{(0,0),(1,0)} + \eta_{(1,0),(1,0)} + \eta_{(0,1),(1,0)} = \tilde\e_{1,0}+\tilde\e_{2,0} + (\tilde \e_{1,1} - \eta_{(1,0),(0,1)} -\eta_{(0,0),(1,1)})\ , 
\end{split}
\ee
where the latter equalities hold term by term in the given order, since $\tilde \e_{1,1} = \eta_{(0,0),(1,1)} + \eta_{(1,0),(0,1)} + \eta_{(0,1),(1,0)}$. It can be checked that the above identifications are compatible with the constraints above.

The equivariant volume for $\mathbb{C}\times\mathrm{dP}_{3}$ is computed via the fixed point formula as
\begin{align}
&\BV_{\mathbb{C}\times\mathrm{dP}_{3}} (\lam,\tilde{\e}) \nonumber \\
&=\frac{e^{
\lambda_{1} (\tilde{\epsilon}_{2,1}-\tilde{\epsilon}_{0,2}-\tilde{\epsilon}_{0,1}+\tilde{\epsilon}_{2,0})
+\lambda_{2} (\tilde{\epsilon}_{1,2}+\tilde{\epsilon}_{0,2}-\tilde{\epsilon}_{1,0}-\tilde{\epsilon}_{2,0})
+\lambda_{7} (\tilde{\epsilon}_{0,2}+2 \tilde{\epsilon}_{0,1}+2 \tilde{\epsilon}_{1,0}+\tilde{\epsilon}_{2,0}+\tilde{\epsilon}_{1,1})
+\lambda_{8} \tilde{\epsilon}_{1}}
}{\tilde{\epsilon}_{1} (\tilde{\epsilon}_{2,1}-\tilde{\epsilon}_{0,2}-\tilde{\epsilon}_{0,1}+\tilde{\epsilon}_{2,0}) (\tilde{\epsilon}_{1,2}+\tilde{\epsilon}_{0,2}-\tilde{\epsilon}_{1,0}-\tilde{\epsilon}_{2,0}) (\tilde{\epsilon}_{0,2}+2 \tilde{\epsilon}_{0,1}+2 \tilde{\epsilon}_{1,0}+\tilde{\epsilon}_{2,0}+\tilde{\epsilon}_{1,1})} \nonumber \\
&\quad -\frac{e^{
\lambda_{1} (\tilde{\epsilon}_{2,1}+\tilde{\epsilon}_{1,2}-\tilde{\epsilon}_{0,1}-\tilde{\epsilon}_{1,0})
-\lambda_{6} (\tilde{\epsilon}_{1,2}+\tilde{\epsilon}_{0,2}-\tilde{\epsilon}_{1,0}-\tilde{\epsilon}_{2,0})
+\lambda_{7} (\tilde{\epsilon}_{1,2}+2 \tilde{\epsilon}_{0,2}+2 \tilde{\epsilon}_{0,1}+\tilde{\epsilon}_{1,0}+\tilde{\epsilon}_{1,1})
+\lambda_{8} \tilde{\epsilon}_{1}}
}{\tilde{\epsilon}_{1} (\tilde{\epsilon}_{2,1}+\tilde{\epsilon}_{1,2}-\tilde{\epsilon}_{0,1}-\tilde{\epsilon}_{1,0}) (\tilde{\epsilon}_{1,2}+\tilde{\epsilon}_{0,2}-\tilde{\epsilon}_{1,0}-\tilde{\epsilon}_{2,0}) (\tilde{\epsilon}_{1,2}+2 \tilde{\epsilon}_{0,2}+2 \tilde{\epsilon}_{0,1}+\tilde{\epsilon}_{1,0}+\tilde{\epsilon}_{1,1})} \nonumber \\
&\quad -\frac{e^{
\lambda_{2} (\tilde{\epsilon}_{2,1}+\tilde{\epsilon}_{1,2}-\tilde{\epsilon}_{0,1}-\tilde{\epsilon}_{1,0})
-\lambda_{3} (\tilde{\epsilon}_{2,1}-\tilde{\epsilon}_{0,2}-\tilde{\epsilon}_{0,1}+\tilde{\epsilon}_{2,0})
+\lambda_{7} (\tilde{\epsilon}_{2,1}+\tilde{\epsilon}_{0,1}+2 \tilde{\epsilon}_{1,0}+2 \tilde{\epsilon}_{2,0}+\tilde{\epsilon}_{1,1})
+\lambda_{8} \tilde{\epsilon}_{1}}
}{\tilde{\epsilon}_{1} (\tilde{\epsilon}_{2,1}+\tilde{\epsilon}_{1,2}-\tilde{\epsilon}_{0,1}-\tilde{\epsilon}_{1,0}) (\tilde{\epsilon}_{2,1}-\tilde{\epsilon}_{0,2}-\tilde{\epsilon}_{0,1}+\tilde{\epsilon}_{2,0}) (\tilde{\epsilon}_{2,1}+\tilde{\epsilon}_{0,1}+2 \tilde{\epsilon}_{1,0}+2 \tilde{\epsilon}_{2,0}+\tilde{\epsilon}_{1,1})} \nonumber \\
&\quad +\frac{e^{
\lambda_{3} (\tilde{\epsilon}_{1,2}+\tilde{\epsilon}_{0,2}-\tilde{\epsilon}_{1,0}-\tilde{\epsilon}_{2,0})
+\lambda_{4} (-\tilde{\epsilon}_{2,1}-\tilde{\epsilon}_{1,2}+\tilde{\epsilon}_{0,1}+\tilde{\epsilon}_{1,0})
+\lambda_{7} (2 \tilde{\epsilon}_{2,1}+\tilde{\epsilon}_{1,2}+\tilde{\epsilon}_{1,0}+2 \tilde{\epsilon}_{2,0}+\tilde{\epsilon}_{1,1})
+\lambda_{8} \tilde{\epsilon}_{1}}
}{
\tilde{\epsilon}_{1} 
(\tilde{\epsilon}_{1,2}+\tilde{\epsilon}_{0,2}-\tilde{\epsilon}_{1,0}-\tilde{\epsilon}_{2,0})
(-\tilde{\epsilon}_{2,1}-\tilde{\epsilon}_{1,2}+\tilde{\epsilon}_{0,1}+\tilde{\epsilon}_{1,0}) 
(2 \tilde{\epsilon}_{2,1}+\tilde{\epsilon}_{1,2}+\tilde{\epsilon}_{1,0}+2 \tilde{\epsilon}_{2,0}+\tilde{\epsilon}_{1,1})
} \nonumber \\
&\quad +\frac{e^{
-\lambda_{4} (\tilde{\epsilon}_{2,1}-\tilde{\epsilon}_{0,2}-\tilde{\epsilon}_{0,1}+\tilde{\epsilon}_{2,0})
-\lambda_{5} (\tilde{\epsilon}_{1,2}+\tilde{\epsilon}_{0,2}-\tilde{\epsilon}_{1,0}-\tilde{\epsilon}_{2,0})
+\lambda_{7} (2 \tilde{\epsilon}_{2,1}+2 \tilde{\epsilon}_{1,2}+\tilde{\epsilon}_{0,2}+\tilde{\epsilon}_{2,0}+\tilde{\epsilon}_{1,1})
+\lambda_{8} \tilde{\epsilon}_{1}}
}{\tilde{\epsilon}_{1} (\tilde{\epsilon}_{2,1}-\tilde{\epsilon}_{0,2}-\tilde{\epsilon}_{0,1}+\tilde{\epsilon}_{2,0}) (\tilde{\epsilon}_{1,2}+\tilde{\epsilon}_{0,2}-\tilde{\epsilon}_{1,0}-\tilde{\epsilon}_{2,0}) (2 \tilde{\epsilon}_{2,1}+2 \tilde{\epsilon}_{1,2}+\tilde{\epsilon}_{0,2}+\tilde{\epsilon}_{2,0}+\tilde{\epsilon}_{1,1})} \nonumber \\
&\quad +\frac{e^{
\lambda_{5} (-\tilde{\epsilon}_{2,1}-\tilde{\epsilon}_{1,2}+\tilde{\epsilon}_{0,1}+\tilde{\epsilon}_{1,0})
+\lambda_{6} (\tilde{\epsilon}_{2,1}-\tilde{\epsilon}_{0,2}-\tilde{\epsilon}_{0,1}+\tilde{\epsilon}_{2,0})
+\lambda_{7} (\tilde{\epsilon}_{2,1}+2 \tilde{\epsilon}_{1,2}+2 \tilde{\epsilon}_{0,2}+\tilde{\epsilon}_{0,1}+\tilde{\epsilon}_{1,1})
+\lambda_{8} \tilde{\epsilon}_{1}}
}{\tilde{\epsilon}_{1} (-\tilde{\epsilon}_{2,1}-\tilde{\epsilon}_{1,2}+\tilde{\epsilon}_{0,1}+\tilde{\epsilon}_{1,0}) (\tilde{\epsilon}_{2,1}-\tilde{\epsilon}_{0,2}-\tilde{\epsilon}_{0,1}+\tilde{\epsilon}_{2,0}) (\tilde{\epsilon}_{2,1}+2 \tilde{\epsilon}_{1,2}+2 \tilde{\epsilon}_{0,2}+\tilde{\epsilon}_{0,1}+\tilde{\epsilon}_{1,1})}\ ,
\end{align}
where the numbering is in accordance with the rows in \eqref{eq:dP3toricvectors}.

Using the above information, one can verify that the following relations hold:
\begin{align}
C^{Q^{1,1,1}}\left(\epsilon (\eta)\right) & =C^{\mathbb{C}\times\mathrm{dP}_{3}}\left(\tilde{\epsilon} (\eta)\right)\ ,\nonumber \\
B^{Q^{1,1,1}}\left(\epsilon (\eta)\right) & =B^{\mathbb{C}\times\mathrm{dP}_{3}}\left(\tilde{\epsilon} (\eta)\right)+\frac{1}{4}\ .
\end{align}
At this stage, \eqref{eq:CY43-C} is satisfied. To evaluate \eqref{eq:CY43-B}, we turn to the quantum curve approach. $B^{\fccz}$ is given as \eqref{eq:CB-A0110} with $\tilde{D}=D_{\mathrm{m}}=0$, which vanishes: $B^{\fccz}\Big|_{\tilde{D} = D_{\mathrm{m}} = 0}=0$. To obtain $B^{\fcczp}$, we use the quantum curve
\begin{equation}
\hat{\mathcal{O}}_{\left(1,1,0\right),\left(0,1,0\right)}^{\fcczp}\left(\hat{x},\hat{p}\right)=\left(\begin{array}{ccc}
+e^{\hat{x}-\hat{p}} & +e^{\hat{x}} & +0\\
+e^{-\hat{p}} & +c & +e^{\hat{p}}\\
+0 & +e^{-\hat{x}} & +e^{-\hat{x}+\hat{p}}
\end{array}\right)\ .
\label{eq:QC-dP3}
\end{equation}
Here $c$ denotes the coefficient associated with the strictly interior point of the Newton polygon. Since it does not affect the perturbative Airy coefficients $C$ and $B$, we leave its explicit value unspecified.
This quantum curve is obtained from $\hat{\mathcal{O}}_{\left(1,1,0\right),\left(0,1,0\right)}^{\fccz}$ with the derivation described in section \ref{subsec:CY43-B}. 
We find that the volume of the Fermi surface for \eqref{eq:QC-dP3} is
\begin{equation}
\mathrm{Vol}=3E^{2}-2\pi^{2}+O\left(e^{-E}\right)\ ,
\end{equation}
and thus the Airy coefficient $B$ obtained from this quantum curve is $B^{\fcczp}=-\frac{1}{4}.$ 
Therefore, \eqref{eq:CY43-B} also holds.

\subsubsection*{Unnamed example} 
The fourth example is the case when the toric diagram of the $\mathrm{CY}_{4}$ consists of a 2d toric diagram of local $\mathbb{P}^{2}$ in the $z=0$ plane and a 2d toric diagram dual to an NS5-brane in the $z=1$ plane. The toric fan vectors for the $\mathrm{CY}_{4}$ and the $\mathrm{CY}_{3}$ are
\begin{equation}
\label{eq:unnamedtoricvectors}
v=\left(\begin{array}{cccc}
1 & 1 & 0 & 0\\
1 & 0 & 1 & 0\\
1 & -1 & -1 & 0\\
1 & 0 & 0 & 0\\
1 & 1 & 0 & 1\\
1 & 0 & 0 & 1
\end{array}\right),\quad \Rightarrow \quad\tilde{v}=\left(\begin{array}{cccc}
1 & 2 & 0 & 0\\
1 & 1 & 1 & 0\\
1 & 0 & 1 & 0\\
1 & -1 & -1 & 0\\
1 & 0 & -1 & 0\\
1 & 1 & 0 & 0\\
1 & 0 & 0 & 0\\
1 & 0 & 0 & 1
\end{array}\right).
\end{equation}
The constraints on the respective equivariant parameters, \eqref{eq:constr1CY4} and \eqref{eq:constr2CY3}, are given by
\be
\begin{split}
	\e^{(0)}_{0,0} &+ \e^{(0)}_{1,0} + \e^{(0)}_{0,1} +\e^{(0)}_{-1,-1} = 1 \ , \qquad \e^{(1)}_{0,0} + \e^{(1)}_{1,0} = 1 \ , \\
	\tilde \e_1 &= 1\ , \qquad \tilde \e_{2,0} + \tilde \e_{1,1}+\tilde \e_{0,1} + \tilde \e_{-1,-1} + \tilde \e_{0,-1}+\tilde \e_{1,0} + \tilde \e_{0,0} = 1\ .
\end{split}
\ee
In terms of the auxiliary variables introduced in \eqref{eq:CY43-Rcorr}, we then find
\be
\begin{split}
	\e^{(0)}_{0,0} &= \eta_{(0,0),(0,0)} + \eta_{(0,0),(1,0)} = \tilde \e_{0,0}+  (\tilde \e_{1,0}- \eta_{(1,0),(0,0)})\ , \\
	\e^{(0)}_{1,0} &= \eta_{(1,0),(0,0)} + \eta_{(1,0),(1,0)} = (\tilde \e_{1,0}- \eta_{(0,0),(1,0)}) + \tilde \e_{2,0}\ , \\
	\e^{(0)}_{0,1} &= \eta_{(0,1),(0,0)} + \eta_{(0,1),(1,0)} = \tilde \e_{0,1} + \tilde \e_{1,1}\ , \\
	\e^{(0)}_{-1,-1} &= \eta_{(-1,-1),(0,0)} + \eta_{(-1,-1),(1,0)} = \tilde \e_{-1,-1} +\tilde \e_{0,-1}\ , \\
\e^{(1)}_{0,0} &= \eta_{(0,0),(0,0)} + \eta_{(1,0),(0,0)} + \eta_{(0,1),(0,0)} + \eta_{(-1,-1),(0,0)} \\ &=\tilde \e_{0,0} +  (\tilde \e_{1,0}- \eta_{(0,0),(1,0)})+ \tilde\e_{0,1} +\tilde\e_{-1,-1}\ , \\
\e^{(1)}_{1,0} &= \eta_{(0,0),(1,0)} + \eta_{(1,0),(1,0)} + \eta_{(0,1),(1,0)} + \eta_{(-1,-1),(1,0)}  \\ &=(\tilde \e_{1,0}- \eta_{(1,0),(0,0)})+\tilde\e_{2,0} + \tilde \e_{1,1} +\tilde\e_{0,-1}\ , 
\end{split}
\ee
where the latter equalities hold term by term in the given order, since $\tilde \e_{1,0} = \eta_{(0,0),(1,0)} + \eta_{1,0),(0,0)}$. It can be checked that the above identifications are compatible with the constraints above.

The equivariant volumes for $\mathrm{CY}_{4}$ and $\mathbb{C}\times\mathrm{CY}_{3}$ are computed via the fixed point formula as
\begin{align}
\BV_{\mathrm{CY}_{4}} (\lam,\e)
&=\frac{e^{
\lambda_{1} (\epsilon_{1,0}^{(0)}-\epsilon_{-1,-1}^{(0)}+\epsilon_{1,0}^{(1)})
+\lambda_{2} (\epsilon_{0,1}^{(0)}-\epsilon_{-1,-1}^{(0)})
+\lambda_{4} (3 \epsilon_{-1,-1}^{(0)}+\epsilon_{0,0}^{(0)}-\epsilon_{1,0}^{(1)})
+\lambda_{6} (\epsilon_{1,0}^{(1)}+\epsilon_{0,0}^{(1)})}
}{
(\epsilon_{1,0}^{(0)}-\epsilon_{-1,-1}^{(0)}+\epsilon_{1,0}^{(1)})
(\epsilon_{0,1}^{(0)}-\epsilon_{-1,-1}^{(0)}) 
(3 \epsilon_{-1,-1}^{(0)}+\epsilon_{0,0}^{(0)}-\epsilon_{1,0}^{(1)})
(\epsilon_{1,0}^{(1)}+\epsilon_{0,0}^{(1)}) 
} \nonumber \\
&+\frac{e^{
\lambda_{1} (\epsilon_{1,0}^{(0)}+2 \epsilon_{-1,-1}^{(0)}+\epsilon_{0,0}^{(0)})
+\lambda_{2} (\epsilon_{0,1}^{(0)}-\epsilon_{-1,-1}^{(0)})
+\lambda_{5} (-3 \epsilon_{-1,-1}^{(0)}-\epsilon_{0,0}^{(0)}+\epsilon_{1,0}^{(1)})
+\lambda_{6} (3 \epsilon_{-1,-1}^{(0)}+\epsilon_{0,0}^{(0)}+\epsilon_{0,0}^{(1)})}
}{
(\epsilon_{1,0}^{(0)}+2 \epsilon_{-1,-1}^{(0)}+\epsilon_{0,0}^{(0)}) 
(\epsilon_{0,1}^{(0)}-\epsilon_{-1,-1}^{(0)}) 
(-3 \epsilon_{-1,-1}^{(0)}-\epsilon_{0,0}^{(0)}+\epsilon_{1,0}^{(1)})
(3 \epsilon_{-1,-1}^{(0)}+\epsilon_{0,0}^{(0)}+\epsilon_{0,0}^{(1)})
} \nonumber \\
&-\frac{e^{
\lambda_{1} (\epsilon_{1,0}^{(0)}-\epsilon_{0,1}^{(0)}+\epsilon_{1,0}^{(1)})
-\lambda_{3} (\epsilon_{0,1}^{(0)}-\epsilon_{-1,-1}^{(0)})
+\lambda_{4} (3\epsilon_{0,1}^{(0)} + \epsilon_{0,0}^{(0)} - \epsilon_{1,0}^{(1)})
+\lambda_{6} (\epsilon_{1,0}^{(1)}+\epsilon_{0,0}^{(1)})}
}{
(\epsilon_{1,0}^{(0)}-\epsilon_{0,1}^{(0)}+\epsilon_{1,0}^{(1)})
(\epsilon_{0,1}^{(0)}-\epsilon_{-1,-1}^{(0)}) 
(3\epsilon_{0,1}^{(0)} + \epsilon_{0,0}^{(0)} - \epsilon_{1,0}^{(1)})
(\epsilon_{1,0}^{(1)}+\epsilon_{0,0}^{(1)}) 
} \nonumber \\
&-\frac{e^{
\lambda_{1} (\epsilon_{1,0}^{(0)}+2 \epsilon_{0,1}^{(0)}+\epsilon_{0,0}^{(0)})
-\lambda_{3} (\epsilon_{0,1}^{(0)}-\epsilon_{-1,-1}^{(0)})
+\lambda_{5} (-(3\epsilon_{0,1}^{(0)} + \epsilon_{0,0}^{(0)} - \epsilon_{1,0}^{(1)}))
+\lambda_{6} (3\epsilon_{0,1}^{(0)} + \epsilon_{0,0}^{(0)} + \epsilon_{0,0}^{(1)})}
}{
(\epsilon_{1,0}^{(0)}+2 \epsilon_{0,1}^{(0)}+\epsilon_{0,0}^{(0)})
(\epsilon_{0,1}^{(0)}-\epsilon_{-1,-1}^{(0)}) 
(-3\epsilon_{0,1}^{(0)} - \epsilon_{0,0}^{(0)} + \epsilon_{1,0}^{(1)}) 
(3\epsilon_{0,1}^{(0)} + \epsilon_{0,0}^{(0)} + \epsilon_{0,0}^{(1)})
} \nonumber \\
&-\frac{e^{
\lambda_{2} (-\epsilon_{1,0}^{(0)}+\epsilon_{0,1}^{(0)}-\epsilon_{1,0}^{(1)})
-\lambda_{3} (\epsilon_{1,0}^{(0)}-\epsilon_{-1,-1}^{(0)}+\epsilon_{1,0}^{(1)})
+\lambda_{4} (3\epsilon_{1,0}^{(0)} + \epsilon_{0,0}^{(0)} + 2\epsilon_{1,0}^{(1)})
+\lambda_{6} (\epsilon_{1,0}^{(1)}+\epsilon_{0,0}^{(1)})}
}{
(-\epsilon_{1,0}^{(0)}+\epsilon_{0,1}^{(0)}-\epsilon_{1,0}^{(1)}) 
(\epsilon_{1,0}^{(0)}-\epsilon_{-1,-1}^{(0)}+\epsilon_{1,0}^{(1)}) 
(3\epsilon_{1,0}^{(0)} + \epsilon_{0,0}^{(0)} + 2\epsilon_{1,0}^{(1)})
(\epsilon_{1,0}^{(1)}+\epsilon_{0,0}^{(1)})
}\ ,
\end{align}
and
\begin{align}
\BV_{\mathbb{C}\times\mathrm{CY}_{3}} (\lam,\tilde{\e})
&=\frac{e^{
\lambda_{1} \mathcal{E}(1, 0, 0, -1, -1, -1, -2) 
+ \lambda_{2} \mathcal{E}(0, 1, 0, 1, 0, -1, -1) 
+\lambda_{6} \mathcal{E}(0, 0, 1, 1, 2, 3, 4) 
+ \lambda_{8} \tilde{\epsilon}_{1}}
}{
\tilde{\epsilon}_{1} 
\mathcal{E}(1, 0, 0, -1, -1, -1, -2)
\mathcal{E}(0, 1, 0, 1, 0, -1, -1)
\mathcal{E}(0, 0, 1, 1, 2, 3, 4)
} \nonumber \\
&-\frac{e^{
\lambda_{1} \mathcal{E}(1, -1, 0, -2, -1, 0, -1) 
- \lambda_{5} \mathcal{E}(0, 1, 0, 1, 0, -1, -1)
+\lambda_{6} \mathcal{E}(0, 3, 1, 4, 2, 0, 1) 
+ \lambda_{8} \tilde{\epsilon}_{1}}
}{
\tilde{\epsilon}_{1} 
\mathcal{E}(1, -1, 0, -2, -1, 0, -1) 
\mathcal{E}(0, 1, 0, 1, 0, -1, -1)
\mathcal{E}(0, 3, 1, 4, 2, 0, 1)
} \nonumber \\
&+\frac{e^{
\lambda_{2} \mathcal{E}(1, 1, 0, 0, -1, -2, -3) 
+ \lambda_{3} \mathcal{E}(-1, 0, 0, 1, 1, 1, 2) 
+ \lambda_{6} \mathcal{E}(1, 0, 1, 0, 1, 2, 2)
+ \lambda_{8} \tilde{\epsilon}_{1}}
}{
\tilde{\epsilon}_{1} 
\mathcal{E}(1, 1, 0, 0, -1, -2, -3) 
\mathcal{E}(-1, 0, 0, 1, 1, 1, 2) 
\mathcal{E}(1, 0, 1, 0, 1, 2, 2)
} \nonumber \\
&-\frac{e^{
\lambda_{3} \mathcal{E}(-2, 0, -1, 1, 0, -1, 0) 
- \lambda_{4} \mathcal{E}(2, 1, 1, 0, 0, 0, -1) 
+ \lambda_{7} \mathcal{E}(5, 2, 3, 0, 1, 2, 0) 
+ \lambda_{8} \tilde{\epsilon}_{1}}
}{
\tilde{\epsilon}_{1}
\mathcal{E}(-2, 0, -1, 1, 0, -1, 0)
\mathcal{E}(2, 1, 1, 0, 0, 0, -1) 
\mathcal{E}(5, 2, 3, 0, 1, 2, 0)
} \nonumber \\
&+\frac{e^{
\lambda_{3} \mathcal{E}(0, 1, 0, 1, 0, -1, -1)
+ \lambda_{6} \mathcal{E}(2, 1, 1, 0, 0, 0, -1) 
+  \lambda_{7} \mathcal{E}(-1, -1, 0, 0, 1, 2, 3)
+ \lambda_{8} \tilde{\epsilon}_{1}}
}{
\tilde{\epsilon}_{1}
\mathcal{E}(0, 1, 0, 1, 0, -1, -1) 
\mathcal{E}(2, 1, 1, 0, 0, 0, -1)
\mathcal{E}(-1, -1, 0, 0, 1, 2, 3)
} \nonumber \\
&+\frac{e^{
\lambda_{4} \mathcal{E}(-1, 1, 0, 2, 1, 0, 1)
+ \lambda_{5} \mathcal{E}(1, -2, 0, -3, -1, 1, 0) 
+ \lambda_{6} \mathcal{E}(1, 2, 1, 2, 1, 0, 0) 
+ \lambda_{8} \tilde{\epsilon}_{1}}
}{
\tilde{\epsilon}_{1}
\mathcal{E}(-1, 1, 0, 2, 1, 0, 1)
\mathcal{E}(1, -2, 0, -3, -1, 1, 0)
\mathcal{E}(1, 2, 1, 2, 1, 0, 0) 
} \nonumber \\
&-\frac{e^{
- \lambda_{4} \mathcal{E}(0, 1, 0, 1, 0, -1, -1)
+ \lambda_{6} \mathcal{E}(2, 0, 1, -1, 0, 1, 0) 
+\lambda_{7} \mathcal{E}(-1, 2, 0, 3, 1, -1, 0) 
+ \lambda_{8} \tilde{\epsilon}_{1}}
}{
\tilde{\epsilon}_{1}
\mathcal{E}(0, 1, 0, 1, 0, -1, -1) 
\mathcal{E}(2, 0, 1, -1, 0, 1, 0)
\mathcal{E}(-1, 2, 0, 3, 1, -1, 0)
}\ ,
\end{align}
where
\begin{align}
 & \mathcal{E}\left(k_{(2,0)},k_{(1,1)},k_{(1,0)},k_{(0,1)},k_{(0,0)},k_{(0,-1)},k_{(-1,-1)}\right) :=\nonumber \\
 & \, \, k_{(2,0)}\tilde{\epsilon}_{2,0}+k_{(1,1)}\tilde{\epsilon}_{1,1}+k_{(1,0)}\tilde{\epsilon}_{1,0}+k_{(0,1)}\tilde{\epsilon}_{0,1}+k_{(0,0)}\tilde{\epsilon}_{0,0}+k_{(0,-1)}\tilde{\epsilon}_{0,-1}+k_{(-1,-1)}\tilde{\epsilon}_{-1,-1}\ .
\end{align}

Given the above information, we find
\begin{align}
C^{\mathrm{CY}_{4}}\left(\epsilon (\eta)\right) & =C^{\mathbb{C}\times\mathrm{CY}_{3}}\left(\tilde{\epsilon} (\eta)\right)\ ,\nonumber \\
B^{\mathrm{CY}_{4}}\left(\epsilon (\eta)\right) & =B^{\mathbb{C}\times\mathrm{CY}_{3}}\left(\tilde{\epsilon} (\eta)\right)+\frac{1}{4}\ .
\end{align}
At this stage, \eqref{eq:CY43-C} is satisfied. 
To evaluate \eqref{eq:CY43-B2}, we turn to the quantum-curve approach. 
The corresponding quantum curves are
\begin{equation}
\hat{\mathcal{O}}^{\mathcal{T}\left[\mathrm{CY}_{4}\right]}\left(\hat{x},\hat{p}\right)=\left(\begin{array}{ccc}
+0 & +e^{\hat{x}} & +0\\
+0 & +c & +e^{\hat{p}}\\
+e^{-\hat{x}-\hat{p}} & +0 & +0
\end{array}\right)\times\left(e^{-\frac{1}{2}\hat{p}}+e^{\frac{1}{2}\hat{p}}\right)\ ,
\end{equation}
and
\begin{equation}
\hat{\mathcal{O}}^{\mathcal{T}\left[\mathbb{C}\times\mathrm{CY}_{3}\right]}\left(\hat{x},\hat{p}\right)=\left(\begin{array}{cccc}
+0 & +e^{\hat{x}-\frac{1}{2}\hat{p}} & +e^{\hat{x}+\frac{1}{2}\hat{p}} & +0\\
+0 & +c_1 e^{-\frac{1}{2}\hat{p}} & +c_2 e^{\frac{1}{2}\hat{p}} & +e^{\frac{3}{2}\hat{p}}\\
+e^{-\hat{x}-\frac{3}{2}\hat{p}} & +e^{-\hat{x}-\frac{1}{2}\hat{p}} & +0 & +0
\end{array}\right)\ .
\end{equation}
Here $c$, $c_1$, and $c_2$ denote the coefficients associated with strictly interior points of the corresponding Newton polygons. Their explicit values do not affect the perturbative Airy coefficients $C$ and $B$, and we therefore leave them unspecified.
We find that the volumes of the Fermi surfaces for these quantum curves are
\begin{align}
\mathrm{Vol}^{\mathcal{T}\left[\mathrm{CY}_{4}\right]}\left(E\right) & =\frac{7}{3}E^{2}-\frac{29}{36}\pi^{2}+O\left(e^{-E}\right)\ ,\nonumber \\
\mathrm{Vol}^{\mathcal{T}\left[\mathbb{C}\times\mathrm{CY}_{3}\right]}\left(E\right) & =\frac{7}{3}E^{2}-\frac{65}{36}\pi^{2}+O\left(e^{-E}\right)\ ,
\end{align}
and thus the Airy functions obtained from these quantum curves are
\begin{equation}
C^{\mathcal{T}\left[\mathrm{CY}_{4}\right]}=\frac{7}{12\pi^{2}},\quad B^{\mathcal{T}\left[\mathrm{CY}_{4}\right]}=-\frac{1}{144}\ ,
\end{equation}
and
\begin{equation}
C^{\mathcal{T}\left[\mathbb{C}\times\mathrm{CY}_{3}\right]}=\frac{7}{12\pi^{2}},\quad B^{\mathcal{T}\left[\mathbb{C}\times\mathrm{CY}_{3}\right]}=-\frac{37}{144}\ .
\end{equation}
Therefore, \eqref{eq:CY43-B2} also holds.

\section{Discussion}
\label{sec:Discussion}
In this paper, we derived the perturbative Airy coefficients $C$ and $B$ for a broad class of holographic three-dimensional SCFTs using the quantum-curve formalism, and found precise agreement, in the examples considered, with the geometric predictions based on equivariant volumes.
Both perspectives are naturally rooted in the underlying brane web constructions. 
Building on earlier results, we significantly extended these methods to include flavored theories, providing new evidence for the validity of the AdS$_4$/CFT$_3$ correspondence beyond the strict large-$N$ limit, 
specifically in the form of the Airy function at finite $N$.
Furthermore, we identified a new structure relating toric Calabi--Yau geometries and formulated it as a conjectural equivariant CY$_4$/CY$_3$ correspondence from the analyses of holography and quantum curves, which we tested in several non-trivial examples.

We conclude with a number of more focused remarks on the implications of our results for the TS/ST correspondence and  more generally for holography, which are discussed in the subsections below.

\subsection{TS/ST correspondence}
Even though we have repeatedly emphasized throughout this work that our results are closely related to, and strongly motivated by, the TS/ST correspondence, we have not made explicit use of it in our derivations. To explain this choice, it is useful to briefly review the correspondence and clarify its scope.

The TS/ST correspondence can be most efficiently formulated as an equivalence between the quantum curve associated with a matrix model—of the type we have considered here, but without the explicit inclusion of the deformation parameters $\Delta$—and the mirror curve of a toric Calabi–Yau threefold. In the context of toric Calabi–Yau threefolds, the mirror curve is a complex plane algebraic curve whose genus is determined by the combinatorial data of the toric diagram. This notion of genus plays a central role in the spectral theory/topological string correspondence: for mirror curves of genus $\mathfrak{g}$, quantization leads to rich spectral structures and generalized spectral determinants that encode topological string amplitudes. The general framework was developed in \cite{Codesido:2015dia}, with the genus-one case and its exact quantization conditions explored in detail in \cite{Grassi:2014zfa}, and extensions to arbitrary genus presented in \cite{Codesido:2015dia,Marino:2016rsq}.

A crucial point for our purposes is that the original TS/ST correspondence is not formulated in equivariant topological string theory. As a consequence, the associated quantum curve does not depend on the deformation parameters $\Delta$. This is not accidental: at present, to the best of our knowledge, an equivariant formulation of the mirror curve itself has not been constructed, and doing so would represent a non-trivial mathematical problem. In contrast, the bulk computations performed in this work rely only on the equivariant generalization of the constant map contributions \cite{Cassia:2025aus}, for which a detailed understanding of the full quantum geometry in the presence of equivariant parameters is not required.

From the perspective of the TS/ST correspondence, however, our results strongly suggest that equivariance does not obstruct the correspondence, but rather provides a natural extension of it. To illustrate this point, consider the original relation between the quantum curve governing the $S^3$ partition function of ABJM theory and the mirror curve of local $\mathbb{P}^1 \times \mathbb{P}^1$ \cite{Grassi:2014zfa}. Using what we have termed the equivariant CY$_4$/CY$_3$ correspondence, we showed that in the presence of deformation parameters $\Delta$, the ABJM partition function is not only holographically related to $\mathbb{C}^4$, as dictated directly by the bulk geometry, but also to $\mathbb{C} \times \cC$. Furthermore, as discussed in appendix \ref{app:An equivariant duality}, there exists a mathematical equivalence—referred to as equivariant duality—between the equivariant volumes of $\cC$ and local $\mathbb{P}^1 \times \mathbb{P}^1$. In this way, we recover the relation between ABJM theory and local $\mathbb{P}^1 \times \mathbb{P}^1$ in the presence of equivariant parameters, albeit without invoking the mirror curve directly. It is then natural to conjecture, although technically challenging to establish, that this relation persists non-perturbatively for arbitrary equivariant deformations.

More generally, we conjecture that the equivariant CY$_4$/CY$_3$ correspondence—possibly supplemented by the equivariant duality described in appendix \ref{app:An equivariant duality}—provides a geometric origin of the TS/ST correspondence itself. In its original form, the TS/ST correspondence may be viewed as fundamentally arising from the mirror curve of a toric manifold, which does not directly encode the bulk physics. By reinstating the missing geometric link through the bulk Calabi–Yau geometry probed by the M2-brane system, we complete the conceptual circle depicted in figure \ref{fig:1}, allowing for a unified and transparent connection between the brane construction, quantum curves, topological strings, and holography.

Pushing this perspective one step further, one may speculate that our results effectively provide an equivariant mirror curve for the relevant Calabi–Yau threefolds. Concretely, this would amount to identifying the equivariant mirror curve $W_X$ with the quantum curve of the corresponding field theory. In the case of ABJM theory, this identification would lead to
\be
	W_{\mathrm{loc.} \mathbb{P}^1 \times \mathbb{P}^1} (x,p; \tilde \e) := \left( \cO^\text{SYM}_{0, (0,0,1)} \right)_W (x,p; D, D_{\mathrm{m}})\ ,  
\ee
where the relation between the equivariant parameters $\tilde{\e}$ and the deformation parameters $D, D_{\mathrm{m}}$ follows from the mapping derived in the previous section. The same reasoning can be applied straightforwardly to all other examples discussed in this work.

\subsection{Holography}

Before discussing the more direct implications of our results for the AdS/CFT correspondence, let us emphasize that our analysis relies on a somewhat unconventional bulk perspective. Instead of focusing on the full gravitational dynamics of the AdS factor, we have worked with the geometry of the transverse space in the brane construction, following the spirit of \cite{Martelli:2005tp,Butti:2005vn,Butti:2005ps,Martelli:2006yb,Amariti:2011uw,Couzens:2018wnk,Gauntlett:2018dpc,Hosseini:2019use,Hosseini:2019ddy,Gauntlett:2019roi,Kim:2019umc,Martelli:2023oqk,Colombo:2023fhu,Cassia:2025aus,Cassia:2025jkr}. There is substantial evidence supporting this approach, and it has been successfully applied in many examples. However, its validity has not been rigorously established in the presence of higher-derivative supergravity corrections, which correspond to finite-$N$ effects. In this sense, our analysis implicitly assumes an additional layer of non-trivial relations beyond those summarized in figure \ref{fig:1}.

With this caveat in mind, and given the network of connections explored in this work, it is natural to ask whether these results bring us closer to a proof of the AdS/CFT correspondence. The matching of the $S^3$ partition function computed by two seemingly independent methods is particularly suggestive. On the one hand, the quantum-curve description produces the area of the polar polygon, weighted by supersymmetric deformation parameters, together with its quantum corrections, see \eqref{eq:VolForm}-\eqref{eq:VolC-K12}. On the other hand, the geometric computation amounts to evaluating the equivariant volume of the cone over the transverse space, see \eqref{eq:Cgeometry}-\eqref{eq:Bgeometry}. As outlined in section \ref{subsec:equiVol}, this reduces to a sum over the areas of triangles in a triangulation of the toric diagram, where each triangle area is weighted by equivariant parameters—mapping precisely the same supersymmetric deformation parameters that appear on the quantum-curve side.

We therefore arrive at a non-trivial identification between two different calculations of weighted areas associated with the same polygon. Although the sequence of correspondences leading to this result is intricate, the final statement is purely geometric. This raises the possibility that a more direct and mathematically precise argument could establish the equivalence of the two computations in general, without relying on the case-by-case checks performed here.

We should note, as emphasized in several places throughout the text, that the aforementioned agreement between the bulk and field theory computations calls for a better understanding of the role played by geometric versus non-geometric phases in the resolution of the corresponding Calabi–Yau singularity. This issue becomes particularly relevant when considering subleading finite-$N$ corrections, which can be shown to depend on the choice of phase. We hope to clarify this point in future work.

Other important holographic predictions, which we have not addressed in the present work, concern contributions that are independent of $N$ in the M-theory description (for instance, in ABJM-like theories we keep $k$ fixed while taking the large-$N$ limit). From the perspective of topological string theory, these terms correspond to higher-genus constant-map contributions. While these are also accessible through the quantum curve description, their analysis is technically more involved. Understanding these corrections represents the next natural step in the present program, prior to the longer-term goal of achieving a fully non-perturbative formulation.

\subsection{Other future directions}

In this subsection, we enumerate additional directions for future research.

First, in describing the relationship between equivariant parameters within the CY$_4$/CY$_3$ correspondence, we introduced the notation $\eta$ in \eqref{eq:CY43-Rcorr} for computational convenience. 
It would be insightful to further explore the physical and geometric meaning of $\eta_{(x_0, y_0),(x_1, y_1)}$. 
For instance, the $\ccz$ family, including its flavored generalizations, admits a perfect-matching description in which each chiral field is a monomial in perfect-matching, or equivalently GLSM, variables. The R-charge of a chiral field is therefore obtained by summing the R-charges assigned to the perfect matchings that contain it. Moreover, several perfect matchings may be mapped to the same toric lattice point, and the number of such perfect matchings is its multiplicity in the perfect-matching/GLSM description \cite{Hanany:2008fj,Benini:2009qs,Amariti:2012tj}.
Comparing these data with the parametrization in \eqref{eq:CY43-Rcorr}, we observe in the simplest examples considered above that the sums of the $\eta_{(x_0,y_0),(x_1,y_1)}$ parameters have the same incidence structure as the perfect-matching decomposition of the matter-field R-charges. Moreover, in these examples, the number of $\eta$ parameters mapped to a point of the Minkowski sum agrees with the perfect-matching multiplicity at the corresponding toric point.
These observations suggest that the $\eta_{(x_0,y_0),(x_1,y_1)}$ parameters may admit an interpretation as R-charge parameters associated with perfect-matching, or equivalently GLSM, variables. Establishing this interpretation systematically, including its dependence on the choice of brane tiling and toric phase, would be an interesting direction for future work.

Second, as demonstrated in this study, the 2d Newton polygon of the quantum curve is intimately related to the 3d toric diagram via the Minkowski sum. 
More precisely, the 3d toric diagram is associated with a 2d toric diagram through the CY$_4$/CY$_3$ correspondence, with the latter directly corresponding to the Newton polygon. 
These findings, however, suggest that a more fundamental description may require the introduction of a "three-dimensional quantum curve." 
Specifically, while we imposed the conditions $\Delta_{\phi_3}=1$ and $\Delta_{\phi_i}=1$ in \eqref{eq:SYM-Rcond} and \eqref{eq:fABJM-Rcond} to facilitate the use of the quantum curve, it would be interesting to investigate whether such a 3d extension allows us to remove these constraints and treat more general cases.

\subsection*{Acknowledgments}
We are grateful to Luca Cassia, Alba Grassi, Tomoki Nosaka, Sanefumi Moriyama, Ali Mert Yetkin and Alberto Zaffaroni for valuable discussions. 
The work of K. H. is financed by the European Union - NextGenerationEU, through the National Recovery and Resilience Plan of the Republic of Bulgaria, project No BG-RRP-2.004-0008-C01. 
The work of N. K. is supported by the JSPS KAKENHI (Grant-in-Aid for Scientific Research (B)) grant number JP24K00628 and the Graduate School of Science, Kyoto University, under the Ginpu Fund 2025. 
Y. P. is supported by the National Natural Science Foundation of China (NSFC) under Grants
No. 12575076 and No. 12247103.

\appendix

\section{From brane configurations to quantum curves}\label{sec:BCtoQC}

In this appendix, we present the framework for relating brane configurations to quantum curves via the $S^{3}$ partition functions. The dictionary was established in \cite{Kubo:2025jxi} for fixed R-charges of the bi-fundamental matters and vanishing bare monopole R-charges. Since we consider non-trivial R-charges here, we generalize their results. While the R-charge deformation can be incorporated without major changes, we revisit the derivation to account for it explicitly.

The strategy for identifying the quantum curve is as follows. First, the brane configuration specifies the quiver gauge theory on its worldvolume. Second, we construct the corresponding matrix models from the $S^{3}$ partition functions via supersymmetric localization. Third, the Fermi gas formalism encodes the partition function in the density matrix. Finally, using the properties of the double sine function appearing in the density matrix, we obtain the quantum curve. The explicit results for the brane configurations \eqref{eq:fSYM-BC}, \eqref{eq:fABJM-BC}, and \eqref{eq:fABJM-BC2} are given in the main text; here, we outline the framework underlying these results.

\subsection{Supersymmetric localization}\label{subsec:Localization}

Here, we explain how the matrix models are obtained from brane configurations using the supersymmetric localization technique \cite{Pestun:2007rz,Kapustin:2009kz}. 
In these models, the integration is performed over the eigenvalues of the scalars in the unitary gauge groups. 
The integrand receives contributions from two sources: the classical part and the 1-loop part. 
The classical factors arise from the Chern-Simons term with the Chern-Simons level $k$ and from the monopole operator with bare R-charge $\Delta_{\mathrm{m}}$. For the $r$-th $\mathrm{U}(N)$ group, they are \cite{Kapustin:2009kz}
\begin{equation}
Z_{\text{CS}}\left(k^{(r)};\mu\right)
=e^{-\frac{i}{4\pi}k^{(r)}\sum_{a=1}^{N}\mu_{a}^{2}}\ ,\qquad
Z_{\mathrm{m}}\left(\Delta_{\mathrm{m}}^{(r)};\mu\right)
=e^{i\Delta_{\mathrm{m}}^{(r)}\sum_{a=1}^{N}\mu_{a}}\ .
\end{equation}
The 1-loop contribution for a vector multiplet is given by
\begin{equation}
Z_{\text{vec}}\left(\mu\right)=\prod_{a<b}^{N}\left(2\sinh\frac{\mu_{a}-\mu_{b}}{2}\right)^{2}\ ,
\end{equation}
while the 1-loop contribution for a chiral multiplet in representation $\mathcal{R}$ is given by
\begin{equation}
Z_{\text{chiral},\mathcal{R}}\left(\Delta;\mu\right)=\prod_{\rho\in\mathcal{R}}s_{1}\left(i\left(1-\Delta\right)-\rho\left(\frac{\mu}{2\pi}\right)\right)\ ,\label{eq:MF-chiral}
\end{equation}
where $\Delta$ is the R-charge of the chiral multiplet.
$s_b(z)$ is the double sine function defined as
\begin{equation}
s_{b}\left(z\right)=\prod_{\ell,m=0}^{\infty}\frac{\ell b+mb^{-1}+\frac{1}{2}\left(b+b^{-1}\right)-iz}{\ell b+mb^{-1}+\frac{1}{2}\left(b+b^{-1}\right)+iz}\ .
\label{eq:DS-Def}
\end{equation}

Using these rules, the matrix models for the flavored SYM theory and the flavored $\ccz$ theory can be directly read off from the quiver diagram in figure \ref{fig:SYM_fABJM_Quiv}. Before presenting the explicit matrix models, we make a few remarks. First, the contribution from the adjoint matters with R-charge one reduces to unity, thanks to the following property of the double sine function
\begin{equation}
s_{b}\left(z\right)s_{b}\left(-z\right)=1\ .\label{eq:DS-Prop}
\end{equation}
Second, the contribution from the two adjoint matters or from the two bi-fundamental matters, satisfying the R-charge conditions given in the first lines of \eqref{eq:SYM-Rcond} or \eqref{eq:fABJM-Rcond}, simplifies to the form $\left(2\cosh\right)^{-1}$, thanks to the following property of the double sine function
\begin{equation}
\frac{s_{b}\left(z+\frac{i}{2}b^{\pm1}\right)}{s_{b}\left(z-\frac{i}{2}b^{\pm1}\right)}=\frac{1}{2\cosh\left(\pi b^{\pm1}z\right)}\ .\label{eq:DS-Cosh}
\end{equation}
Third, the contributions can be reorganized into matrix factors associated with the brane configuration \cite{Assel:2014awa,Kubo:2025jxi}, assuming the R-charge constraints discussed in the previous remarks. Specifically, we introduce a matrix factor for each $\left(p,q\right)$ web, allowing the matrix model to be expressed in terms of these factors. Another advantage of introducing these matrix factors is that they are naturally compatible with the Fermi gas formalism, as we will see below. Concretely, for the typical $\left(p,q\right)$ web depicted in figure \ref{fig:BW-gen}, the matrix factor is given by~\footnote{Note that the contributions from monopoles with non-trivial bare R-charges will be incorporated directly into the matrix model.}
\begin{align}
\mathcal{Z}_{F_{+},F_{-}}^{\left(1,q\right)}\left(D;\mu,\nu\right) & 
=\mathcal{Z}^{\left(1,q\right)}\left(D;\mu,\nu\right)
\mathcal{Z}_{\mathrm{D5}^{+}}\left(D;\mu,\nu\right)^{F_{+}}
\mathcal{Z}_{\mathrm{D5}^{-}}\left(D;\mu,\nu\right)^{F_{-}}\ .
\label{eq:1qD5-MF}
\end{align}
The matrix factor $\mathcal{Z}^{\left(1,q\right)}$ corresponds to the $\left(1,q\right)$5-branes,
\begin{equation}
\mathcal{Z}^{\left(1,q\right)}\left(D;\mu,\nu\right)=\frac{1}{\left(2\pi\right)^{N}}e^{\frac{iq}{4\pi}\sum_{a=1}^{N}\left(\mu_{a}^{2}-\nu_{a}^{2}\right)}\frac{\prod_{a<b}^{N}2\sinh\frac{\mu_{a}-\mu_{b}}{2}\prod_{a<b}^{N}2\sinh\frac{\nu_{a}-\nu_{b}}{2}}{\prod_{a,b}^{N}2\cosh\frac{\mu_{a}-\nu_{b}+2\pi iD}{2}}\ .
\label{eq:1q-MF}
\end{equation}
The matrix factor $\mathcal{Z}_{\mathrm{D5}^{\pm}}$ corresponds to the D5-branes extending in the directions $\pm x^{9}$,
\begin{align}
\mathcal{Z}_{\mathrm{D5^{+}}}\left(D;\mu,\nu\right) & =e^{\frac{i}{8\pi}\sum_{a=1}^{N}\left(\mu_{a}^{2}-\nu_{a}^{2}\right)}\frac{\prod_{a=1}^{N}s_{1}\left(\frac{\nu_{a}}{2\pi}+\left(\frac{1}{4}-\frac{D}{2}\right)i\right)}{\prod_{a=1}^{N}s_{1}\left(\frac{\mu_{a}}{2\pi}-\left(\frac{1}{4}-\frac{D}{2}\right)i\right)}\ ,\nonumber \\
\mathcal{Z}_{\mathrm{D5}^{-}}\left(D;\mu,\nu\right) & =e^{-\frac{i}{8\pi}\sum_{a=1}^{N}\left(\mu_{a}^{2}-\nu_{a}^{2}\right)}\frac{\prod_{a=1}^{N}s_{1}\left(\frac{\mu_{a}}{2\pi}+\left(\frac{1}{4}+\frac{D}{2}\right)i\right)}{\prod_{a=1}^{N}s_{1}\left(\frac{\nu_{a}}{2\pi}-\left(\frac{1}{4}+\frac{D}{2}\right)i\right)}\ .
\label{eq:D5on1q-MF}
\end{align}

Also, the matrix factor $\mathcal{Z}^{\left(0,1\right)}$ corresponds to the isolated D5-branes,
\begin{equation}
\mathcal{Z}^{\left(0,1\right)}\left(\mu\right)=\frac{1}{\prod_{a=1}^{N}2\cosh\frac{\mu_{a}}{2}}\ .
\label{eq:D5-MF}
\end{equation}
We include one such factor in the integrand for
each isolated D5-brane.

\subsection{Fermi gas formalism}\label{subsec:FGF}

In this section, we demonstrate how to apply the Fermi gas formalism to matrix models. Additionally, we show that the inverse of the density matrix takes the form of the quantum curve.

Before discussing the computation of the quantum curve, we summarize the notation used in quantum mechanics. In this paper, $\hat{x}$ and $\hat{p}$ denote the position and momentum operators, respectively. Their commutation relation is given by $\left[\hat{x},\hat{p}\right] = i\hbar$. The ket $\ket{\cdot}$ represents a position eigenvector, while we also introduce the symbol $\kket{\cdot}$ to denote a momentum eigenvector. The inner products of these vectors are normalized as
\begin{align}
 & \braket{x_{1}|x_{2}}=\delta\left(x_{1}-x_{2}\right)\ ,\qquad\bbrakket{p_{1}|p_{2}}=\delta\left(p_{1}-p_{2}\right)\ ,\nonumber \\
 & \brakket{x|p}=\frac{1}{\sqrt{2\pi\hbar}}e^{\frac{i}{\hbar}xp}\ ,\qquad\bbraket{p|x}=\frac{1}{\sqrt{2\pi\hbar}}e^{-\frac{i}{\hbar}xp}\ .
 \label{eq:Normalization}
\end{align}
The following formulas are useful:
\begin{align}
 & e^{-\frac{ia}{\hbar}\hat{x}}f\left(\hat{p}\right)e^{\frac{ia}{\hbar}\hat{x}}=f\left(\hat{p}+a\right)\ ,\qquad e^{-\frac{ia}{\hbar}\hat{p}}f\left(\hat{x}\right)e^{\frac{ia}{\hbar}\hat{p}}=f\left(\hat{x}-a\right)\ ,\nonumber \\
 & e^{-\frac{ia}{2\hbar}\hat{x}^{2}}f\left(\hat{p}\right)e^{\frac{ia}{2\hbar}\hat{x}^{2}}=f\left(\hat{p}+a\hat{x}\right)\ .
 \label{eq:OpSim}
\end{align}
The Baker-Campbell-Hausdorff formula for the position and the momentum operators is given by
\begin{equation}
e^{c_{1}\hat{x}}e^{c_{2}\hat{p}}=e^{\frac{i}{2}c_{1}c_{2}\hbar}e^{c_{1}\hat{x}+c_{2}\hat{p}}=e^{ic_{1}c_{2}\hbar}e^{c_{2}\hat{p}}e^{c_{1}\hat{x}}\ .
\label{eq:CBHform}
\end{equation}

Next, we discuss the Fermi gas formalism. Although this formalism cannot always be applied, in \cite{Kubo:2025jxi} it was explicitly employed to analyze the class of models where the associated brane configurations consist of brane webs of the type shown in figure \ref{fig:BW-gen} (with isolated D5-branes). We perform a similar computation. The Fermi gas formalism can be applied to the matrix factors individually. In particular, we examine the matrix factor given in \eqref{eq:1qD5-MF}. First, we utilize the Cauchy determinant formula \cite{Kapustin:2010xq,Marino:2011eh}.
\begin{equation}
\frac{\prod_{a<b}^{N}2\sinh\frac{\mu_{a}-\mu_{b}}{2}\prod_{a<b}^{N}2\sinh\frac{\nu_{a}-\nu_{b}}{2}}{\prod_{a,b}^{N}2\cosh\frac{\mu_{a}-\nu_{b}}{2}}=\det\left(\left[\frac{1}{2\cosh\frac{\mu_{a}-\nu_{b}}{2}}\right]_{a,b}^{N\times N}\right)\ .
\label{eq:1Loop-Det}
\end{equation}
By applying this formula to the matrix factor ${\cal Z}^{\left(1,q\right)}\left(D;\mu,\nu\right)$ in \eqref{eq:1q-MF}, we find
\begin{equation}
\mathcal{Z}^{\left(1,q\right)}\left(D;\mu,\nu\right)=\frac{1}{\left(2\pi\right)^{N}}e^{\frac{iq}{4\pi}\sum_{a=1}^{N}\left(\mu_{a}^{2}-\nu_{a}^{2}\right)}\det\left(\left[\frac{1}{2\cosh\frac{\mu_{a}-\nu_{b}+2\pi iD}{2}}\right]_{a,b}^{N\times N}\right)\ .
\end{equation}
We then apply the Fourier transform,
\begin{equation}
\frac{1}{2\cosh\frac{\mu-\nu+c}{2}}=\frac{1}{2\pi}\int dp\frac{e^{\frac{i}{2\pi}p\left(\mu-\nu+c\right)}}{2\cosh\frac{p}{2}}=2\pi\braket{\mu|\frac{e^{\frac{ic}{2\pi}\hat{p}}}{2\cosh\left(\frac{\pi}{\hbar}\hat{p}\right)}|\nu}\ ,
\label{eq:Cosh-op}
\end{equation}
such that
\begin{equation}
\mathcal{Z}^{\left(1,q\right)}\left(D;\mu,\nu\right)=\det\left(\left[\braket{\mu_{a}|\frac{e^{-D\left(\hat{p}-q\hat{x}\right)}}{2\cosh\frac{\hat{p}-q\hat{x}}{2}}|\nu_{b}}\right]_{a,b}^{N\times N}\right)\ ,
\end{equation}
where we fixed the commutation relation $\hbar=2\pi$ and used \eqref{eq:OpSim}. The matrix factor \eqref{eq:1qD5-MF} is
\begin{equation}
{\cal Z}_{F_{+},F_{-}}^{\left(1,q\right)}\left(D;\mu,\nu\right)
=\det\left(\left[\braket{\mu_{a}|\hat{\rho}_{F_{+},F_{-}}^{\left(1,q\right)}\left(D\right)|\nu_{b}}\right]_{a,b}^{N\times N}\right)\ ,
\end{equation}
where $\hat{\rho}_{F_{+},F_{-}}^{\left(1,q\right)}$ is the density matrix expressed as
\begin{equation}
\hat{\rho}_{F_{+},F_{-}}^{\left(1,q\right)}\left(D\right)
=\frac{s_{1}\left(\frac{\hat{x}}{2\pi}+\left(\frac{1}{4}+\frac{D}{2}\right)i\right)^{F_{-}}}{s_{1}\left(\frac{\hat{x}}{2\pi}-\left(\frac{1}{4}-\frac{D}{2}\right)i\right)^{F_{+}}}
\frac{e^{-D\left(\hat{p}-\left(q+\frac{1}{2}F_{+}-\frac{1}{2}F_{-}\right)\hat{x}\right)}}{2\cosh\frac{\hat{p}-\left(q+\frac{1}{2}F_{+}-\frac{1}{2}F_{-}\right)\hat{x}}{2}}
\frac{s_{1}\left(\frac{\hat{x}}{2\pi}+\left(\frac{1}{4}-\frac{D}{2}\right)i\right)^{F_{+}}}{s_{1}\left(\frac{\hat{x}}{2\pi}-\left(\frac{1}{4}+\frac{D}{2}\right)i\right)^{F_{-}}}\ .
\label{eq:DM-1q}
\end{equation}
The associated quantum curve is the inverse of this operator
\begin{equation}
\hat{\mathcal{O}}_{F_{+},F_{-}}^{\left(1,q\right)}\left(D\right)
=\hat{\rho}_{F_{+},F_{-}}^{\left(1,q\right)}\left(D\right)^{-1}\ .
\end{equation}
We use the relations,
\begin{align}
 & \frac{1}{s_{1}\left(\frac{\hat{x}}{2\pi}+\left(\frac{1}{4}-\frac{D}{2}\right)i\right)}e^{D\hat{p}}\left(e^{\frac{1}{2}\hat{p}}+e^{-\frac{1}{2}\hat{p}}\right)s_{1}\left(\frac{\hat{x}}{2\pi}-\left(\frac{1}{4}-\frac{D}{2}\right)i\right)\nonumber \\
 & ~~~ =e^{\frac{1}{2}\left(\frac{1}{2}+D\right)\hat{p}}\left(e^{\frac{1}{2}\hat{x}}+e^{-\frac{1}{2}\hat{x}}\right)e^{\frac{1}{2}\left(\frac{1}{2}+D\right)\hat{p}}+e^{\left(-\frac{1}{2}+D\right)\hat{p}}\ ,\nonumber \\
 & s_{1}\left(\frac{\hat{x}}{2\pi}-\left(\frac{1}{4}+\frac{D}{2}\right)i\right)e^{D\hat{p}}\left(e^{\frac{1}{2}\hat{p}}+e^{-\frac{1}{2}\hat{p}}\right)\frac{1}{s_{1}\left(\frac{\hat{x}}{2\pi}+\left(\frac{1}{4}+\frac{D}{2}\right)i\right)}\nonumber \\
 &~~~ =e^{\left(\frac{1}{2}+D\right)\hat{p}}+e^{\frac{1}{2}\left(-\frac{1}{2}+D\right)\hat{p}}\left(e^{\frac{1}{2}\hat{x}}+e^{-\frac{1}{2}\hat{x}}\right)e^{\frac{1}{2}\left(-\frac{1}{2}+D\right)\hat{p}}\ ,\label{eq:DS-sim}
\end{align}
where we used \eqref{eq:DS-Cosh} and \eqref{eq:OpSim}. Then, we find
\begin{equation}
{\cal Z}_{F_{+},F_{-}}^{\left(1,q\right)}\left(D;\mu,\nu\right)
=\det\left(\left[\braket{\mu_{a}|\hat{\mathcal{O}}_{F_{+},F_{-}}^{\left(1,q\right)}\left(D\right)^{-1}|\nu_{b}}\right]_{a,b}^{N\times N}\right)\ ,
\end{equation}
where
\begin{align}
\hat{\mathcal{O}}_{F_{+},F_{-}}^{\left(1,q\right)}\left(D\right) & =e^{\frac{1}{2}\left(\frac{1}{2}+D\right)\hat{p}}\left\{ e^{-\left(\frac{1}{2}+D\right)\left(q+\frac{1}{2}F_{+}-\frac{1}{2}F_{-}\right)\hat{x}}\left(e^{\frac{1}{2}\hat{x}}+e^{-\frac{1}{2}\hat{x}}\right)^{F_{+}}\right\} e^{\frac{1}{2}\left(\frac{1}{2}+D\right)\hat{p}}\nonumber \\
 & \quad+e^{-\frac{1}{2}\left(\frac{1}{2}-D\right)\hat{p}}\left\{ e^{\left(\frac{1}{2}-D\right)\left(q+\frac{1}{2}F_{+}-\frac{1}{2}F_{-}\right)\hat{x}}\left(e^{\frac{1}{2}\hat{x}}+e^{-\frac{1}{2}\hat{x}}\right)^{F_{-}}\right\} e^{-\frac{1}{2}\left(\frac{1}{2}-D\right)\hat{p}}\ .
 \label{eq:QC-1q}
\end{align}

If the isolated-D5 matrix factor in \eqref{eq:D5-MF} is present, it can be absorbed into the matrix element, where it becomes the operator $\left(2\cosh\hat{x}/2\right)^{-1}$.
Hence, in terms of the quantum curve, the contribution corresponds to adding the following operator
\begin{equation}
\hat{\mathcal{O}}^{\left(0,1\right)}=e^{\frac{1}{2}\hat{x}}+e^{-\frac{1}{2}\hat{x}}\ .
\label{eq:QC-01}
\end{equation}

After applying the Fermi gas formalism to the matrix factors separately, we can glue them by using the following formula for arbitrary operators $\hat{A},\hat{B}$:
\begin{align}
 & \frac{1}{N!}\int\prod_{a=1}^{N}d\alpha_{a}\det\left(\left[\braket{\mu_{a}|\hat{A}|\alpha_{b}}\right]_{a,b}^{N\times N}\right)\det\left(\left[\braket{\alpha_{a}|\hat{B}|\nu_{b}}\right]_{a,b}^{N\times N}\right)\nonumber \\
 & =\det\left(\left[\braket{\mu_{a}|\hat{A}\hat{B}|\nu_{b}}\right]_{a,b}^{N\times N}\right)\ .
 \label{eq:GlueForm}
\end{align}
This means that the total quantum curve can be obtained simply by multiplying all the quantum curves for 5-brane webs.
We therefore obtain \eqref{eq:SYM-QC} and \eqref{eq:fABJM-QC} for the flavored SYM theory and flavored $\ccz$ theory, respectively.

We have discussed the technical aspects of the dictionary relating brane configurations to quantum curves. 
In appendix \ref{sec:Poly-Web}, we review the more pictorial correspondence between them.

\section{Polygons and webs}\label{sec:Poly-Web}
\begin{figure}
\begin{centering}
\includegraphics[scale=0.14]{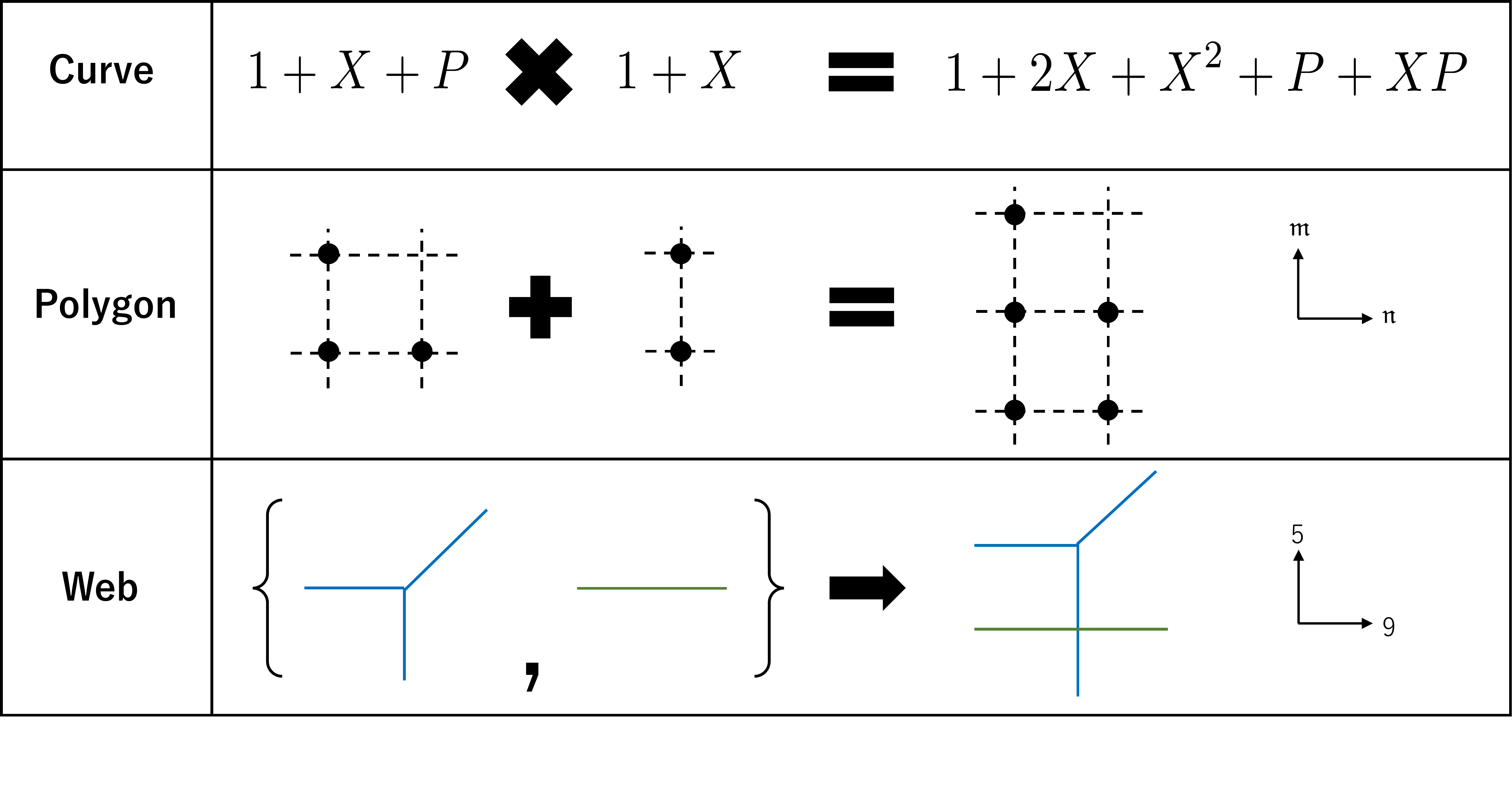}
\par\end{centering}
\caption{
An example of the correspondence among curves, Newton polygons, and webs. In each of the three columns, the curve, its Newton polygon, and the dual web correspond to one another. The rightmost column is obtained from the two left columns by multiplying the curves, taking the Minkowski sum of the polygons, and combining the webs, respectively. Thus, the figure illustrates that the correspondence among curves, polygons, and webs is compatible with these respective operations.
}\label{fig:Brane-TD}
\end{figure}

In this appendix, we provide the necessary details regarding convex lattice polygons (hereafter simply polygons) and webs, which are concepts used throughout this paper.
Throughout this section, please refer to figure \ref{fig:Brane-TD} as an illustrative example.

In this paper, quantum curves, toric diagrams, and brane constructions appear as key elements. 
Among these, toric diagrams can be viewed directly as polygons, while brane
constructions are represented by collections of brane webs. 
Furthermore, quantum curves are related to polygons, as discussed in detail in section \ref{subsec:QC-NP}.

An intriguing and fundamental property is the duality between quantum curves or toric diagrams and brane webs \cite{Cremonesi:2010ae,Kubo:2025jxi}. 
In this context, duality is defined as follows.
As previously noted, quantum curves and toric diagrams are represented by polygons, while brane constructions are represented by webs.
Then, the duality can be stated in terms of them.
Namely, polygons and webs are dual in the sense that the surfaces of the web correspond to the vertices of the polygon, and the asymptotic directions of the web correspond to the transverse directions of the polygon.~\footnote{
In contrast to the ordinary duality, here we are not concerned with whether the web has additional internal surfaces or not.
} 

In the following, we provide a summary of the concepts and definitions related to polygons and webs that are employed in this paper.

\subsection{Quantum curves and Newton polygons}\label{subsec:QC-NP}

In this subsection, we explain the relation between the curves and polygons.
The key idea is the concept called the Newton polygon.
The Newton polygon can be depicted by putting dots on the $(\mathfrak{m}, \mathfrak{n})$ plane at positions where the coefficients of the quantum curve $c_{\mathfrak{m}, \mathfrak{n}} \neq 0$.

This concludes the basic explanation of the Newton polygon. However, we provide several additional remarks relevant to this paper below.
For concrete examples, please refer to figure \ref{fig:Brane-TD}.

In general, the coordinates $(\mathfrak{m}, \mathfrak{n})$ are not integers. 
In particular, they take continuous values under R-charge deformations. 
However, we do not want to distinguish between theories that differ only by their R-charge assignments. 
Therefore, in this paper, we regard Newton polygons that differ only by a translation in the $(\mathfrak{m}, \mathfrak{n})$-plane as being identical. 
Note that the Newton polygon also does not encode detailed information about the coefficients $c_{\mathfrak{m}, \mathfrak{n}}$ of the curve.
Note also that since the relative differences between the points within a single Newton polygon are integers, it can still be thought of as a polygon residing on a 2d integer lattice in this sense.

We often neglect the strictly interior points (where "strictly" means that points on the edges are excluded).
In fact, these interior points do not affect the calculation of the coefficients $C$ and $B$ of the Airy function. 
In contrast, the vertices and points on the edges play a crucial role both in the computation of $C$ and $B$ and in the duality with the brane web.

When considering the duality between Newton polygons and 5-brane webs, the horizontal and vertical directions correspond to $\mathfrak{n}$ and $\mathfrak{m}$, respectively.
This convention differs from the usual coordinate notation in the sense that we use $(p,x)$ rather than $(x,p)$.

For quantum curves consisting of only two terms, the corresponding Newton polygon becomes a line.
However, in this paper, we still employ the term Newton polygon even in such cases.
Note that this type of quantum curve corresponds to a single $(p,q)$ 5-brane (as the middle example in figure \ref{fig:Brane-TD}).

\subsection{Minkowski sum, quantum curves and dual webs}\label{subsec:Web-Poly}

In this appendix, we briefly explain the realization of the Minkowski sum in the quantum-curve and dual-web pictures.
Although the statements in this appendix are not proved rigorously, they can be understood intuitively from the example in figure \ref{fig:Brane-TD}.

Let us first briefly look at the Minkowski sum. 
This is an operation that generates a single polygon from two polygons.
Let $P_{1}$ and $P_{2}$ be convex lattice polygons in $\mathbb{Z}^{2}$. Their Minkowski sum is
\begin{equation}
P_{1}+P_{2}=\left\{ p_{1}+p_{2}\in\mathbb{Z}^{2}|p_{1}\in P_{1},p_{2}\in P_{2}\right\}\ ,
\end{equation}
which is again a convex lattice polygon, see the middle of figure \ref{fig:Brane-TD}.

Next, let $P_{1}$ and $P_{2}$ be convex lattice polygons. 
We introduce curves whose Newton polygons are $P_{1}$ and $P_{2}$:
\begin{equation}
f_{i}\left(x,p\right)=\sum_{(\mathfrak{m},\mathfrak{n})\in P_{i}}e^{\mathfrak{m}x+\mathfrak{n}p}\ .
\label{eq:NP-Curve}
\end{equation}
Then, one finds that the Minkowski sum corresponds to the product of the curves. 
Namely, if $P_{12}$ is the Newton polygon of the curve $f_{1}f_{2}$, then $P_{1}+P_{2}=P_{12}$,  see the top of figure \ref{fig:Brane-TD}.
Remark that since we regard Newton polygons that differ only by a translation in the $(\mathfrak{m}, \mathfrak{n})$-plane as being identical as explained above, the curves in \eqref{eq:NP-Curve} are not completely specified.
Nevertheless, the statement is independent of the choice of the curve since $P_{12}$ also does not encode the overall exponents of $e^{x}$ and $e^{p}$.

Lastly, we observe that the Minkowski sum corresponds to bringing the two associated webs together. 
Namely, let $P_{1}$ and $P_{2}$ be convex lattice polygons and let $\mathsf{w}_{1}$ and $\mathsf{w}_{2}$ be the dual webs. 
Let $P_{12}:= P_{1}+P_{2}$ and let $\mathsf{w}_{12}$ be the collection of $\mathsf{w}_{1}$ and $\mathsf{w}_{2}$. 
Then, $P_{12}$ and $\mathsf{w}_{12}$ are again dual to each other, see the bottom of figure \ref{fig:Brane-TD}.
In the main text, this corresponds to the brane configuration 
\begin{equation}
\mathsf{w}_{1}-\mathsf{w}_{2}-_{\mathrm{p}}\ ,
\end{equation}
as \eqref{eq:fSYM-BC} or \eqref{eq:fABJM-BC}.

\section{Wigner transform}\label{sec:WT}

In this appendix we review the Wigner transform. We keep the commutation relation general, $\left[\hat{x},\hat{p}\right]=i\hbar$.

The Wigner transform of an operator $\hat{A}\left(\hat{x},\hat{p}\right)$ is given by
\begin{equation}
A_{{\rm W}}\left(x,p\right)=\int dy\braket{x-\frac{y}{2}|\hat{A}\left(\hat{x},\hat{p}\right)|x+\frac{y}{2}}e^{\frac{i}{\hbar}py}\ .
\label{eq:WT-Def}
\end{equation}
The Wigner transform of a product is given by
\begin{equation}
\left(\hat{A}\hat{B}\right)_{{\rm W}}=A_{{\rm W}}\star B_{{\rm W}}\ ,
\end{equation}
where the Moyal product $\star$ is defined as
\begin{equation}
\star=\exp\left[\frac{i\hbar}{2}\left(\overleftarrow{\partial}_{x}\overrightarrow{\partial}_{p}-\overleftarrow{\partial}_{p}\overrightarrow{\partial}_{x}\right)\right]\ .
\end{equation}
An important motivation for introducing the Wigner transform is the fact that the trace of an operator $\hat{A}$ is obtained by integrating $A_{{\rm W}}$ over the phase space
\begin{equation}
{\rm tr}\hat{A}\left(\hat{x},\hat{p}\right)=\int\frac{dxdp}{2\pi\hbar}A_{{\rm W}}\left(x,p\right)\ .
\label{eq:Tr-Wig-Form}
\end{equation}

In the main section we need to obtain the small $\hbar$ expansion of a function of an operator $\hat{f}\left(\hat{A}\left(\hat{x},\hat{p}\right)\right)$ up to the subleading order, $O\left(\hbar^{2}\right)$. $\hat{f}\left(\hat{A}\left(\hat{x},\hat{p}\right)\right)$ can be expanded around $A_{{\rm W}}\left(x,p\right)$ as
\begin{equation}
\hat{f}\left(\hat{A}\right)=\sum_{r=0}^{\infty}\frac{1}{r!}f^{\left(r\right)}\left(A_{{\rm W}}\right)\left(\hat{A}-A_{{\rm W}}\left(x,p\right)\right)^{r}\ .
\end{equation}
Hence
\begin{equation}
f\left(\hat{A}\right)_{{\rm W}}=\sum_{r=0}^{\infty}\frac{1}{r!}f^{\left(r\right)}\left(A_{{\rm W}}\right){\cal F}_{r}\ ,
\label{eq:WT-Taylor}
\end{equation}
where
\begin{equation}
{\cal F}_{r}=\left[\left(\hat{A}-A_{{\rm W}}\left(x,p\right)\right)^{r}\right]_{{\rm W}}\ .
\end{equation}
It is known that \cite{Marino:2011eh},
\begin{align}
\mathcal{F}_{0} & =1\ ,
\mathcal{F}_{1}=0\ ,\nonumber \\
\mathcal{F}_{2} & =-\frac{\hbar^{2}}{4}\left[\frac{\partial^{2}A_{{\rm W}}}{\partial x^{2}}\frac{\partial^{2}A_{{\rm W}}}{\partial p^{2}}-\left(\frac{\partial^{2}A_{{\rm W}}}{\partial x\partial p}\right)^{2}\right]+O\left(\hbar^{4}\right)\ ,\nonumber \\
\mathcal{F}_{3} & =-\frac{\hbar^{2}}{4}\left[\left(\frac{\partial A_{{\rm W}}}{\partial x}\right)^{2}\frac{\partial^{2}A_{{\rm W}}}{\partial p^{2}}+\left(\frac{\partial A_{{\rm W}}}{\partial p}\right)^{2}\frac{\partial^{2}A_{{\rm W}}}{\partial x^{2}}-2\frac{\partial A_{{\rm W}}}{\partial x}\frac{\partial A_{{\rm W}}}{\partial p}\frac{\partial^{2}A_{{\rm W}}}{\partial x\partial p}\right]+O\left(\hbar^{4}\right)\ ,\nonumber \\
{\cal F}_{r} & =0+O\left(\hbar^{4}\right)\ ,\quad\left(r\geq4\right)\ .
\label{eq:F0123r}
\end{align}

We can show that the Wigner transform of the quantum curve is exactly given by the classical version of itself. Namely, the Wigner transform of the quantum curve \eqref{eq:QCform} is~\footnote{We assume that $c_{m,n}$ are independent of $\hbar$.}
\begin{equation}
\mathcal{O}_{\mathrm{W}}\left(x,p\right)=\sum_{\mathfrak{m},\mathfrak{n}}c_{\mathfrak{m}\mathfrak{n}}e^{\mathfrak{m}x+\mathfrak{n}p}\ .
\label{eq:QC-W}
\end{equation}
Note that this clearly means the Newton polygon is also the same. This elegant property was pointed out in \cite{Hatsuda:2015oaa}, and we follow the discussion there. To prove \eqref{eq:QC-W}, it is sufficient to show that
\begin{equation}
\left(e^{\mathfrak{m}\hat{x}+\mathfrak{n}\hat{p}}\right)_{\mathrm{W}}\left(x,p\right)=e^{\mathfrak{m}x+\mathfrak{n}p}\ .
\end{equation}
From the definition \eqref{eq:WT-Def}, the left-hand side is given as
\begin{align}
\left(e^{\mathfrak{m}\hat{x}+\mathfrak{n}\hat{p}}\right)_{\mathrm{W}}\left(x,p\right) & =\int dy\braket{x-\frac{y}{2}|e^{\frac{1}{2}\mathfrak{m}\hat{x}}e^{\mathfrak{n}\hat{p}}e^{\frac{1}{2}\mathfrak{m}\hat{x}}|x+\frac{y}{2}}e^{\frac{i}{\hbar}py}\nonumber \\
 & =e^{\mathfrak{m}x}\int dy\braket{x-\frac{y}{2}|e^{\mathfrak{n}\hat{p}}|x+\frac{y}{2}}e^{\frac{i}{\hbar}py}\ .
\end{align}
By inserting the identity operator $1=\int dq\kket q\bbra q$, we finally find
\begin{equation}
\left(e^{\mathfrak{m}\hat{x}+\mathfrak{n}\hat{p}}\right)_{\mathrm{W}}\left(x,p\right)=e^{\mathfrak{m}x}\int\frac{dqdy}{2\pi\hbar}e^{\frac{i}{\hbar}\left(p-q\right)y}e^{\mathfrak{n}q}=e^{\mathfrak{m}x}e^{\mathfrak{n}p}\ .
\end{equation}
Here the normalization of the operator is given in \eqref{eq:Normalization}.

\section{An equivariant duality}
\label{app:An equivariant duality}

In this appendix, we comment in more detail on an aspect that was largely set aside in the main body of this work. As mentioned there, the conical spaces under consideration generally admit multiple resolutions, which are encoded in the different possible triangulations of their toric diagrams. In particular, triangulations in which every vertex of the toric diagram appears as the tip of at least one triangle correspond to geometric phases. These geometric phases are naturally associated with the holographically dual field theories, as they reproduce the correct coefficients $C$ and $B$ in the $S^3$ partition function, cf.\ \eqref{eq:S3pf}.

By contrast, non-geometric phases—corresponding to triangulations in which some vertices do not appear as tips of triangles—still reproduce the same leading coefficient $C$ via \eqref{eq:Cgeometry}, but generically yield different results for the subleading corrections encoded in $B$, see \eqref{eq:Bgeometry}. It therefore remains an open problem to identify the precise holographic field-theoretic interpretation of these non-geometric phases.

Nevertheless, we observe here an interesting duality which, in this case, holds exactly at the level of the full equivariant volume of two distinct manifolds. As illustrated in figure \ref{fig:localP2} and figure \ref{fig:localP1xP1}, toric manifolds containing a central lattice point admit triangulations into non-geometric phases in which this central point does not appear as the tip of any triangle. Due to the one-to-one correspondence between a triangulation and the associated equivariant volume expressed in terms of the effective $\nu$ variables (see section \ref{subsec:TCY-Loc}), it follows immediately that the corresponding equivariant volumes are exactly equal.

\begin{figure}
\centering
\begin{tikzpicture}[scale=2]
\begin{scope}[shift={(-4,0)}]
\begin{scope}
  \coordinate (v1) at (-1,-1);
  \coordinate (v2) at (1,0);
  \coordinate (v3) at (0,1);
  \coordinate (v4) at (0,0);
  \draw[thick] (v1) -- (v2) -- (v3) -- cycle;
  \draw[thick] (v2) -- (v4);
  \draw[thick] (v1) -- (v4);
\draw[thick] (v3) -- (v4);
  \foreach \p in {v1,v2,v3,v4}{\fill (\p) circle (1.5pt);}
  \node at (0.5,-1) {Geometric phase};
\end{scope}
\end{scope}

\begin{scope}
  \coordinate (v1) at (-1,-1);
  \coordinate (v2) at (1,0);
  \coordinate (v3) at (0,1);
  \coordinate (v4) at (0,0);
  \draw[thick] (v1) -- (v2) -- (v3) -- cycle;
   \foreach \p in {v1,v2,v3,v4}{\fill (\p) circle (1.5pt);}
  \node at (0.5,-1) {Non-geom. phase};
\end{scope}

\end{tikzpicture}
\caption{The toric diagram of local $\mathbb{P}^2$, together with its two possible triangulations. The latter one corresponds to a non-geometric phase formally matching the unique triangulation of $\mathbb{C}^3$.}
\label{fig:localP2}
\end{figure}
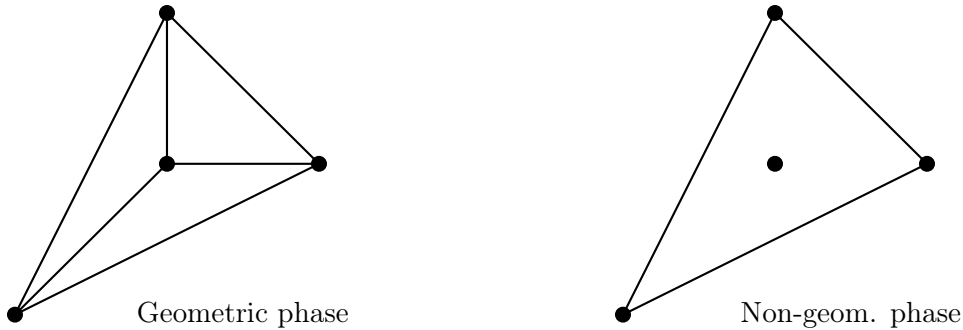

\begin{figure}
\centering
\begin{tikzpicture}[scale=2]
\begin{scope}[shift={(-2.2,0)}]
\begin{scope}
  \coordinate (v1) at (0,0);
  \coordinate (v2) at (1,0);
  \coordinate (v3) at (0,1);
  \coordinate (v4) at (1,1);
\coordinate (v5) at (0.5,0.5);
  \draw[thick] (v1) -- (v2) -- (v4) -- (v3) -- cycle;
  \draw[thick] (v2) -- (v3);
  \draw[thick] (v1) -- (v4);
  \foreach \p in {v1,v2,v3,v4,v5}{\fill (\p) circle (1.5pt);}
  \node at (0.5,-0.35) {Geometric phase};
\end{scope}
\end{scope}

\begin{scope}
  \coordinate (v1) at (0,0);
  \coordinate (v2) at (1,0);
  \coordinate (v3) at (0,1);
  \coordinate (v4) at (1,1);
\coordinate (v5) at (0.5,0.5);
  \draw[thick] (v1) -- (v2) -- (v4) -- (v3) -- cycle;
  \draw[thick] (v2) -- (v3);
   \foreach \p in {v1,v2,v3,v4,v5}{\fill (\p) circle (1.5pt);}
  \node at (0.5,-0.35) {Non-geom. phase $1$};
\end{scope}

\begin{scope}[shift={(2.2,0)}]
  \coordinate (v1) at (0,0);
  \coordinate (v2) at (1,0);
  \coordinate (v3) at (0,1);
  \coordinate (v4) at (1,1);
\coordinate (v5) at (0.5,0.5);
  \draw[thick] (v1) -- (v2) -- (v4) -- (v3) -- cycle;
  \draw[thick] (v1) -- (v4);
  \foreach \p in {v1,v2,v3,v4,v5}{\fill (\p) circle (1.5pt);}
  \node at (0.5,-0.35) {Non-geom. phase $2$};
\end{scope}

\end{tikzpicture}
\caption{The toric diagram of local $\mathbb{P}^1 \times \mathbb{P}^1$, together with its three possible resolutions. The latter two correspond to non-geometric phases formally matching the two triangulations of the conifold, cf.\ figure \ref{fig:conifold}.}
\label{fig:localP1xP1}
\end{figure}

In the two simplest examples depicted in the figures, it follows that
\be
	\BV_{\mathbb{C}^3} (\lam, \nu) = \BV_{\mathrm{loc.} \mathbb{P}^2}^\text{non-geom.} (\lam, \nu) = \frac{\mathe^{\sum_{i=1}^3 \lam^i \nu_i}}{\nu_1 \nu_2 \nu_3}\ ,
\ee
as well as
\be
	\BV_{\cC}^- (\lam, \nu) = \BV_{\mathrm{loc.} \mathbb{P}^1 \times  \mathbb{P}^1}^\text{non-geom. 1} (\lam, \nu)\ , \qquad \BV_{\cC}^+ (\lam, \nu) = \BV_{\mathrm{loc.} \mathbb{P}^1 \times  \mathbb{P}^1}^\text{non-geom. 2} (\lam, \nu)\ ,
\ee
cf.\ the calculations of the conifold in section \ref{subsec:TCY-Loc}.

More generally, there are exactly sixteen reflexive Newton polygons corresponding to toric three-folds with a single central point, see e.g.\ figure 3.1 in \cite{Gu:2017ppx} for the explicit form and geometric triangulations. These correspond to genus-one mirror curve manifolds, while their duals are clearly of genus-zero type, such that the duality can be written as
\be
	\BV_{\mathfrak{g}=0}^\text{geom.} (\lam, \nu) = \BV_{\mathfrak{g} = 1}^\text{non-geom.} (\lam, \nu)\ .
\ee 
Even more generally, we expect that this sort of equivariant duality always features manifolds of different genera mirror curves. It would be interesting to explore this type of duality in a more mathematically rigorous framework, and generalize the discussion to an arbitrary dimension of the CY manifold. We leave this for future work.

\bibliographystyle{JHEP}

\bibliography{References}

\end{document}